\documentclass[11pt,a4paper]{article}
\pdfoutput=1
\usepackage{jheppub}
\usepackage[T1]{fontenc}
\usepackage{fix-cm}
\usepackage{lmodern}
\usepackage{amsmath,amssymb,mathtools,mathrsfs,empheq}

\newcommand\tx{\tilde x}
\newcommand\tu{\tilde u}
\newcommand\mn{\mathfrak n}

\newcommand\tB{\tilde B}
\newcommand\tn{\tilde n}
\newcommand\tv{\tilde v}

\newcommand\mm{\mathfrak m}

\def\ri{{\rm i}}

\usepackage{tikz}

\usepackage{xcolor}

\newcommand{\mg}{\mathfrak{g}}

\newcommand\fft[2]{\frac{#1}{#2}}

\def\U{{\rm U}}

\usepackage{tensor}

\title{Large $N$ Partition Functions of the ABJM Theory}
\author[a]{Nikolay Bobev,}
\author[a]{Junho Hong,}
\author[a]{and Valentin Reys}

\affiliation[a]{Instituut voor Theoretische Fysica, KU Leuven, \\
	Celestijnenlaan 200D, B-3001 Leuven, Belgium}

\emailAdd{nikolay.bobev@kuleuven.be}
\emailAdd{junho.hong@kuleuven.be}
\emailAdd{valentin.reys@kuleuven.be}

\abstract{We study the large $N$ limit of some supersymmetric partition functions of the $\U(N)_{k}\times \U(N)_{-k}$ ABJM theory computed by supersymmetric localization. We conjecture an explicit expression, valid to all orders in the large $N$ limit, for the partition function on the $\U(1)\times \U(1)$ invariant squashed sphere in the presence of real masses in terms of an Airy function. Several non-trivial tests of this conjecture are presented. In addition, we derive an explicit compact expression for the topologically twisted index of the ABJM theory valid at fixed $k$  to all orders in the $1/N$ expansion. We use these results to derive the topologically twisted index and the sphere partition function in the 't Hooft limit which correspond to genus $\tt g$ type IIA string theory free energies to all orders in the $\alpha'$ expansion. We discuss the implications of our results for holography and the physics of AdS$_4$ black holes.}

\begin{document}
	
	\maketitle

\section{Introduction}
\label{sec:Introduction}
	
Supersymmetric localization provides a powerful tool for the exact calculation of the path-integral of strongly interacting supersymmetric QFTs. In the context of AdS/CFT one can apply supersymmetric localization to holographic CFTs in the large $N$ limit with great efficacy both to test the holographic duality and to uncover features of the dual gravitational dynamics.\footnote{See \cite{Pestun:2016zxk} for a review on supersymmetric localization and its many applications} Our goal in this work is to apply supersymmetric localization to the ABJM theory in the large $N$ limit and obtain results for its partition function on compact Euclidean manifolds that are valid to all orders in the $1/N$ expansion. Using a combination of analytic results and several consistency checks we conjecture a closed form expression for the partition functions of the theory on $S^3$ in presence of a $\U(1)\times \U(1)$ invariant squashing deformation together with three arbitrary real masses. In addition, we employ precise numerical tools to find a simple analytic formula for the partition function of the theory on $S^1\times \Sigma_{\mathfrak{g}}$ with a partial topological twist on the smooth genus-$\mathfrak{g}$ Riemann surface $\Sigma_{\mathfrak{g}}$, known as the topologically twisted index (TTI). The main motivation for our work comes from holography where the $1/N$ corrections to the leading $N^\fft32$ behavior of the ABJM partition function offer invaluable insights into the perturbative quantum gravity corrections to string and M-theory which in turn are of particular interest in the context of black hole physics in AdS.

The main character in our story is the ABJM theory~\cite{Aharony:2008ug}. This Chern-Simons-matter theory at level $k$ describes $N$ M2-branes on $\mathbb{C}^4/\mathbb{Z}_k$ and is therefore specified by the two parameters $(N,k)$. The fields that specify the Lagrangian of the theory are in the adjoint representation of one of the two U$(N)$ gauge groups or in the bifundamental representation of U$(N) \times \text{U}(N)$. They carry color indices and one can perform a 't Hooft expansion of the partition function in $1/N$. Since $k$ plays the role of the inverse gauge coupling, the 't~Hooft parameter is
\begin{equation}
	\lambda = \frac{N}{k} \, .
\end{equation}
At large $N$, ABJM is dual to the M-theory background AdS$_4 \times S^7/\mathbb{Z}_k$, where the sphere orbifold is freely acting. The radius $L$ common to the AdS and $S^7/\mathbb{Z}_k$ factors in this Freund-Rubin background is related to the gauge theory parameters as
\begin{equation}
	\label{eq:L-kN}
	\Bigl(\frac{L}{\ell_\mathrm{P}}\Bigr)^6 = 32\pi^2\,k\,N \, ,
\end{equation}
where $\ell_\mathrm{P}$ is the 11d Planck length. The 11d supergravity approximation to the M-theory description emerges in the limit where $k$ is held fixed and $N \rightarrow \infty$. We will call this limit the \textit{M-theory limit}.

The metric on $S^7/\mathbb{Z}_k$ can be written as a Hopf fibration over $\mathbb{CP}^3$ as
\begin{equation}
	\label{eq:S7-Hopf}
	ds^2_{S^7/\mathbb{Z}_k} = \frac1{k^2}\,\bigl(d\varphi + k\,\omega\bigr)^2 + ds^2_{\mathbb{CP}^3} \, ,
\end{equation}
where $\varphi$ is a $2\pi$-periodic coordinate and $d\omega = J$ with $J$ the K\"ahler form on $\mathbb{CP}^3$. Using~\eqref{eq:S7-Hopf}, one can perform a reduction from M-theory to type IIA string theory and obtain a background of the latter of the form AdS$_4 \times \mathbb{CP}^3$. To do perturbation theory in the type IIA frame we require the radius $L/k$ of the fibration circle to be small. Using~\eqref{eq:L-kN} this amounts to the condition $k \gg N^{1/5}$, and thus $k$ must be taken large as we take $N \rightarrow \infty$. The type IIA string coupling constant is given by
\begin{equation}
	\label{eq:gst-L}
	g_\mathrm{st}^2 = \frac{1}{k^2}\,\Bigl(\frac{L}{\ell_\mathrm{s}}\Bigr)^2 \, ,
\end{equation}
where $\ell_\mathrm{s}$ is the string length. The 't Hooft coupling can then be written in terms of the dimensionless ratio $L/\ell_\mathrm{s}$ as
\begin{equation}
	\label{eq:lambda-L}
	\lambda = \frac{1}{32\pi^2}\,\Bigl(\frac{L}{\ell_\mathrm{s}}\Bigr)^4 \, .
\end{equation}
Therefore, the perturbative regime in type IIA string theory corresponds to the limit where $\lambda$ is fixed and $N \rightarrow \infty$. We will call this limit the \textit{IIA limit}.

We apply results from supersymmetric localization to study two types of large $N$ partition functions of the ABJM theory. First we use the matrix model pioneered in  \cite{Kapustin:2009kz} to study the free energy of the theory on a squashed $S^3$ in the presence of three supersymmetry preserving real mass parameters. This partition function has been widely studied in the literature, see \cite{Marino:2016new} for a review. In the absence of squashing and real mass deformations in the  M-theory limit there is a closed form expression for the perturbative part of the free energy in terms of an Airy function. Building on various recent results in the literature we conjecture that there is a similar Airy function expression that captures the full perturbative result for the free energy in the $1/N$ expansion with general deformation parameters. In addition, we show how one can perform a judicial resummation of this large $N$ result in the M-theory limit to derive closed form expressions in the type IIA limit of the partition function valid at fixed order in the worldsheet genus expansion and to all orders in the perturbative $\alpha'$ expansion. The second partition function we study is the so-called topologically twisted index on $S^1\times \Sigma_{\mathfrak{g}}$ introduced in \cite{Benini:2015noa,Benini:2016hjo,Closset:2016arn}. The supersymmetric localization matrix model arising from this path-integral has so far evaded analytic treatment beyond the $N^\fft32$ leading order term in the large $N$ limit \cite{Benini:2015eyy}. We use this leading order result together with the Bethe Ansatz Equations arising from a residue calculation of the matrix integrals to set up a precise numerical algorithm that evaluates the TTI for large but finite values of $N$. Based on our numerical results we are able to deduce a closed form analytic expression for the ABJM TTI for general background flavor magnetic fluxes and real masses valid to all orders in the large $N$ expansion in the M-theory limit, which improves the previous numerical approach focused on the universal logarithmic correction \cite{Liu:2017vll}. We then use this analytic formula to arrive at analogous results in the type IIA limit. A short summary of our main results appeared in the prequel to this work \cite{Bobev:2022jte}. Our goal here is to provide further details on the derivation of these results accompanied by a more in-depth discussion of their interpretation. 

We continue in the next section with a summary of some previous results for the $S^3$ free energy of the ABJM theory as well as our conjecture for this partition function in the presence of a squashing deformation and general real masses. In Section~\ref{sec:expand} we study how these results can be expanded in two different large $N$ limits suitable for M-theory and type IIA string theory. In Section~\ref{sec:TTI} we shift our attention to the TTI and present a compact closed form expression for this partition function valid to all orders in the large $N$ expansion. We study the implications of our results for holography and AdS$_4$ black holes in Section~\ref{sec:holography} and finish with a discussion of some open problems in Section~\ref{sec:discussions}. The four appendices contain some technical details of our calculations as well as a summary of the numerical data that supports our results.

\section{Sphere partition functions: an Airy tale}
\label{sec:sphere}

The goal of this section is to study the partition function of the $\U(N)_{k}\times \U(N)_{-k}$  ABJM theory on $S^3$ in the M-theory limit of fixed Chern-Simons (CS) level $k$ and large $N$. In this limit, the perturbative part of the partition function on the round sphere is given by an Airy function to all orders in the $1/N$ expansion~\cite{Fuji:2011km,Marino:2011eh}. In addition, there are non-perturbative corrections of order $\mathcal{O}(e^{-\sqrt{kN}})$ and $\mathcal{O}(e^{-\sqrt{N/k}})$ that are not captured by the Airy function. These corrections are most conveniently described in the IIA frame, where the former type arises from D2-branes wrapping $\mathbb{RP}^3 \subset \mathbb{CP}^3$ while the latter type comes from fundamental strings wrapping $\mathbb{CP}^1 \subset \mathbb{CP}^3$, see e.g.~\cite{Hatsuda:2012dt}. Both effects contribute at the same non-perturbative order at large $N$ in the M-theory limit when $k$ is held fixed.

The round $S^3$ partition function can be deformed in two natural ways. First, one can keep the geometry to be that of the round $S^3$ and add three real masses associated with the Cartan generators of the flavor symmetry group. When one of the real masses vanishes, it was shown in \cite{Nosaka:2015iiw} that the perturbative part of the partition function is again given by an Airy function. Second, one can deform the metric on the round $S^3$ to a squashed sphere with $\U(1)\times \U(1)$ isometry while still preserving part of the supersymmetry. The metric on the squashed sphere is specified by the following embedding in $\mathbb{R}^4$ with coordinates $(x_1,x_2,x_3,x_4)$,
\begin{equation}
	\label{eq:squash-sphere}
	\omega_1^2(x_1^2+x_2^2)+\omega_2^2(x_3^2+x_4^2)=1\,.
\end{equation}
The corresponding partition function then depends on the squashing parameter $b^2 = \omega_1/\omega_2$ and should be symmetric under the exchange $\omega_1 \leftrightarrow \omega_2$ or equivalently $b \leftrightarrow 1/b$. So far, a perturbatively exact expression for the partition function of ABJM theory on such squashed spheres has only been obtained when $k=1$ and $b^2=3$, in which case~\cite{Hatsuda:2016uqa} showed using topological string theory that it is again given by an Airy function. Below, we will argue that this form generalizes to arbitrary squashing parameter. We will present similar arguments regarding the general case where both mass and squashing deformations are turned on.

\subsection{Summary of known results}
\label{subsec:summary}

We begin by summarizing some known results and setting up the notation. We will say that the large $N$ partition function of a 3d $\mathcal{N} \geq 2$ SCFT at fixed $k$ is of ``\textit{Airy-type}''  if it takes the form
\begin{equation}
	\label{eq:Z}
	Z(N,k\,;\mathfrak{a}) = C(k,\mathfrak{a})^{-1/3}\,e^{\mathcal{A}(k,\mathfrak{a})}\,\text{Ai}(z) \, , \qquad z(N,k\,;\mathfrak{a}) = C(k,\mathfrak{a})^{-1/3}\bigl(N - B(k,\mathfrak{a})\bigr) \, ,
\end{equation}
where $\mathfrak{a}$ denotes a set of parameters that $Z$ can depend on, such as real masses or geometric deformations. In what follows, we parametrize the $k$-dependence of the functions $C$ and $B$ in~\eqref{eq:Z} as
\begin{equation}
	\label{eq:C-B}
	C(k,\mathfrak{a}) = \frac{\gamma(\mathfrak{a})}{k} \, , \qquad \text{and} \qquad B(k,\mathfrak{a}) = \frac{\alpha(\mathfrak{a})}{k} + \beta(\mathfrak{a})\,k\, .
\end{equation}
It has been established in the literature that a number of interesting partition functions are of Airy-type. We collect the relevant results below.\\

In~\cite{Fuji:2011km,Marino:2011eh}, the large $N$ partition function of the ABJM theory on the round $S^3$ was shown to be of the form~\eqref{eq:Z}, \eqref{eq:C-B} with the following parameters:
\begin{equation}
	\label{eq:c-a-b-ABJM-round}
	\gamma = \frac{2}{\pi^2} \, , \qquad \alpha = \frac13\, , \qquad \beta = \frac{1}{24} \, .
\end{equation}
Furthermore, the function $\mathcal{A}(k)$ entering \eqref{eq:Z} is given by\footnote{We replaced the factor of $(-1)^{n+1}$ with $(-1)^n$ in the last line of \eqref{eq:A-ABJM} from the previous arXiv version of this paper. In \cite{Marino:2016new} this replacement corresponds to the correction $(-1)^g \to (-1)^{g-1}$ in (2.60) required for a consistent comparison with the higher genus free energy (2.51). We would like to thank Matteo Beccaria for bringing this issue to our attention.}
\begin{equation}
	\begin{split}
		\label{eq:A-ABJM}
		\mathcal{A}(k) =&\; \frac{2\zeta(3)}{\pi^2 k}\Bigl(1 - \frac{k^3}{16}\Bigr) + \frac{k^2}{\pi^2}\int_0^{\infty}dx\,\frac{x}{e^{kx} - 1}\log\bigl(1 - e^{-2x}\bigr)\\[1mm]
		=&\; -\frac{\zeta(3)}{8\pi^2}\,k^2 + \frac12\log(2) + 2\,\zeta'(-1) + \frac16\log\Bigl(\frac{\pi}{2k}\Bigr) \\
		&\; + \sum_{n\geq2}\Bigl(\frac{2\pi}{k}\Bigr)^{2n-2}\,\frac{(-1)^{n}4^{n-1}|B_{2n}B_{2n-2}|}{n(2n-2)(2n-2)!} \, ,
	\end{split}
\end{equation}
where the closed form integral expression in the first line was first found in \cite{Hatsuda:2014vsa} and in the second line we have also displayed its large $k$ expansion in terms of the Bernoulli numbers $B_n$.

From an $\mathcal{N}=2$ perspective, ABJM theory has an $\mathrm{SO}(4) \times \mathrm{U}(1)$ flavor symmetry group.\footnote{For $k=1,2$ there is supersymmetry enhancement and the ABJM theory has $\mathcal{N}=8$ supersymmetry. This will not be essential in the discussion below.} As briefly mentionned above, one can deform the theory by turning on three real masses $\{m_1,m_2,m_3\}$, where $m_1$ corresponds to the Cartan of U(1) and $m_2,m_3$ to the Cartans of $\mathrm{SO}(4)$. When one of the latter vanishes, say $m_3=0$, it was shown in~\cite{Nosaka:2015iiw} (see also~\cite{Chester:2021gdw}) that the large $N$ partition function of the mass-deformed ABJM theory on the round $S^3$ is of Airy-type with
\begin{equation}
	\label{eq:c-a-b-ABJM-m}
	\gamma(m_\pm) = \frac{2}{\pi^2(1+m_+^2)(1+m_-^2)} \, , \quad \alpha(m_\pm) = \frac{2 - m_+^2 - m_-^2}{6(1+m_+^2)(1+m_-^2)} \, , \quad \beta = \frac{1}{24} \, ,
\end{equation}
where $m_\pm = m_2 \pm m_1$. In addition, the function $\mathcal{A}(k,m_\pm)$ entering \eqref{eq:Z} is related to the function $\mathcal{A}(k)$ in the undeformed theory~\eqref{eq:A-ABJM} according to
\begin{equation}
	\label{eq:A-ABJM-m}
	\mathcal{A}(k,m_\pm) = \frac{\mathcal{A}[k(1+ \mathrm{i}m_+)] + \mathcal{A}[k(1- \mathrm{i}m_+)] + \mathcal{A}[k(1+ \mathrm{i}m_-)] + \mathcal{A}[k(1- \mathrm{i}m_-)]}{4} \, .
\end{equation}
Observe that while $\gamma$ and $\alpha$ in \eqref{eq:c-a-b-ABJM-m} acquire a dependence on the mass parameters, $\beta$ does not.

Another case of interest concerns ABJM theory at CS level $k=1$ and on a squashed sphere specified by~\eqref{eq:squash-sphere} with squashing parameter $b^2 = 3$. In~\cite{Hatsuda:2016uqa}, the corresponding large $N$ partition function was shown to be of Airy-type with
\begin{equation}
	\label{eq:c-a-b-ABJM-b-3}
	\gamma(b)\big\vert_{b^2=3} = \frac{9}{8\pi^2} \, , \qquad \alpha(b)\big\vert_{b^2=3} + \beta(b)\big\vert_{b^2=3} = \frac{1}{8} \, ,
\end{equation}
and
\begin{equation}
	\label{eq:A-ABJM-b-3}
	\mathcal{A}(k,b)\big\vert_{k=1,b^2=3} = -\frac{\zeta(3)}{3\pi^2} + \frac16\log(3) \approx 0.1425041054 \, .
\end{equation}

In view of these results, it is natural to ask whether other large $N$ partition functions of the ABJM theory on $S^3$ are of Airy-type. In the following, we will argue that this is indeed the case for general mass and squashing deformations. We will first state our conjecture for the perturbative part of the partition functions and then provide evidence for its validity.

\subsection{ABJM theory on the squashed sphere}
\label{subsec:Airyb}

We conjecture that the large $N$ partition function of the ABJM theory on the squashed $S_b^3$ is of Airy-type~\eqref{eq:Z} for any fixed CS level $k$ and squashing parameter $b$ with
\begin{equation}
	\label{eq:c-a-b-ABJM-b}
	\gamma(b) = \frac{32}{\pi^2}\,\Bigl(b + \frac{1}{b}\Bigr)^{-4} \, , \qquad \alpha(b) = -\frac23\,\Bigl(b^2 - 4 + \frac{1}{b^2}\Bigr)\Bigl(b + \frac{1}{b}\Bigr)^{-2} \, , \qquad \beta = \frac1{24} \, ,
\end{equation}
where we have determined the coefficients $\alpha,\beta,\gamma$ based on the first two leading terms of order $N^\fft32$ and $N^\fft12$ recently determined in \cite{Bobev:2021oku}. In analogy with the mass-deformed case~\eqref{eq:A-ABJM-m}, it is then natural to expect that the function $\mathcal{A}(k,b)$ entering~\eqref{eq:Z} is related to the function $\mathcal{A}(k)$ in~\eqref{eq:A-ABJM} as
\begin{equation}
	\label{eq:A-ABJM-b}
	\mathcal{A}(k,b) =  \frac{\mathcal{A}[k(1+ \mathrm{i}b_+)] + \mathcal{A}[k(1- \mathrm{i}b_+)] + \mathcal{A}[k(1+ \mathrm{i}b_-)] + \mathcal{A}[k(1- \mathrm{i}b_-)]}{4} \, , 
\end{equation}
where we have defined
\begin{equation}
	\label{eq:b-pm}
	b_\pm = \frac{1}{2\sqrt{2}}\Bigl(b - \frac{1}{b}\Bigr)\sqrt{b^2 +\frac{1}{b^2} \pm \sqrt{b^4 + 14 + \frac{1}{b^4}}} \, .
\end{equation}
It is worth distinguishing, however, that \eqref{eq:A-ABJM-b} may not be the unique choice for the $N$-independent function $\mathcal A(k,b)$ unlike the coefficients $\alpha,\beta,\gamma$ that are completely fixed as \eqref{eq:c-a-b-ABJM-b} by applying the Airy structure \eqref{eq:Z} to the $S^3_b$ partition function with the known large $N$ expansion \cite{Bobev:2021oku}.

Our proposal is manifestly symmetric under $b \leftrightarrow 1/b$ and agrees with~\eqref{eq:c-a-b-ABJM-b-3} when $k=1$ and $b^2=3$. The comparison between~\eqref{eq:A-ABJM-b} and~\eqref{eq:A-ABJM-b-3} is somewhat subtle since we must use the integral representation of the function $\mathcal{A}(k)$ in~\eqref{eq:A-ABJM} and analytically continue it into the complex $k$-plane. Leaving aside potential issues with this continuation, we obtain the following numerical estimate from our proposal~\eqref{eq:A-ABJM-b} at $k=1$ and $b^2=3$,
\begin{equation} 
\mathcal{A}(k,b)\big\vert_{k=1,b^2=3} \approx 0.13119908407 \, .
\end{equation}
This is not in a perfect agreement with the numerical value in~\eqref{eq:A-ABJM-b-3} and different by about eight percent. This implies either a non-trivial issue in the analytic continuation of the $\mathcal A(k)$ function over a complex $k$ or that the proposal \eqref{eq:A-ABJM-b} is incomplete and therefore does not capture the $b$ dependence precisely. We hope to return to this issue in the future.

Besides being compatible with the results of~\cite{Hatsuda:2016uqa}, additional evidence for our Airy conjecture comes from an exact relation satisfied by the free energy of mass-deformed ABJM theory on the squashed sphere $F(b,m_i) = -\log Z(b,m_i)$ for specific values of the parameters. Indeed, the authors of~\cite{Chester:2021gdw} (see also \cite{Minahan:2021pfv} for related work) used methods inspired by the Fermi gas approach to show that the free energy on the squashed sphere $S^3_b$ with $m_3 = \mathrm{i}\tfrac{b - b^{-1}}{2}$ must be related to the free energy on the round sphere $S^3$ with two mass deformations turned on as follows:
\begin{equation}
	\label{eq:Shai-rel}
	F\Bigl(b,m_1,m_2,\mathrm{i}\frac{b-b^{-1}}{2}\Bigr) = F\Bigl(1,\frac{b^{-1}m_+ - b\,m_-}{2},\frac{b^{-1}m_+ + b\,m_-}{2},0\Bigr) \, ,
\end{equation}
with $m_\pm = m_2 \pm m_1$. Expanding both sides around the massless theory on the round sphere, this relation yields infinitely many constraints between mass and squashing derivatives at every order. Leveraging superconformal Ward identities, it was shown in~\cite{Chester:2021gdw} that the following equalities follow from~\eqref{eq:Shai-rel}:\footnote{Strictly speaking, the last two relations have been obtained with $\mathcal{N}=8$ supersymmetry but we found that our proposal satisfies them even in the $\mathcal N=6$ ABJM setting with a generic $k$. Note that it is still possible for non-perturbative corrections to break the last two relations in the $\mathcal N=6$ ABJM setting.}
\begin{align}
	\label{eq:m-b-rel}
	\partial_b^2 F =&\; 2\,\partial_{m_\pm}^2 F + 2\,\partial_{m_+}\partial_{m_-} F \, , \nonumber \\
	\partial_b^3 F =&\; -6\,\partial_{m_\pm}^2 F - 6\,\partial_{m_+}\partial_{m_-} F \, , \\
	\partial_b^4 F =&\; 78\,\partial_{m_\pm}^2 F + 10\,\partial_{m_\pm}^4 F - 18\,\partial_{m_+}^2\partial_{m_-}^2 F - 8\,\partial_{m_\pm}^3\partial_{m_\mp} F - 30\,\partial_{m_+}\partial_{m_-} F \, , \nonumber \\
	\partial_b^5 F =&\; - 660\,\partial_{m_\pm}^2 F - 100\,\partial_{m_\pm}^4 F + 180\,\partial_{m_+}^2\partial_{m_-}^2 F \, , \nonumber
\end{align}
where all derivatives are evaluated at $b=1$ and $m_\pm = 0$. The right-hand sides above can be computed using the result for the free energy of mass-deformed ABJM theory on the round sphere~\cite{Nosaka:2015iiw} summarized in~\eqref{eq:c-a-b-ABJM-m} and \eqref{eq:A-ABJM-m}. It is then straightforward to check that our proposal for the squashed sphere partition function indeed gives the correct result for the derivatives with respect to $b$, and thus satisfies all the relations~\eqref{eq:m-b-rel}. This confirmation also includes the proposed $b$-dependence of the function $\mathcal{A}(k,b)$ given in~\eqref{eq:A-ABJM-b}. We view this as non-trivial evidence supporting our Airy-type conjecture on the squashed sphere.

\subsection{Mass-deformed ABJM theory on the squashed sphere}
\label{sec:ABJM-m-gen}

We now consider general mass and squashing deformations of the ABJM theory at fixed CS level $k$. This means that we turn on three independent real masses and put the theory on a squashed sphere with arbitrary squashing parameter $b$. The partition function $Z(b,m_i)$ can be computed by supersymmetric localization and yields the following matrix model~(see for instance \cite{Willett:2016adv,Closset:2019hyt,Chester:2021gdw}),
\begin{align}
\label{eq:MM-b-m}
Z(b,m_i) =&\; \int\frac{d^N\mu d^N\nu}{(N!)^2} e^{\mathrm{i}\pi k \sum_i(\nu_i^2 - \mu_i^2)} \prod_{i>j}4\sinh[\pi b(\mu_i - \mu_j)]\sinh[\pi b^{-1}(\mu_i - \mu_j)] \nonumber\\
&\;\times \prod_{i>j}4\sinh[\pi b(\nu_i - \nu_j)]\sinh[\pi b^{-1}(\nu_i - \nu_j)] \\
&\;\times \prod_{i,j}\Bigl[s_b\Bigl(\frac{\mathrm{i}Q}{4} - \mu_j + \nu_i - \frac{m_1+m_2+m_3}{2}\Bigr)s_b\Bigl(\frac{\mathrm{i}Q}{4} - \mu_j + \nu_i - \frac{m_1-m_2-m_3}{2}\Bigr) \nonumber\\
&\qquad \times s_b\Bigl(\frac{\mathrm{i}Q}{4} + \mu_j - \nu_i + \frac{m_1+m_2-m_3}{2}\Bigr)s_b\Bigl(\frac{\mathrm{i}Q}{4} + \mu_j - \nu_i + \frac{m_1-m_2+m_3}{2}\Bigr)\Bigr] \nonumber\, ,
\end{align}
where $Q = b + b^{-1}$, the $\mu_i,\nu_i$ ($i=1,\ldots,N$) correspond to the Cartans of the gauge group, and the function $s_b(x)$ is the double sine function (see~\cite{Chester:2021gdw} for definitions and details). Observe that~\eqref{eq:MM-b-m} is symmetric under $b \leftrightarrow 1/b$ as expected. The large $N$ evaluation of this matrix model is a complicated task. In the limit of vanishing $m_3$ (or $m_2$) and on the round sphere where $b=1$, the Cauchy determinant formula can be used to simplify the integrand, ultimately giving the Airy-type result summarized in~\eqref{eq:c-a-b-ABJM-m} and \eqref{eq:A-ABJM-m}. As far as we know, no similar simplifications have been found for generic values of the parameters $(b,m_i)$.

Despite the lack of an analytic result for the matrix model~\eqref{eq:MM-b-m}, we conjecture that the large $N$ partition function $Z(b,m_i)$ is again of Airy-type~\eqref{eq:Z}. To write down the parameters $(\gamma,\alpha,\beta)$ it will be convenient to express the mass deformations $\{m_1,m_2,m_3\}$ in terms of four quantities $\Delta_a$ defined as:
\begin{equation}
	\begin{split}
		\Delta_1 =&\; \frac12 - \mathrm{i}\,\frac{m_1 + m_2 + m_3}{b + b^{-1}} \, , \qquad \Delta_2 = \frac12 - \mathrm{i}\,\frac{m_1 - m_2 - m_3}{b + b^{-1}} \, , \\
		\Delta_3 =&\; \frac12 + \mathrm{i}\,\frac{m_1 + m_2 - m_3}{b + b^{-1}} \, , \qquad \Delta_4 = \frac12 + \mathrm{i}\,\frac{m_1 - m_2 + m_3}{b + b^{-1}} \, .
	\end{split}
\end{equation}
Note that $\sum_a \Delta_a = 2$ so that we have three independent mass parameters. We then propose the following form for the functions $C$ and $B$ entering~\eqref{eq:Z}:
\begin{equation}
	\begin{split}
	\label{eq:c-a-b-ABJM-m-gen}
	\gamma(b,\Delta_a) =&\; \frac{2}{\pi^2\Delta_1\Delta_2\Delta_3\Delta_4}\,\Bigl(b + \frac{1}{b}\Bigr)^{-4} \, , \\[1mm]
	\alpha(b,\Delta_a) =&\; -\frac1{12}\sum_a\Delta_a^{-1} + \frac{1 - \frac14\,\sum_a\Delta_a^2}{3\,\Delta_1\Delta_2\Delta_3\Delta_4}\,\Bigl(b + \frac{1}{b}\Bigr)^{-2} \, , \qquad \beta = \frac{1}{24} \, . 
	\end{split}
\end{equation}
This proposal is easily seen to be consistent with all cases discussed previously upon taking various limits. First, setting $b=1$ and $m_{1,2,3} = 0$ in~\eqref{eq:c-a-b-ABJM-m-gen}, we recover~\eqref{eq:c-a-b-ABJM-round}. Setting $b=1$, $m_3 = 0$ and $m_\pm = m_2 \pm m_1$ gives back~\eqref{eq:c-a-b-ABJM-m}, and we recover~\eqref{eq:c-a-b-ABJM-b} by setting $m_{1,2,3} = 0$. 

Currently, we do not have a conjecture for the analytic form of the function $\mathcal{A}(k,b,\Delta_a)$ for generic values of the parameters. We note however that it should reduce to the corresponding $\mathcal{A}(k)$, $\mathcal{A}(k,m_\pm)$ and $\mathcal{A}(k,b)$ functions encountered previously in the above limits. Another remark is that while the quantities $(\gamma,\alpha,\beta)$ are symmetric under full permutations of the $\Delta_a$, the function $\mathcal{A}(k,b,\Delta_a)$ is expected to break this symmetry. This is apparent from the matrix model~\eqref{eq:MM-b-m} which is only symmetric under $\Delta_1 \leftrightarrow \Delta_2$, $\Delta_3 \leftrightarrow \Delta_4$ and $(\Delta_1,\Delta_2) \leftrightarrow (\Delta_3,\Delta_4)$. This breaking of the full permutation symmetry was also emphasized in~\cite{Chester:2021gdw}.

As in the previous subsection, we can make use of the exact relation~\eqref{eq:Shai-rel} to provide evidence for the conjecture. Upon setting $m_3 = \mathrm{i}\frac{b - b^{-1}}{2}$ in~\eqref{eq:c-a-b-ABJM-m-gen}, we obtain
\begin{equation}
\gamma = \frac{2}{\pi^2(1 + b^{-2}m_+^2)(1 + b^2 m_-^2)} \, , \qquad \alpha = \frac{2 - b^{-2} m_+^2 - b^2 m_-^2}{6(1+b^{-2}m_+^2)(1 + b^2 m_-^2)} \, ,
\end{equation}
which are precisely the parameters $\gamma(b^{\mp 1}m_\pm)$ and $\alpha(b^{\mp 1}m_\pm)$ in~\eqref{eq:c-a-b-ABJM-m} for the mass-deformed theory on the round sphere at $m_3 = 0$. Thus, our proposal satisfies the relation derived in~\cite{Chester:2021gdw} provided that
\begin{equation}\label{eq:mathcalAsqmass}
\mathcal{A}\Bigl(k,b,m_1,m_2,\mathrm{i}\frac{b-b^{-1}}{2}\Bigr) = \mathcal{A}(k,b^{\mp 1}m_\pm) \, ,
\end{equation}
where the right-hand side is given by~\eqref{eq:A-ABJM-m}. While we cannot check \eqref{eq:mathcalAsqmass} explicitly at the moment, our proposal still passes an infinite number of consistency checks at each order in the $1/N$ expansion beyond $\mathcal{O}(N^0)$. A similar ``bootstrapping'' reasoning for the functions $C$ and $B$ using~\eqref{eq:Shai-rel} was used recently in~\cite{Hristov:2022lcw} (which appeared simultaneously with~\cite{Bobev:2022jte}) to arrive at the result~\eqref{eq:c-a-b-ABJM-m-gen}, although the function $\mathcal{A}$ was not discussed in~\cite{Hristov:2022lcw}. It would be most interesting to have access to the function $\mathcal{A}(k,b,\Delta_a)$ analytically in order to fully characterize the partition function of mass-deformed ABJM theory on the squashed sphere. In this context, our Airy conjecture~\eqref{eq:c-a-b-ABJM-m-gen} can help shed new light on the matrix model~\eqref{eq:MM-b-m} for general values of the parameters $(b,m_i)$, from which the function $\mathcal{A}$ could be extracted. \\

Our conjecture also has important implications for the study of the dynamics of ABJM theory, as it allows one to consider integrated correlators of flavor currents and/or stress tensors over $S^3$. As a simple example, consider the coefficient $\mathcal{C}_T$ of the stress tensor two-point function,
\begin{equation}
\langle T_{\mu\nu}(\vec{x})T_{\rho\sigma}(0)\rangle = \frac{\mathcal{C}_T}{64\pi}\bigl(P_{\mu\rho}P_{\nu\sigma} + P_{\nu\rho}P_{\mu\sigma} - P_{\mu\nu}P_{\rho\sigma}\bigr)\frac{1}{16\pi^2\vec{x}{\,}^2} \, , \quad P_{\mu\nu} = \eta_{\mu\nu}\nabla^2 - \partial_\mu\partial_\nu \, .
\end{equation}
This coefficient can be computed by taking two derivatives of the squashed-sphere free energy with respect to the parameter $b$, as follows from $\mathcal{N}=2$ Ward identities~\cite{Closset:2012ru}:
\begin{equation}
	\label{eq:cT-b-rel}
	\mathcal{C}_T = \frac{32}{\pi^2}\,\partial^2_b F(b,m_i)\big\vert_{b=1,m_i=0} \, .
\end{equation}
Using our Airy conjecture to compute the derivatives, we obtain the following result
\begin{equation}
	\label{eq:cT-ABJM}
	\mathcal{C}_T = \frac{32k^2}{\pi^2}\mathcal{A}''(k) - \frac{128}{3\pi^2} + \frac{8\cdot2^{\frac23}(k^2 - 24k N - 28)}{9k^\frac23\pi^\frac43}\,\frac{\text{Ai}'(f_{k,N})}{\text{Ai}(f_{k,N})} \, ,
\end{equation}
where
\begin{equation}
	f_{k,N} = -\frac{\pi^\frac23}{24\cdot2^\frac13}\,\frac{k^2 -24 k N + 8}{k^\frac23} \, ,
\end{equation}
and $\mathcal{A}(k)$ is given in~\eqref{eq:A-ABJM}. This matches the result obtained previously in~\cite{Chester:2020jay}, which is a consequence of the first relation in~\eqref{eq:m-b-rel}.

Taking more derivatives of the free energy with respect to mass and squashing parameters, our conjecture yields explicit expressions for integrated correlators of flavor currents and/or stress tensors. By taking an appropriate flat space limit, one can then use these results to study scattering amplitudes in M-theory~\cite{Chester:2018aca}. This approach has proven to be particularly powerful to probe the details of the effective action of M-theory (see e.g.~\cite{Binder:2018yvd}) and we hope that our conjecture can provide new results for correlators in the ABJM theory that can serve as a starting point for such studies.\\

Having presented our conjecture for the perturbative part of the partition function of mass-deformed ABJM theory on the squashed sphere, we now proceed with a discussion of its large $N$ expansion, both in the M-theory and the IIA limits.

\section{Two expansions for Airy-type partition functions}
\label{sec:expand}

In this section, we discuss two types of large $N$ expansions for the free energy associated to an Airy-type partition function. The first type is the M-theory expansion at fixed CS level $k$ and large $N$. Using the results presented above, we can obtain explicit expressions for the leading and subleading terms in the $1/N$ expansion of the free energy and (in some cases) compare them to results derived in the dual supergravity theory on AdS$_4$. The second type is the IIA expansion where we expand the free energy at fixed $\lambda = N/k$ and large $N$, which corresponds to the double expansion of Type IIA string theory in the string coupling $g_{\mathrm{st}}$ and in $\alpha'$.

\subsection{M-theory expansion}
\label{sec:M-expand}

Given a partition function of Airy-type, we can expand the free energy
\begin{equation}
	\label{eq:F}
	F(N,k,\mathfrak{a}) = -\log Z(N,k,\mathfrak{a}) \, ,
\end{equation}
at fixed $k$ and large $N$ using the known asymptotic expansion of the Airy function for large argument:
\begin{equation}
	\label{eq:Ai-expand}
	\text{Ai}(z) = \frac{\text{exp}\bigl[-\frac23\,z^{3/2}\bigr]}{2\sqrt{\pi}\,z^{1/4}}\,\sum_{n = 0}^{+\infty}\,\left(-\tfrac32\right)^n\,u_n\,z^{-3n/2} \, .
\end{equation}
In this expansion, the coefficients $u_n$ are numbers given recursively by $u_0 = 1$ and 
\begin{equation}
	\label{eq:u}
	u_n = \frac{(6n-5)(6n-3)(6n-1)}{216(2n-1)n}\,u_{n-1}\,, \qquad \forall \;\; n \geq 1 \, .
\end{equation}
It is then straightforward to expand the partition function of the ABJM theory on the squashed sphere in the M-theory limit to arbitrarily high order using our proposal~\eqref{eq:c-a-b-ABJM-b}-\eqref{eq:A-ABJM-b}. The result up to $\mathcal{O}(N^0)$ reads
\begin{equation}
	\label{eq:F-Mexp-b}
	F(N,k,b) = \frac{\pi\sqrt{2k}}{12}(b + b^{-1})^2N^{\frac32} - \frac{\pi\sqrt{2k}}{12}\Bigl(\frac{k^2 - 16}{16k}(b+b^{-1})^2 + \frac{6}{k}\Bigr)N^{\frac12} + \frac14\log N \, .
\end{equation}
The leading $\mathcal{O}(N^{\frac32})$ term has been obtained previously from a computation of the regularized on-shell action of the two-derivative gauged supergravity solution dual to the ABJM theory on the squashed sphere~\cite{Martelli:2011fu}. It was further shown in~\cite{Bobev:2020egg,Bobev:2021oku} that the subleading $\mathcal{O}(N^{\frac12})$ term arises from four-derivative corrections to the bulk supergravity action. The result~\eqref{eq:F-Mexp-b} also implies that the $\log N$ term is independent of the squashing parameter $b$. This term was derived at $b=1$ from a one-loop computation in 11d supergravity in~\cite{Bhattacharyya:2012ye}. This shows that our Airy conjecture for the partition function of ABJM theory on the squashed sphere is consistent with known holographic results in the dual supergravity theory up to order $\mathcal{O}(N^0)$. The subleading corrections to~\eqref{eq:F-Mexp-b} are readily obtained from~\eqref{eq:c-a-b-ABJM-b} and~\eqref{eq:Ai-expand}, and it would be interesting to derive them from similar bulk considerations. We will comment further on this in Section~\ref{sec:discussions}.\\

By using~\eqref{eq:Ai-expand}, we can also expand the partition function of the mass-deformed ABJM theory on the round sphere in the M-theory limit using~\eqref{eq:c-a-b-ABJM-m-gen} and setting $b=1$. Up to $\mathcal{O}(N^0)$, we obtain
\begin{equation}
	\begin{split}
	\label{eq:delta-conj}
	F(N,k,\Delta) =&\; \frac{\pi(16\,N^{\frac32} - k\,N^{\frac12})}{12}\sqrt{2k\Delta_1\Delta_2\Delta_3\Delta_4} \\
	&\; + \frac{\pi\,N^{\frac12}}{12\sqrt{2k\Delta_1\Delta_2\Delta_3\Delta_4}}\left({\textstyle\sum_a} \Delta_a^2 - \bigl({\textstyle\sum_a} \Delta_a\bigr)^2 + 8\,\frac{\Delta_1\Delta_2\Delta_3\Delta_4}{\sum_a \Delta_a}\,{\textstyle\sum_a}\Delta_a^{-1}\right) \\[1mm]
	&\; + \frac14\log N \, ,
	\end{split}
\end{equation}
where we have used the constraint $\sum_a \Delta_a = 2$ to group the $\mathcal{O}(N^\frac32)$ and $\mathcal{O}(N^\frac12)$ terms together with homogeneous functions of degree two and zero in the $\Delta_a$, respectively. Observe that the leading term in the large $N$ limit matches the two-derivative regularized on-shell action studied in~\cite{Freedman:2013oja}. In this form, we can follow the reasoning explained in~\cite{Bobev:2021oku} (see Section 7.2 there) and compare~\eqref{eq:delta-conj} to the action of Euclidean AdS$_4$ where the 4d $\mathcal{N}=2$ gravity multiplet is put on-shell while keeping the three vector multiplets dual to the real mass deformations off-shell. This ``partially off-shell'' action is governed by the prepotential of the gauged supergravity theory, which for the STU model at the two-derivative level is
\begin{equation}
F^{\text{STU}}_{2\partial}(X) = \sqrt{X^0X^1X^2X^3}\,,
\end{equation}
where $X^I$ with $I=0,1,2,3$ are the scalars of the vector multiplets (including the conformal compensator $X^0$). Our result~\eqref{eq:delta-conj} then predicts a higher-derivative correction to the STU prepotential, 
\begin{equation}
	\label{eq:4der-STU}
	F^{\text{STU}}_{4\partial}(X) = \frac{(\sum_I X^I)^2 - \sum_I (X^I)^2}{8\sqrt{\prod_I X^I}} - \sqrt{\textstyle{\prod_I X^I}}\,\frac{\sum_I (X^I)^{-1}}{\sum_I X^I} \, ,
\end{equation}
which corrects a previous prediction made in~\cite{Bobev:2021oku}. Going further, we can expand the Airy-type partition function to arbitrarily high order in $1/N$. Doing so, we observe that the free energy always contains a homogeneous term of degree two $\sqrt{\Delta_1\Delta_2\Delta_3\Delta_4}$, whose coefficient only contains monomials of the form $k^p N^q$ with $p+q = 2$. As explained in~\cite{Bobev:2021oku}, this coefficient should be interpreted as the \emph{renormalized} dimensionless ratio $L^2/(2G_N)$ in the bulk gauged supergravity theory. Collecting the relevant terms, we find
\begin{equation}
	\begin{split}
	\frac{L^2}{2G_N} =&\; \frac{\sqrt{2k}}{3}\Bigl(N^{\tfrac32} - \frac{k}{16}N^{\tfrac12} + \frac{k^2}{1536}N^{-\tfrac12} + \frac{k^3}{221184}N^{-\tfrac32} + \frac{k^4}{14155776}N^{-\tfrac52} \\
	&\qquad\;\; + \frac{k^5}{679477248}N^{-\frac72} + \frac{7k^6}{195689447424}N^{-\frac92}\Bigr) + \mathcal{O}(N^{-\tfrac{11}{2}}) \, .
	\end{split}
\end{equation}
It is remarkable that the above expression admits a simple resummation as
\begin{equation}
	\label{eq:L/G-all-order}
	\frac{L^2}{2G_N} = \frac{\sqrt{2k}}{3}\Bigl(N - \frac{k}{24}\Bigr)^{3/2} \, .
\end{equation}
Therefore, our conjecture for the partition function of the mass-deformed ABJM theory on the round sphere gives a prediction for the quantum corrected holographic dictionary relating the bulk quantity $L^2/(2G_N)$ to the field theory quantities $(N,k)$ to all orders in the $1/N$ expansion.\footnote{A similar expression has been put forward in~\cite{Hristov:2022lcw} using the same methods.}

Besides the term homogeneous of degree two, expanding the free energy~\eqref{eq:delta-conj} also produces terms of lower homogeneity degree in the $\Delta_a$. We find that it is possible to rearrange the large $N$ expansion of the Airy function in the following form:
\begin{align}
\label{eq:F-m-homog}
	F =&\; \frac{4\pi\sqrt{2k}}{3}\Bigl(N - \frac{k}{24}\Bigr)^{3/2}\sqrt{\Delta_1\Delta_2\Delta_3\Delta_4} - \mathcal{A}(k,1,\Delta_a) - \frac14\log (k\Delta_1\Delta_2\Delta_3\Delta_4) + \frac14\log 2 \nonumber\\[1mm]
	&\; + \sum_{n=0}^3 C_n\Bigl(N - \frac{k}{24}\Bigr)^{1/2-n}\frac{F_{4\partial}^{\text{STU}}(\Delta)^{n+1}}{(\Delta_1\Delta_2\Delta_3\Delta_4)^{n/2}} + \sum_{n=1}^3 D_n\Bigl(N-\frac{k}{24}\Bigr)^{-1/2-n}\frac{F_{4\partial}^{\text{STU}}(\Delta)^{n-1}}{(\Delta_1\Delta_2\Delta_3\Delta_4)^{n/2}} \nonumber\\[1mm]
	&\;  + \frac{1105}{147456\pi^3\sqrt{2k}}\frac{1}{k}\Bigl(N - \frac{k}{24}\Bigr)^{-9/2}\frac{1}{(\Delta_1\Delta_2\Delta_3\Delta_4)^{3/2}} \nonumber\\[1mm]
	&\; + \frac14\log\Bigl(N - \frac{k}{24}\Bigr) - \frac{1}{24k}\Bigl(N - \frac{k}{24}\Bigr)^{-1}\frac{F_{4\partial}^{\text{STU}}(\Delta)}{\sqrt{\Delta_1\Delta_2\Delta_3\Delta_4}} \\[1mm]
	&\; - \frac{1}{288k^2}\Bigl(N - \frac{k}{24}\Bigr)^{-2}\frac{F_{4\partial}^{\text{STU}}(\Delta)^2}{\Delta_1\Delta_2\Delta_3\Delta_4}  - \frac{5}{512\pi^2 k}\Bigl(N - \frac{k}{24}\Bigr)^{-3}\frac{1}{\Delta_1\Delta_2\Delta_3\Delta_4} \nonumber\\[1mm]
	&\; - \frac{1}{2592k^3}\Bigl(N - \frac{k}{24}\Bigr)^{-3}\frac{F_{4\partial}^{\text{STU}}(\Delta)^3}{(\Delta_1\Delta_2\Delta_3\Delta_4)^{3/2}} - \frac{5}{1024\pi^2 k^2}\Bigl(N - \frac{k}{24}\Bigr)^{-4}\frac{F_{4\partial}^{\text{STU}}(\Delta)}{(\Delta_1\Delta_2\Delta_3\Delta_4)^{3/2}} + \ldots \, , \nonumber
\end{align}
where the function $F^{\text{STU}}_{4\partial}$ is given in~\eqref{eq:4der-STU} and we have defined\footnote{With some pattern recognition, this suggests $C_n = \frac{(2n-3)!!}{(n+1)!(12k)^n}\,\frac{2\pi}{3\sqrt{2k}}$ for all $n \geq 0$.}
\begin{equation}
	C_0 = -\frac{2\pi}{3\sqrt{2k}} \, , \;\;\; C_1 = \frac{2\pi}{72\sqrt{2k}}\frac{1}{k} \, , \;\;\; C_2 = \frac{2\pi}{2592\sqrt{2k}}\frac{1}{k^2} \, , \;\;\; C_3 = \frac{2\pi}{41472\sqrt{2k}}\frac{1}{k^3} \, ,
\end{equation}
and
\begin{equation}
	D_1 = \frac{5}{96\pi\sqrt{2k}} \, , \quad D_2 = \frac{5}{384\pi\sqrt{2k}}\frac{1}{k} \, , \quad D_3 = \frac{25}{9216\pi\sqrt{2k}}\frac{1}{k^2} \, .
\end{equation}
The expression~\eqref{eq:F-m-homog} contains all terms of homogeneity degree $2$, $0$, $-2$, $-4$ and $-6$ in the $\Delta_a$, to all orders in the $1/N$ expansion. From a field theory perspective, it seems a priori unnatural to expand the free energy $F(N,k,\Delta_a)$ according to the homogeneity of the functions of the mass deformation parameters $\Delta_a$. But from the bulk perspective, this formulation is well-suited to read off the quantities entering the gauged supergravity prepotential. In particular, using~\eqref{eq:L/G-all-order} and the map $\{\Delta_1,\Delta_2,\Delta_3,\Delta_4\} = \{X^0,X^1,X^2,X^3\}$,~\eqref{eq:F-m-homog} suggests that the partially on-shell action of the Euclidean AdS$_4$ solution takes the form\footnote{Note that one should be careful about the map between the supergravity scalars in the STU model and the real masses in the ABJM theory. The two-derivative results in \cite{Freedman:2013oja} show that one has to carefully evaluate the supergravity on-shell action, taking into account finite boundary counterterms as well as alternate quantization, in order to derive the correct holographic relation.}
\begin{equation}
	\begin{split}
	\label{eq:I-os}
	I_{\text{EAdS}_4}^{\text{STU}} =&\; \frac{2\pi L^2}{G_N}F_{2\partial}^{\text{STU}}(X) - \frac{2\pi}{3\sqrt{2k}}\Bigl(N - \frac{k}{24}\Bigr)^{1/2}F_{4\partial}^{\text{STU}}(X) \\[1mm]
	&\; +\frac16\log\Bigl(\frac{L^2}{G_N}\Bigr) - \mathcal{A}(k,1,X) - \frac14\sum_{I=0}^3\log X^I - \frac{5}{12}\log k + \frac16\log 3 \\[1mm]
	&\; + \frac{\Bigl(N - \frac{k}{24}\Bigr)^{-1/2}}{\sqrt{X^0X^1X^2X^3}}\Biggl[\frac{2\pi k^{-3/2}}{72\sqrt{2}}F_{4\partial}^{\text{STU}}(X)^2 - \frac{k^{-1}}{24}\Bigl(N - \frac{k}{24}\Bigr)^{-1/2}F_{4\partial}^{\text{STU}}(X) \\
	&\qquad\qquad\qquad\qquad\;\; + \frac{5k^{-1/2}}{96\pi\sqrt{2}}\Bigl(N - \frac{k}{24}\Bigr)^{-1}\Biggr] + \ldots \, ,
	\end{split}
\end{equation}
where we have focused on homogeneity degree $2$, $0$ and $-2$ for conciseness. From a Wilsonian effective action point of view, the four-derivative term on the first line comes with a renormalized coefficient, whose all-order expression in the $1/N$ expansion we were able to read off from~\eqref{eq:F-m-homog}. The second line contains inhomogeneous and constant terms which can usually be associated with loop effects in the bulk supergravity theory. The third line of~\eqref{eq:I-os} should come from six-derivative couplings in the gauged supergravity action, and it is clear that these couplings come in three different types with independent renormalized Wilsonian coefficients. Unfortunately, we currently lack an independent gravitational computation of the partially on-shell action including such higher-derivative terms, and so we cannot precisely disentangle the various contributions to the STU prepotential. It is however striking that the four-derivative correction $F_{4\partial}^{\text{STU}}$ seems to control six- and higher-derivative corrections. This may be a consequence of the high amount of supersymmetry preserved by the EAdS$_4 \times S^7/\mathbb{Z}_k$ solution of 11d supergravity. It would be most interesting to investigate the bulk consequences of our Airy conjecture along these lines in more detail, and understand the origin of the above higher-derivative corrections from M-theory.

We end this foray into the M-theory expansion of the ABJM theory free energy and its holographic implications by noting that we can study the mass-deformed theory on the squashed sphere along the same lines. In Appendix~\ref{app:squashed-prepot}, we present some of these results and show how our Airy conjecture is compatible with an earlier conjecture made in~\cite{Bobev:2021oku} regarding the four-derivative couplings in the dual gauged supergravity action. In particular, similarly to \eqref{eq:L/G-all-order}, we derive the relation between the 4-derivative supergravity coefficients $c_{1,2}$ defined in \cite{Bobev:2020egg,Bobev:2021oku} and the field theory parameters $(k,N)$ to all orders in the large $N$ expansion.

\subsection{Type IIA expansion}
\label{sec:IIA-expand}

We now study the Type IIA expansion of the free energy~\eqref{eq:F} where the 't~Hooft coupling $\lambda$ is kept fixed while taking $N \rightarrow \infty$. Generically, we write this expansion as
\begin{equation}
	\label{eq:IIA-expand}
	F(N,k,\mathfrak{a}) = -\sum_{{\tt g} \geq 0}\,(2\pi \mathrm{i})^{2{\tt g}-2}\,F_{\tt g}(\lambda,\mathfrak{a})\,\left(\frac{N}{\lambda}\right)^{2 - 2{\tt g}} \, .
\end{equation}
Using~\eqref{eq:gst-L}, this can be written as an expansion in the string coupling $g_{\mathrm{st}}$ whose coefficients are the genus-${\tt g}$ free energies with ${\tt g}$ the genus of the worldsheet. Each such coefficient can then be expanded in $\ell_s \sim \sqrt{\alpha'}$ using~\eqref{eq:lambda-L}.

Interestingly, for Airy-type partition functions of the form~\eqref{eq:Z}, the M-theory and IIA expansions can be related as follows. Using~\eqref{eq:Ai-expand}, we can obtain a useful rewriting of the free energy~\eqref{eq:F} in the M-theory limit. Some details are presented in Appendix~\ref{app:expand} and the result reads
\begin{equation}
	\label{eq:M-th-F}
	\begin{split}
		F(N,k,\mathfrak{a}) =&\; \sum_{n\geq0}\,\sum_{{\tt g}\geq0}\,\left[\mathcal{F}_{{\tt g},n}(\gamma,\alpha,\beta)\,\frac{2}{3\sqrt{\gamma}}\,\frac{(-\beta)^n}{n!}\,\prod_{r=0}^{n-1}\left(\frac{3}{2} - r\right)\right] N^{3/2 - n}k^{n + 1/2 - 2{\tt g}} \\[1mm]
		+&\; \sum_{n\geq1}\,\sum_{{\tt g}\geq0}\,\mathcal{G}_{{\tt g},n}(\gamma,\alpha,\beta)\,N^{-n}k^{n + 2 - 2{\tt g}} \\
		-&\; \mathcal{A}(k,\mathfrak{a}) + \frac14\log\left(\frac{16\pi^2\gamma}{k}\right) + \frac14\log N \, .
	\end{split}
\end{equation}
The definition of the various quantities entering~\eqref{eq:M-th-F} are as follows. The function controlling the $N^{3/2-n} k^{n+1/2 - 2{\tt g}}$ term in the expansion is given by\footnote{Here and below, we suppress the explicit $\mathfrak{a}$ dependence on the parameters $(\gamma,\alpha,\beta)$ entering~\eqref{eq:Z} to lighten the notation.}
\begin{equation}
	\label{eq:cal-F}
	\mathcal{F}_{{\tt g},n}(\gamma,\alpha,\beta) = \begin{cases} 1 &\text{for} \;\; {\tt g}=0 \\[1mm] n\,\alpha\,\beta^{-1} &\text{for} \;\; {\tt g}=1 \\[1mm] \sum_{m=0}^{\lfloor\frac{{\tt g}}{2}\rfloor}\,f^{(m)}_{{\tt g},n}(\gamma,\alpha,\beta) &\text{for} \;\; {\tt g} > 1 \, , \end{cases}
\end{equation}
where
\begin{equation}
	f^{(m)}_{{\tt g},n}(\gamma,\alpha,\beta) = \mathcal{P}^{(m)}(u)\,\frac{n(n-1)\ldots(n-{\tt g}+1-m)}{({\tt g}-2m)!}\,\gamma^m\,\alpha^{{\tt g} - 2m}\,\beta^{-{\tt g}-m} \, ,
\end{equation}
and $\mathcal{P}^{(m)}(u)$ is a set of polynomials of degree $2m - 1$ in the $u_n$ defined in~\eqref{eq:u}. For low values of $m$, they read
\begin{equation}
		\mathcal{P}^{(0)}(u) = 1 \, , \quad \mathcal{P}^{(1)}(u) = 6\,u_1 \, , \quad \mathcal{P}^{(2)}(u) = \frac{12}{35}\,\Bigl(u_1^3 - 3\,u_1\,u_2 + 3\,u_3\Bigr) \, .
\end{equation}
For higher $m$, the $\mathcal{P}^{(m)}$ can be obtained recursively and we give more values in Appendix~\ref{app:expand}. At present, we do not have a closed form expression or a generating function for these numbers, and we will come back to this point below. The function controlling the coefficient of $N^{-n}\,k^{n + 2 - 2{\tt g}}$ in~\eqref{eq:M-th-F} takes a similar form, namely
\begin{equation}
	\label{eq:cal-G}
	\mathcal{G}_{{\tt g},n}(\gamma,\alpha,\beta) = \begin{cases} 0 &\text{for} \;\; {\tt g}=0 \\[1mm] -\frac{\beta^n}{4n} &\text{for} \;\; {\tt g}=1 \\[1mm] \sum_{m=1}^{\lceil\frac{{\tt g}}{2}\rceil}\,g^{(m)}_{{\tt g},n}(\gamma,\alpha,\beta) &\text{for} \;\; {\tt g} > 1 \, , \end{cases}
\end{equation}
where
\begin{equation}
	g^{(m)}_{{\tt g},n}(\gamma,\alpha,\beta) = \mathcal{Q}^{(m)}(u)\,\frac{(n-1)\ldots(n-{\tt g}+3-m)}{({\tt g}-2m+1)!}\,\gamma^{m-1}\,\alpha^{{\tt g} - 2m+1}\,\beta^{n-{\tt g}+2-m} \, ,
\end{equation}
and $\mathcal{Q}^{(m)}(u)$ is a set of polynomials of degree $2m-2$ in the $u_n$. The first three are
\begin{equation}
	\begin{split}
		\mathcal{Q}^{(1)}(u) =&\; -\frac14 \, , \qquad \mathcal{Q}^{(2)}(u) = \frac{9}{16}\,\Bigl(u_1^2 - 2\,u_2\Bigr) \, , \\[1mm]
		\mathcal{Q}^{(3)}(u) =&\; \frac{27}{2560}\,\Bigl(u_1^4 - 4\,u_1^2\,u_2 + 4\,u_1\,u_3 + 2\,u_2^2 - 4\,u_4\Bigr) \, . \\[1mm]
	\end{split}
\end{equation}
Similarly to the constants $\mathcal{P}^{(m)}$, we currently do not have a closed form expression or a generating function for all $\mathcal{Q}^{(m)}$. We give more values at higher $m$ in Appendix~\ref{app:expand}.

The expansion~\eqref{eq:M-th-F} can be rewritten as a type IIA expansion of the form~\eqref{eq:IIA-expand} by expressing $k$ in terms of the 't Hooft parameter $\lambda$ and collecting the relevant powers of $N$. To do so, we will assume that the function $\mathcal{A}(k,\mathfrak{a})$ in~\eqref{eq:Z} has a large $k$ expansion,
\begin{equation}
	\label{eq:A-expand}
	\mathcal{A}(k,\mathfrak{a}) = \sum_{{\tt g}\geq0}\,\mathcal{A}_{\tt g}(\mathfrak{a})\,k^{2 - 2{\tt g}} + \widehat{\mathcal{A}}\,\log k \, .
\end{equation} 
Note that we do not include a dependence on $\mathfrak{a}$ in the $\log k$ term since the latter will be related to the $\log N$ term in the free energy (see below in~\eqref{eq:IIA-F}), which is expected to be universal~\cite{Bhattacharyya:2012ye,Liu:2017vbl,Liu:2017vll}. We stress that the assumption~\eqref{eq:A-expand} is satisfied in the undeformed and mass-deformed ABJM theory on the round three-sphere, as is clear from~\eqref{eq:A-ABJM} and~\eqref{eq:A-ABJM-m}. The expansion~\eqref{eq:M-th-F} can then be written as
\begin{align}
	\label{eq:IIA-F}
	F(N,\lambda,\mathfrak{a}) =&\; \sum_{{\tt g}\geq0}\,\Bigl[\,\sum_{n\geq0}\,\lambda^{-n - 1/2 + 2{\tt g}}\,\mathcal{F}_{{\tt g},n}\,\frac{2}{3\sqrt{\gamma}}\,\frac{(-\beta)^n}{n!}\,\prod_{r=0}^{n-1}\Bigl(\frac{3}{2} - r\Bigr)\Bigr] N^{2-2{\tt g}} \nonumber \\[1mm]
	+&\; \sum_{{\tt g}\geq0}\,\Bigl[\,\sum_{n\geq1} \lambda^{-n-2+2{\tt g}}\,\mathcal{G}_{{\tt g},n}\Bigr] N^{2-2{\tt g}} \\[1mm]
	-&\; \sum_{{\tt g}\geq0}\,\Bigl[\lambda^{-2+2{\tt g}}\,\mathcal{A}_{\tt g}\Bigr] N^{2-2{\tt g}} + \frac14\log(16\pi^2\gamma) + \Bigl(\widehat{\mathcal{A}} + \frac14\Bigr)\log\lambda - \widehat{\mathcal{A}}\,\log N \, . \nonumber
\end{align}
Comparing with~\eqref{eq:IIA-expand}, we read off the genus-${\tt g}$ free energies\footnote{We have slightly extended the definition~\eqref{eq:cal-G} to include the case $\mathcal{G}'_{{\tt g},0} = 0$ along with $\mathcal{G}'_{{\tt g},n\neq0} = \mathcal{G}_{{\tt g},n}$ to obtain a more compact expression.}
\begin{align}
	\label{eq:Fg}
	F_{\tt g}(\lambda,\mathfrak{a}) =&\; (2\pi \mathrm{i})^{2-2{\tt g}}\,\Bigl[\mathcal{A}_{\tt g} - \sum_{n\geq0}\,\Bigl\{\mathcal{F}_{{\tt g},n}\,\frac{2}{3\sqrt{\gamma}}\frac{(-\beta)^n}{n!}\,\prod_{r=0}^{n-1}\Bigl(\frac{3}{2} - r\Bigr) + \mathcal{G}'_{{\tt g},n}\,\lambda^{-3/2}\Bigr\}\,\lambda^{3/2 - n}\Bigr] \nonumber \\[1mm] 
	&\; - \delta_{{\tt g},1}\Bigl[\,\frac14\log(16\pi^2\gamma) + \Bigl(\widehat{\mathcal{A}} + \frac14\Bigr)\log\lambda\Bigr] \, .
\end{align}
It is important to note that this result is valid for any large $N$ partition function of Airy-type, provided the function $\mathcal{A}(k,\mathfrak{a})$ can be expanded as in \eqref{eq:A-expand}. Depending on the theory of interest, one must simply specify the set of parameters $(\gamma,\alpha,\beta)$ and $(\mathcal{A}_{\tt g},\widehat{\mathcal{A}})$ to be used in~\eqref{eq:Fg}. Furthermore, and quite remarkably, this series representation can be \emph{resummed} to yield simple functions of the 't Hooft coupling $\lambda$. In other words, we are able to resum the full series of $\alpha'$ corrections to the genus-${\tt g}$ free energies. We now illustrate this for the ABJM theory. Along the way, we will be able to compare our result to other approaches used in the literature.\\

Let us first review some known facts about the genus-${\tt g}$ free energies for ABJM theory on the round sphere. By carefully analyzing the large $N$ limit of the matrix model resulting from localization of the partition function, explicit expressions for some $F_{\tt g}(\lambda)$ of low genera have been obtained previously (see~\cite{Marino:2016new} for a review and references). The full non-perturbative genus-0 free energy takes the form
\begin{equation}
	\label{eq:F0-ABJM-Marino}
	F_0(\hat{\lambda}) = \frac{4\pi^3\sqrt{2}}{3}\,\hat{\lambda}^{3/2} + \frac{\zeta(3)}{2} + \sum_{\ell\geq 1} e^{-2\pi\ell\sqrt{2\hat{\lambda}}}\,f_\ell\left(\frac{1}{\pi\sqrt{2\hat{\lambda}}}\right) \, ,
\end{equation}
where $f_\ell$ is a polynomial of degree $2\ell - 3$ for $\ell \geq 2$, and we have introduced the shifted 't~Hooft coupling\footnote{This shift is naturally associated to an eight-derivative correction in the M-theory frame~\cite{Bergman:2009zh}, and we will comment further on this in Section~\ref{sec:discussions}.}
\begin{equation}
	\hat{\lambda} = \lambda - \frac{1}{24} \, .
\end{equation}
The series of exponentially suppressed terms in $F_0$ has the interpretation of coming from IIA worldsheet instantons wrapping the $\mathbb{CP}^1$ cycle inside $\mathbb{CP}^3$~\cite{Drukker:2010nc}. As a non-perturbative effect, they are not captured by~\eqref{eq:Z} and so we will discard them in what follows. Using topological string methods on local $\mathbb{P}^1 \times \mathbb{P}^1$~\cite{Huang:2006si,Drukker:2010nc}, the non-perturbative genus-1 free energy has been found to take the following form:
\begin{equation}
	\label{eq:F1-ABJM-Marino}
	F_1(\lambda) = -\log\eta(\tau-1) + \frac16\log\lambda + 2\,\zeta'(-1) + \frac16\log\frac{\pi}{2}\, ,
\end{equation}
where $\eta$ is the Dedekind function. Note that we have included the contribution from constant maps at genus one obtained in~\cite{Marino:2009dp} and confirmed numerically in~\cite{Hanada:2012si}. We have also discarded a constant $\fft16\log N$ term since it should be thought of as being part of the full partition function prior to the genus expansion. Lastly, the auxiliary function $\tau$ (which plays the role of the modular parameter for a family of elliptic curves encoding the mirror geometry of local $\mathbb{P}^1 \times \mathbb{P}^1$) is defined in terms of the elliptic integral of the first kind $K$ as
\begin{equation}
	\label{eq:tau-k}
	\tau = \mathrm{i}\,\frac{K'\Bigl(\frac{\mathrm{i}\kappa}{4}\Bigr)}{K\Bigl(\frac{\mathrm{i}\kappa}{4}\Bigr)} \, ,
\end{equation}
where $\kappa$ is related to the 't Hooft coupling via an inverse hypergeometric function,
\begin{equation}
	\label{eq:k-lambda}
	\lambda = \frac{\kappa}{8\pi}\,{}_3F_2\Bigl(\frac12,\,\frac12,\,\frac12;\,1,\,\frac32;\,-\frac{\kappa^2}{16}\Bigr) \, .
\end{equation}
The non-perturbative genus-2 free energy can also be given explicitly in terms of Eisenstein series and Jacobi theta functions upon solving the holomorphic anomaly equation (see again~\cite{Marino:2016new} for a review):
\begin{equation}
	\label{eq:F2-ABJM-Marino}
	F_2(\lambda) = \frac{1}{432\,t_4^2t_2}\,\left(-\frac{5E_2(\tau)^3}{3} + 3\,t_2\,E_2(\tau)^2 - 2\,E_4(\tau)E_2(\tau) + \frac{16\,t_2^3 + 15\,t_2^2t_4 - 15\,t_4^2t_2 + 2\,t_4^3}{30}\right) \, ,
\end{equation}
where
\begin{equation}
	t_2 = \vartheta_2(\tau)^4 \, , \quad t_4 = \vartheta_4(\tau)^4 \, .
\end{equation}
Evidently, the above results provide rather implicit functions of the 't Hooft coupling (aside from the genus-0 result). Moreover, while the higher-genus free energies can be obtained recursively, there is currently no closed form expression or generating function available in the literature. In contrast, our IIA expansion applied to ABJM theory on the round sphere allows us to derive the perturbative parts of the genus-${\tt g}$ free energies in a simple form.\\

\begin{figure}[t!]
	\centering
		\includegraphics[width=0.48\textwidth]{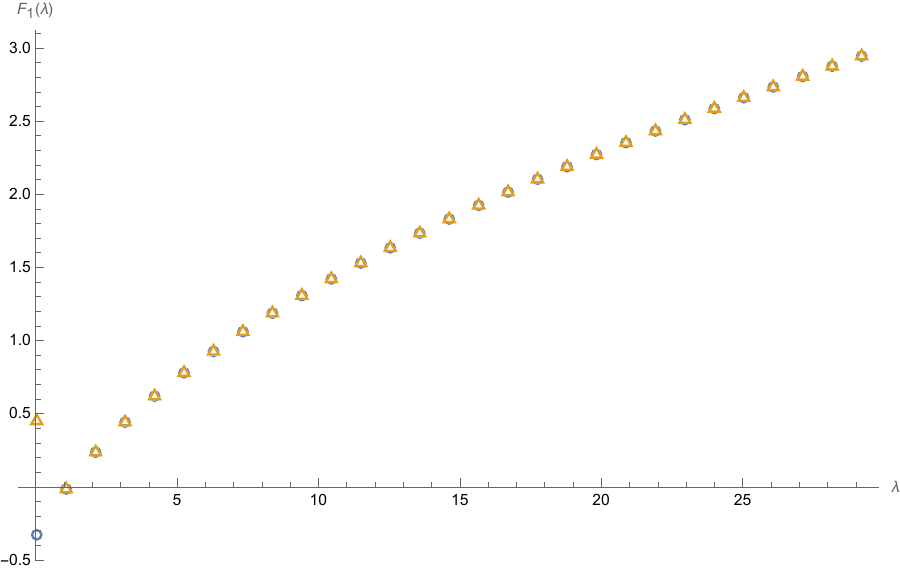}~~\includegraphics[width=0.48\textwidth]{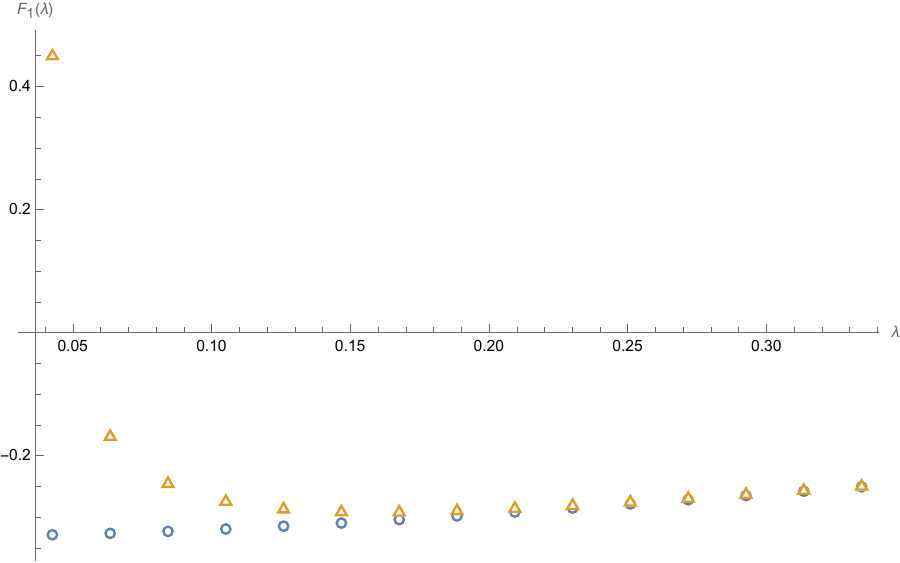}
		\caption{The genus-1 free energies~\eqref{eq:F1-ABJM-Marino} (blue circles) and~\eqref{eq:F1-ABJM-Us} (yellow triangles) for different values of the 't Hooft coupling: $\frac{1}{24} < \lambda \leq 30$ (left) and $\frac{1}{24} < \lambda \leq \frac13$ (right).}
		\label{fig:F1-compare}
\end{figure}

Using~\eqref{eq:c-a-b-ABJM-round} and~\eqref{eq:A-ABJM} in~\eqref{eq:Fg}, we obtain the genus-0 free energy
\begin{equation}
	\label{eq:F0-ABJM-Us}
	F_0(\hat{\lambda}) = \frac{4\pi^3\sqrt{2}}{3}\,\hat{\lambda}^{3/2} + \frac{\zeta(3)}{2} \, ,
\end{equation}
which precisely matches the perturbative part of~\eqref{eq:F0-ABJM-Marino}. At genus 1, we find a simple explicit function of the 't Hooft coupling,
\begin{equation}
	\label{eq:F1-ABJM-Us}
	F_1(\lambda) = \frac{\pi}{3\sqrt{2}}\,\hat{\lambda}^{1/2} - \frac14\log\hat{\lambda} - \frac34\log 2 + \frac16\log\lambda + 2\,\zeta'(-1) + \frac16\log\frac{\pi}{2} \, .
\end{equation}
This can be compared to~\eqref{eq:F1-ABJM-Marino} by numerically inverting the relations~\eqref{eq:tau-k} and~\eqref{eq:k-lambda}. We plot the results in Figure~\ref{fig:F1-compare}, where we see that there is excellent agreement up to the expected non-perturbative corrections included in~\eqref{eq:F1-ABJM-Marino} at small 't Hooft coupling.

\begin{figure}[t!]
	\centering
		\includegraphics[width=0.48\textwidth]{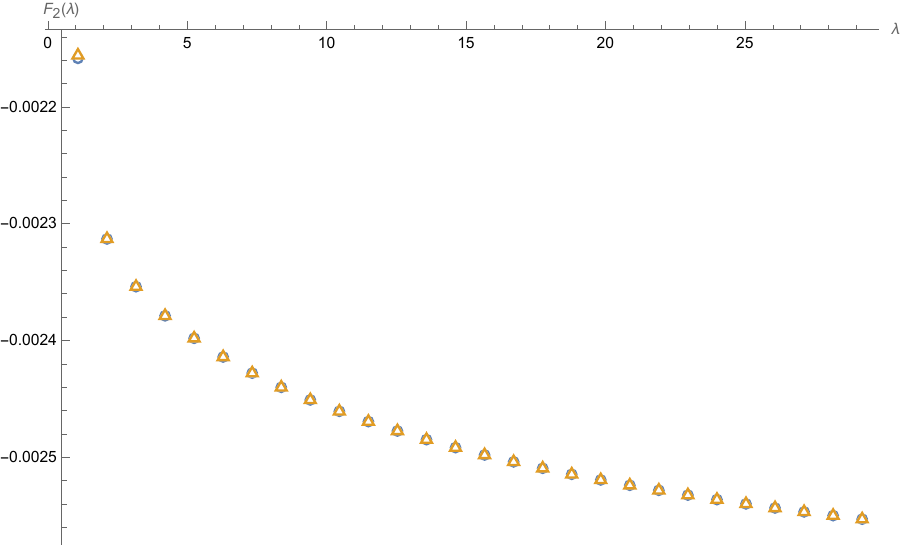}~~\includegraphics[width=0.48\textwidth]{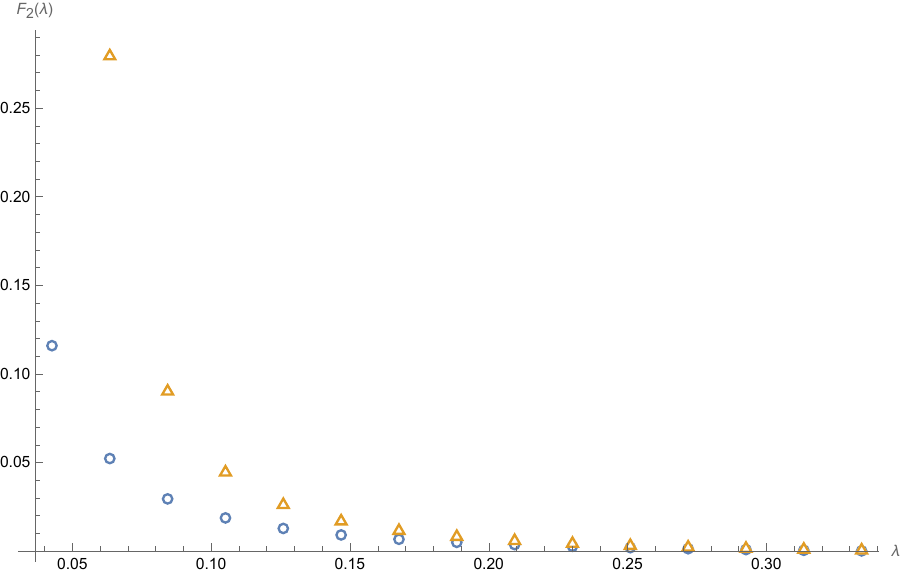}
		\caption{The genus-2 free energies~\eqref{eq:F2-ABJM-Marino} (blue circles) and~\eqref{eq:F2-ABJM-Us} (yellow triangles) for different values of the 't Hooft coupling: $\frac{1}{24} < \lambda \leq 30$ (left) and $\frac{1}{24} < \lambda \leq \frac13$ (right).}
		\label{fig:F2-compare}
\end{figure}

From~\eqref{eq:Fg}, we also easily find the genus-2 free energy
\begin{equation}
	\label{eq:F2-ABJM-Us}
	F_2(\hat{\lambda}) = \frac{5}{96\pi^3\sqrt{2}}\,\hat{\lambda}^{-3/2} - \frac{1}{48\pi^2}\,\hat{\lambda}^{-1} + \frac{1}{144\pi\sqrt{2}}\,\hat{\lambda}^{-1/2} - \frac{1}{360} \, .
\end{equation}
This can be compared to~\eqref{eq:F2-ABJM-Marino} as shown in Figure~\ref{fig:F2-compare}. Just as in the genus-1 case, our result is in perfect agreement at large 't Hooft coupling, and differs at small $\lambda$ due to non-perturbative effects that are not captured by~\eqref{eq:Z} and~\eqref{eq:Fg}.
The advantage of~\eqref{eq:Fg} is that it is straightforward to go to higher genera. We find that the perturbative free energies always resum to simple polynomial functions of the shifted `t Hooft coupling $\hat{\lambda}$. We give the next few examples here:
\begin{align}
	F_3(\hat{\lambda}) =&\; \frac{5}{512\pi^6}\,\hat{\lambda}^{-3} - \frac{5}{768\pi^5\sqrt{2}}\,\hat{\lambda}^{-5/2} + \frac{1}{1152\pi^4}\,\hat{\lambda}^{-2} - \frac{1}{10368\pi^3\sqrt{2}}\,\hat{\lambda}^{-3/2} - \frac{1}{22680} \, , \nonumber \\[1mm]
	F_4(\hat{\lambda}) =&\; \frac{1105}{147456\pi^9\sqrt{2}}\,\hat{\lambda}^{-9/2} - \frac{5}{2048\pi^8}\,\hat{\lambda}^{-4} + \frac{25}{36864\pi^7\sqrt{2}}\,\hat{\lambda}^{-7/2} - \frac{1}{20736\pi^6}\,\hat{\lambda}^{-3} \nonumber \\ 
	&\; + \frac{1}{331776\pi^5\sqrt{2}}\,\hat{\lambda}^{-5/2} - \frac{1}{340200} \, , \\[1mm]
	F_5(\hat{\lambda}) =&\; \frac{565}{131072\pi^{12}}\,\hat{\lambda}^{-6} - \frac{1105}{393216\pi^{11}\sqrt{2}}\,\hat{\lambda}^{-11/2} + \frac{5}{12288\pi^{10}}\,\hat{\lambda}^{-5} - \frac{175}{2654208\pi^9\sqrt{2}}\,\hat{\lambda}^{-9/2} \nonumber \\
	&\;+ \frac{1}{331776\pi^8}\,\hat{\lambda}^{-4} - \frac{1}{7962 624\pi^7\sqrt{2}}\,\hat{\lambda}^{-7/2} - \frac{1}{2494800} \, . \nonumber
\end{align}

Let us remark that we do not have a closed form expression for the genus-${\tt g}$ free energies since they depend on the pure numbers $\mathcal{P}^{(m)}$ and $\mathcal{Q}^{(m)}$ entering~\eqref{eq:cal-F} and~\eqref{eq:cal-G}. However, it is a simple matter of comparing the expansion~\eqref{eq:M-th-F} to that of the Airy function~\eqref{eq:Ai-expand} to obtain these constants algorithmically. Thus, we see our result~\eqref{eq:Fg} as a significant improvement on previous studies since it yields the perturbative genus-${\tt g}$ free energies without relying on an auxiliary topological string theory on non-compact Calabi-Yau manifolds and using modularity to solve the holomorphic anomaly equation order by order.\\

The case of ABJM theory on the round sphere just discussed serves as a benchmark, and we can also study the free energies in theories for which there is no known relation to a topological string theory. For instance, we can use~\eqref{eq:Fg} with the data~\eqref{eq:c-a-b-ABJM-m} and~\eqref{eq:A-ABJM-m} to obtain the perturbative genus-${\tt g}$ free energies in mass-deformed ABJM theory on the round $S^3$ with $m_3 = 0$. Just as in the massless ABJM theory, we find that they can be resummed to give simple functions of the 't Hooft coupling and mass parameters~$m_\pm$. At genus 0, we obtain
\begin{equation}
	\label{eq:F0-ABJM-m}
	F_0(\hat{\lambda},m_\pm) = \frac{4\pi^3\sqrt{2}}{3}\sqrt{(1+m_+^2)(1+m_-^2)}\;\hat{\lambda}^{3/2} + \frac{\zeta(3)}{2}\,\frac{2 - m_+^2 - m_-^2}{2} \, .
\end{equation}
Observe that when~$m_\pm = 0$, this reduces to~\eqref{eq:F0-ABJM-Us}. At genus 1, we find
\begin{equation}
	\label{eq:F1-ABJM-m}
	\begin{split}
		F_1(\lambda,m_\pm) =&\; \frac{\pi}{6\sqrt{2}}\,\frac{2 - m_+^2 - m_-^2}{\sqrt{(1+m_+^2)(1+m_-^2)}}\,\hat{\lambda}^{1/2} - \frac14\log\hat{\lambda} + \frac16\log\lambda \\
		&\;+ \frac{5}{24}\log[(1+m_+^2)(1+m_-^2)] + 2\,\zeta'(-1) + \frac1{12}\log\left(\frac{\pi^2}{2048}\right) \, .
	\end{split}
\end{equation}
Again, this is in agreement with~\eqref{eq:F1-ABJM-Us} when~$m_\pm = 0$. Going up to genus 2,~\eqref{eq:Fg} can be resummed to yield
\begin{align}
	\label{eq:F2-ABJM-m}
	F_2(\hat{\lambda},m_\pm) =&\; \frac{5}{96\pi^3\sqrt{2}}\,\frac{1}{(1+m_+^2)^{1/2}(1+m_-^2)^{1/2}}\,\hat{\lambda}^{-3/2} - \frac{1}{48\pi^2}\,\frac{2 - m_+^2 - m_-^2}{2(1+m_+^2)(1+m_-^2)}\,\hat{\lambda}^{-1} \nonumber \\[1mm]
	&\; + \frac{1}{144\pi\sqrt{2}}\,\frac{(2 - m_+^2 - m_-^2)^2}{4(1+m_+^2)^{3/2}(1+m_-^2)^{3/2}}\,\hat{\lambda}^{-1/2} \\[1mm]
	&\; - \frac{1}{360}\,\frac14\Bigl[\frac{1}{(1+\mathrm{i} m_+)^2} + \frac{1}{(1-\mathrm{i} m_+)^2} + \frac{1}{(1+\mathrm{i} m_-)^2} + \frac{1}{(1-\mathrm{i} m_-)^2}\Bigr] \, . \nonumber
\end{align}
It is straightforward to obtain the free energies at higher genera. Inserting them in~\eqref{eq:IIA-expand} then gives the 't Hooft expansion of the perturbative part of the partition function for the mass-deformed ABJM theory. To the best of our knowledge, these results have not appeared in the literature before. It would be interesting to obtain them from a topological string reformulation of the relevant matrix model analyzed in~\cite{Nosaka:2015iiw}.

We can obtain similar results for the genus-${\tt g}$ free energies in the ABJM theory on an arbitrarily squashed sphere. With the data~\eqref{eq:c-a-b-ABJM-b}-\eqref{eq:A-ABJM-b}, the genus-0 free energy obtained from~\eqref{eq:Fg} can be resummed to yield
\begin{equation}
	\label{eq:F0-ABJM-b}
	F_0(\hat{\lambda},b) = \frac14\Bigl(b + \frac{1}{b}\Bigr)^2\,\Bigl(\frac{4\pi^3\sqrt{2}}{3}\,\hat{\lambda}^{3/2} + \frac{\zeta(3)}{2}\Bigr) - \frac{1}{16}\Bigl(b^2- \frac{1}{b^2}\Bigr)^2\,\zeta(3)\, .
\end{equation}
As expected, this reduces to~\eqref{eq:F0-ABJM-Us} when setting~$b=1$. At genus 1, we find
\begin{equation}
	\label{eq:F1-ABJM-b}
	\begin{split}
		F_1(\lambda,b) =&\; -\frac{\pi}{6\sqrt{2}}\,\Bigl(b^2 - 4 + \frac{1}{b^2}\Bigr)\,\hat{\lambda}^{1/2} - \frac14\log\hat{\lambda} + \frac16\log\lambda \\
		&\;+ \frac{5}{6}\log\Bigl(b + \frac{1}{b}\Bigr) - \frac56\log(2) + 2\,\zeta'(-1) + \frac1{12}\log\Bigl(\frac{\pi^2}{2048}\Bigr) \, ,
	\end{split}
\end{equation}
which reduces to~\eqref{eq:F1-ABJM-Us} when~$b=1$. At genus 2, the resummed free energy reads
\begin{align}
	\label{eq:F2-ABJM-b}
	F_2(\hat{\lambda},b) =&\; \frac{5}{24\pi^3\sqrt{2}}\,\Bigl(b + \frac{1}{b}\Bigr)^{-2}\,\hat{\lambda}^{-3/2} + \frac{1}{24\pi^2}\,\Bigl(b^2 - 4 + \frac{1}{b^2}\Bigr)\Bigl(b + \frac{1}{b}\Bigr)^{-2}\,\hat{\lambda}^{-1} \nonumber \\[1mm]
	&\; + \frac{1}{144\pi\sqrt{2}}\,\,\Bigl(b^2 - 4 + \frac{1}{b^2}\Bigr)^2\Bigl(b + \frac{1}{b}\Bigr)^{-2}\,\hat{\lambda}^{-1/2} \\[1mm]
	&\; - \frac{1}{1440}\Bigl(7b^8 - 34b^6 + 168b^4 - 446 b^2 + 738 - 446 b^{-2} + 168 b^{-4} - 34 b^{-6} + 7 b^{-8}\Bigr) \nonumber\\
	&\qquad\qquad \times \Bigl(b+ \frac{1}{b}\Bigr)^{-8} \, , \nonumber
\end{align}
which matches~\eqref{eq:F2-ABJM-Us} upon setting~$b=1$. For generic values of $b$, the above perturbative genus-${\tt g}$ free energies are new. For the specific value $b = \sqrt{3}$, the matrix model obtained by localization of the partition function $Z(b)\vert_{b^2=3}$ has been related to the topological string on local $\mathbb{P}^2$~\cite{Hatsuda:2016uqa}. It would be interesting to compare our perturbative results to the full non-perturbative genus-$\tt g$ free energies obtained from this topological string theory, similar to the local $\mathbb{P}^1\times\mathbb{P}^1$ case discussed above.

Our results in the Type IIA frame suggest that the genus-${\tt g}$ free energies for a variety of string backgrounds of the form $M_4 \times \mathbb{CP}_3$ can be obtained to all orders in $\alpha'$. It would be very interesting to understand how this calculation can be performed from the point of view of the worldsheet string theory. \\

This concludes our study of Airy-type partition functions. In the rest of the paper, we study another interesting observable in the ABJM theory called the topologically twisted index. While this quantity is not expressed in terms of an Airy function, we will show using a detailed numerical investigation that its perturbative part takes an even simpler form. 

\section{Topologically twisted index}
\label{sec:TTI}

In this section, we explore the topologically twisted index (TTI) of the U$(N)_k\times$U$(N)_{-k}$ ABJM theory on $S^1\times\Sigma_\mg$ with $\Sigma_{\mg}$ a Riemann surface of genus $\mg$~\cite{Benini:2015noa,Benini:2016hjo}. Unlike the $S^3$ partition function discussed previously, whose perturbative contributions are captured by an Airy function as in~\eqref{eq:Z}, there is at present no known closed form expression for the TTI. Instead, the TTI of the ABJM theory has been computed in the M-theory limit (large $N$ with fixed CS level $k$) using the Bethe Ansatz (BA) formulation, and subsequently used to reproduce the entropy of the dual magnetically charged AdS$_4$ black hole at the leading $\mathcal{O}(N^\fft32)$ order~\cite{Benini:2015eyy}. Subleading corrections of order $N^\fft12$ and $\log N$ have been investigated numerically in~\cite{Liu:2017vll} and confirmed analytically by a bulk supergravity analysis~\cite{Bobev:2020egg,Bobev:2021oku} and a one-loop calculation~\cite{Liu:2017vbl}, respectively. The TTI has also been studied in the IIA limit of large $N$ and fixed $\lambda=N/k$ using a similar numerical approach~\cite{PandoZayas:2019hdb}, which provided results consistent with the M-theory limit in the overlapping regime of validity. In this section, we propose an analytic expression for the all-order perturbative $1/N$ expansion of the TTI of the ABJM theory in the M-theory limit. This proposal is based on a detailed numerical analysis, which greatly improves on the results just mentioned.

\subsection{Bethe Ansatz formulation}
\label{sec:TTI:BA}

We begin with a review of the BA formulation for the TTI of ABJM theory on $S^1\times\Sigma_\mg$ following~\cite{Benini:2015eyy,Benini:2016hjo,Closset:2017zgf}. The matrix model obtained from supersymmetric localization takes the form\footnote{The CS contribution in fact involves an extra phase factor as $x_i^{k\mm_i}\to(-x_i)^{k\mm_i}$ if one chooses the periodic boundary condition for fermions along $S^1$, see Appendix C of \cite{Closset:2017zgf} for example. Here we did not include the factor of $(-1)^{k\mm_i}$ explicitly, however, since the fugacity associated with the U(1) topological symmetry \cite{Benini:2015noa} can be shifted appropriately to absorb such an extra phase factor.}
\begin{align}
\label{TTI:1}
		&Z_{S^1\times\Sigma_\mg}(N,k,\Delta,\mn) \nonumber \\
		&=\fft{1}{(N!)^2}\sum_{\mm,\tilde{\mm}\in\mathbb Z^N}\oint_{\mathcal C}\,\prod_{i=1}^N\fft{dx_i}{2\pi \mathrm{i}\,x_i}\,\prod_{j=1}^N\fft{d\tx_j}{2\pi \mathrm{i}\,\tx_j}\,\prod_{i=1}^Nx_i^{k\mm_i}\,\prod_{j=1}^N\tx_j^{-k\tilde{\mm}_j} \nonumber \\
		&\kern5em\times\bigl(\det\mathbb B(N,k,u,\tu,\Delta)\bigr)^{\mg}\,\prod_{i\neq j}^N\bigg(1-\fft{x_i}{x_j}\bigg)^{1-\mg}\bigg(1-\fft{\tx_i}{\tx_j}\bigg)^{1-\mg} \\
		&\kern5em\times\prod_{i,j=1}^N\prod_{a=1,2}\left(\fft{\sqrt{\fft{x_i}{\tx_j}y_a}}{1-\fft{x_i}{\tx_j}y_a}\right)^{\mm_i-\tilde{\mm}_j+1-\mg-\mn_a}\prod_{a=3,4}\left(\fft{\sqrt{\fft{\tx_j}{x_i}y_a}}{1-\fft{\tx_j}{x_i}y_a}\right)^{-\mm_i+\tilde{\mm}_j+1-\mg-\mn_a} \, . \nonumber
\end{align}
Let us briefly recall the meaning of various parameters and implicit expressions entering~\eqref{TTI:1} (see~\cite{Benini:2015noa,Benini:2015eyy,Benini:2016hjo} for more details):
\begin{itemize}
	\item The TTI (\ref{TTI:1}) is a function of chemical potentials $\Delta_a~(\text{with fugacities }y_a=e^{\mathrm{i}\pi\Delta_a})$\footnote{Compared to the convention of \cite{Benini:2015eyy}, we have $\Delta_a^\text{(there)}=\pi\Delta_a^\text{(here)}$.} and magnetic fluxes $\mn_a$ associated with certain linear combinations of the U(1)${}^3$ Cartan of the flavor symmetries together with the superconformal $R$-symmetry. They satisfy the constraints
	\begin{equation}
		\sum_{a=1}^4\Delta_a=2\mathbb{Z} \,,\qquad \sum_{a=1}^4\mn_a=2(1-\mg) \,, \label{constraints}
	\end{equation}
where the first comes from requiring invariance of the ABJM superpotential under the global symmetries, and the second comes from supersymmetry. Below we will make a specific choice for the $\Delta_a$ such that $\sum_a \Delta_a = 2$, see~\eqref{eq:delta-choice}. We refer to~\cite{Benini:2015eyy} for the precise linear combinations of global symmetries and the corresponding constraints on chemical potentials and magnetic fluxes. In what follows, we will use the shorthand notation $\Delta$ and $\mn$ for the sets $\{\Delta_a|a=1,2,3,4\}$ and $\{\mn_a|a=1,2,3,4\}$ and simply call them flavor chemical potentials and flavor magnetic fluxes, respectively. We will also work with real $\Delta$, although the results we will derive can be analytically continued to complex $\Delta$ since the TTI is a meromorphic function of the flavor fugacities $y_a$~\cite{Benini:2016rke}.
	\item The quantities $x_i=e^{\mathrm{i}u_i},\tx_j=e^{\mathrm{i}\tu_j}$ and $\mm_i,\tilde{\mm}_j$ parametrize the BPS configurations upon which the path-integral formulation of the TTI localizes. They correspond to gauge zero-modes and gauge magnetic fluxes for the two U($N$) gauge groups, respectively. The integrand of~\eqref{TTI:1} consists of a classical and a one-loop contribution around the BPS configurations. Following the convention of~\cite{Benini:2015eyy}, we take the sums over $\mm_i$ (resp. $\tilde{\mm}_j$) with the upper (resp. lower) bound $\mm_i\leq M-1$ (resp. $\tilde{\mm}_j\geq1-M$) for some large positive integer $M$. For the $x_i,\tx_j$ integration variables, the contour $\mathcal C$ is a particular choice picking up the so-called Jeffrey-Kirwan residues, and we refer to~\cite{Benini:2015noa,Benini:2015eyy} for more details.
	
	\item The $2N\times 2N$ Jacobian matrix $\mathbb B$, which appears in the one-loop contribution to the integrand of~\eqref{TTI:1}, comes from integrating out fermionic zero-modes~\cite{Benini:2016hjo}. It is given explicitly as
	\begin{equation}
		\begin{split}
			\mathbb B(N,k,u,\tu,\Delta)&=\fft{\partial(B_1,\ldots,B_N,\tB_1,\ldots,\tB_N)}{\partial(u_1,\ldots,u_N,\tu_1,\ldots,\tu_N)} \, ,
		\end{split}\label{Jacobian}
	\end{equation}
	where we have used shorthand expressions $u$ and $\tu$ for $\{u_i|i=1,\cdots,N\}$ and $\{\tu_j|j=1,\cdots,N\}$ respectively and also introduced the BA operators as
	\begin{equation}
		\begin{split}
			e^{\mathrm{i}B_i}&=\sigma_i\,x_i^k\,\prod_{j=1}^N\fft{(1-y_3\fft{\tx_j}{x_i})(1-y_4\fft{\tx_j}{x_i})}{(1-y_1^{-1}\fft{\tx_j}{x_i})(1-y_2^{-1}\fft{\tx_j}{x_i})}\, ,\\
			e^{\mathrm{i}\tB_j}&=\tilde\sigma_j\,\tx_j^k\,\prod_{i=1}^N\fft{(1-y_3\fft{\tx_j}{x_i})(1-y_4\fft{\tx_j}{x_i})}{(1-y_1^{-1}\fft{\tx_j}{x_i})(1-y_2^{-1}\fft{\tx_j}{x_i})}\, .
		\end{split}\label{BA:operators}
	\end{equation}
	In the BA operators (\ref{BA:operators}), the sign ambiguities $\sigma_i,\tilde\sigma_j$ are given explicitly as
	\begin{equation}
	\begin{split}
		\sigma_i&=\prod_{j=1}^N\fft{\sqrt{\fft{x_i}{\tx_j}y_1}}{-\fft{x_i}{\tx_j}y_1}\fft{\sqrt{\fft{x_i}{\tx_j}y_2}}{-\fft{x_i}{\tx_j}y_2}\fft{1}{\sqrt{\fft{\tx_j}{x_i}y_3}}\fft{1}{\sqrt{\fft{\tx_j}{x_i}y_4}}\in\{\pm1\} \, ,\\
		\tilde\sigma_j&=\prod_{i=1}^N\fft{\sqrt{\fft{x_i}{\tx_j}y_1}}{-\fft{x_i}{\tx_j}y_1}\fft{\sqrt{\fft{x_i}{\tx_j}y_2}}{-\fft{x_i}{\tx_j}y_2}\fft{1}{\sqrt{\fft{\tx_j}{x_i}y_3}}\fft{1}{\sqrt{\fft{\tx_j}{x_i}y_4}}\in\{\pm1\} \, ,
	\end{split}\label{BA:sign}
	\end{equation}
	where $\sigma_i,\tilde\sigma_j\in\{\pm1\}$ comes from the constraints (\ref{constraints}). Later in Section~\ref{sec:TTI:pert:all-order} we will check that the ambiguities are fixed to $\sigma_i=\tilde\sigma_j=(-1)^N$ in our case, which is independent of the gauge holonomies $u_i,\tu_j$. The Jacobian matrix $\mathbb B$ defined in~\eqref{Jacobian} can then be written more explicitly as
	\begin{equation}
		\begin{split}
			\mathbb B(N,k,u,\tu,\Delta)&=\begin{pmatrix}
				\delta_{jl}\Big(k-\sum_{m=1}^NG_{jm}\Big) & G_{jl}\\
				-G_{lj} & \delta_{jl}\Big(k+\sum_{m=1}^NG_{mj}\Big)
			\end{pmatrix} \, ,
		\end{split}
	\end{equation}
	in terms of an $N\times N$ matrix $G_{ij}$ defined as
	\begin{equation}
		G_{ij} = \fft{\partial\log D(z)}{\partial\log z}\bigg|_{z=\tx_j/x_i} \, , \qquad D(z) = \fft{(1-zy_3)(1-zy_4)}{(1-zy_1^{-1})(1-zy_2^{-1})} \, .
	\end{equation}
	Note that for $\Delta_a\in\mathbb R$ we have $\overline{G}_{ij}=-G_{ji}$ and therefore $\det\mathbb{B}\in\mathbb R$.
\end{itemize}

The first step towards a convenient rewriting of the TTI is to sum over the gauge magnetic fluxes in the integral expression~\eqref{TTI:1}. This yields
\begin{align}
\label{TTI:2}
		&Z_{S^1\times\Sigma_\mg}(N,k,\Delta,\mn) \\
		&=\fft{1}{(N!)^2}\oint_{\mathcal C}\,\prod_{i=1}^N\fft{dx_i}{2\pi \mathrm{i}x_i}\prod_{j=1}^N\fft{d\tx_j}{2\pi \mathrm{i}\tx_j}\,\prod_{i\neq j}^N\bigg(1-\fft{x_i}{x_j}\bigg)^{1-\mg}\bigg(1-\fft{\tx_i}{\tx_j}\bigg)^{1-\mg}\prod_{i=1}^N\fft{(e^{\mathrm{i}B_i})^M}{1-e^{\mathrm{i}B_i}}\prod_{j=1}^N\fft{(e^{\mathrm{i}\tB_j})^M}{1-e^{\mathrm{i}\tB_j}} \nonumber \\
		&\kern5em\times\bigl(\det\mathbb B(N,k,u,\tu,\Delta)\bigr)^\mg\prod_{i,j=1}^N\prod_{a=1,2}\left(\fft{\sqrt{\fft{x_i}{\tx_j}y_a}}{1-\fft{x_i}{\tx_j}y_a}\right)^{1-\mg-\mn_a}\prod_{a=3,4}\left(\fft{\sqrt{\fft{\tx_j}{x_i}y_a}}{1-\fft{\tx_j}{x_i}y_a}\right)^{1-\mg-\mn_a}. \nonumber
\end{align}
The integral~\eqref{TTI:2} now receives contributions from specific poles of the new integrand (refer to~\cite{Benini:2015noa,Benini:2015eyy,Benini:2016rke} for details), whose locations are given as solutions to the Bethe Ansatz Equations (BAE)\footnote{In previous studies of the TTI \cite{Benini:2015eyy,Liu:2017vll} the sign ambiguities were fixed as $\sigma_i=\tilde\sigma_j=1$, but in Section~\ref{sec:TTI:pert:all-order} we will show that $\sigma_i=\tilde\sigma_j=(-1)^N$ for the BAE solutions of interest.}
\begin{equation}
\begin{split}
	1&=e^{\mathrm{i}B_i}=\sigma_i\,x_i^k\,\prod_{j=1}^N\fft{(1-y_3\fft{\tx_j}{x_i})(1-y_4\fft{\tx_j}{x_i})}{(1-y_1^{-1}\fft{\tx_j}{x_i})(1-y_2^{-1}\fft{\tx_j}{x_i})}\, ,\\ 1&=e^{\mathrm{i}\tB_j}=\tilde\sigma_j\,\tx_j^k\,\prod_{i=1}^N\fft{(1-y_3\fft{\tx_j}{x_i})(1-y_4\fft{\tx_j}{x_i})}{(1-y_1^{-1}\fft{\tx_j}{x_i})(1-y_2^{-1}\fft{\tx_j}{x_i})}\, .\label{BAE}
\end{split}
\end{equation}
Picking up the appropriate residues, the resulting TTI then reads\footnote{Here we focus explicitly on the case $\mathfrak{g}\neq 1$ and comment on the torus case below.}
\begin{align}
\label{TTI:3}
		&Z_{S^1\times\Sigma_\mg}(N,k,\Delta,\mn) \\
		&=\prod_{a=1}^4y_a^{-\fft{N^2}{2}\mn_a}\sum_{\{x_i,\tx_j\}\in\text{BAE}}\left[\fft{1}{\det\mathbb B}\fft{\prod_{i=1}^Nx_i^N\tx_i^N\prod_{i\neq j}^N(1-\fft{x_i}{x_j})(1-\fft{\tx_i}{\tx_j})}{\prod_{i,j=1}^N\prod_{a=1}^2(\tx_j-x_iy_a)^{1-\fft{\mn_a}{1-\mg}}\prod_{a=3}^4(x_i-\tx_jy_a)^{1-\fft{\mn_a}{1-\mg}}}\right]^{1-\mg}. \nonumber
\end{align}
We call~\eqref{TTI:3} the BA formulation for the TTI of the ABJM theory on $S^1\times\Sigma_\mg$. To compute the TTI using this formulation, one should first solve the BAE~\eqref{BAE}, evaluate the summand in~\eqref{TTI:3}, and finally sum over all contributions from each BAE solution. Let us summarize some of the key properties of the BA formulation:
\begin{itemize}
	\item Any BAE solution has a $k$-fold degeneracy since the BAE~\eqref{BAE} is invariant under the shift $\{x_i,\tx_j\}\to\{x_i e^{2\pi \mathrm{i}/k},\tx_je^{2\pi \mathrm{i}/k}\}$~\cite{Benini:2015eyy}. Since the RHS of the BA formula~\eqref{TTI:3} is invariant under such a constant phase shift, the contribution from these $k$-fold degenerate BAE solutions can be obtained by multiplying individual contributions by a factor of $k$. 
	\item To obtain the exact TTI using~\eqref{TTI:3}, one should in principle find the most general solutions to the BAE~\eqref{BAE}. Due to the transcendental nature of the equations, this is a daunting task. We will therefore assume that the contribution to the free energy from a particular BAE solution that will be reviewed in Section~\ref{sec:TTI:summary:analytic} (together with its $k$-fold degeneracy) is dominant in the M-theory limit. We expect the contributions from other BAE solutions, if any, to be exponentially suppressed compared to the dominant one. We will comment further on this assumption in Section~\ref{sec:discussions}.

	\item The logarithm of the contribution from a particular BAE solution to the TTI in~\eqref{TTI:3} is linear in the flavor magnetic fluxes $\mn_a$. Hence the M-theory limit of the logarithm of the TTI will also be linear in the flavor magnetic fluxes provided the TTI is dominated by a particular BAE solution at large $N$, as per our assumption above.

	\item By examination, the TTI~\eqref{TTI:3} is invariant under the following permutations
	\begin{equation}
	\begin{split}
		(\mn_1,\Delta_1)~&\leftrightarrow~(\mn_2,\Delta_2)\,,\\
		(\mn_3,\Delta_3)~&\leftrightarrow~(\mn_4,\Delta_4)\,,\\
		(\mn_1,\mn_2,\Delta_1,\Delta_2)~&\leftrightarrow~(\mn_3,\mn_4,\Delta_3,\Delta_4)\,.\label{TTI:sym}
	\end{split}
	\end{equation}
	Note that~\eqref{TTI:sym} is not a complete set of permutations. See~\cite{Chester:2021gdw} and the discussion below~\eqref{eq:c-a-b-ABJM-m-gen} for a similar remark regarding the $S^3$ partition function of ABJM theory.
	\item The BAE~\eqref{BAE} are independent of the genus $\mg$. This allows one to obtain a simple relation between the TTI on $S^1\times\Sigma_\mg$ for an arbitrary Riemann surface of genus $\mg > 1$ and the TTI on $S^1\times S^2$,
	\begin{equation}
		\log Z_{S^1\times\Sigma_\mg}(N,k,\Delta,\mn) = (1-\mg)\,\log Z_{S^1\times S^2}\Bigl(N,k,\Delta,\fft{\mn}{1-\mg}\Bigr) \, .\label{TTI:Sigma:S^2}
	\end{equation}
	In view of this, we will analyze the case with $\mg=0$ and restore the dependence on the genus $\mg > 1$ at the end. 
	\item In this paper we focus on the case with $\mg\neq1$. The $\mg=1$ case where the Riemann surface is a torus $T^2$ should be treated more carefully due to the presence of additional fermionic zero-modes when localizing the path-integral, as was remarked in~\cite{Benini:2016hjo}.
\end{itemize}
%

\subsection{Summary of known results}
\label{sec:TTI:summary}

We now summarize known analytic and numerical results for the TTI based on the BA formulation reviewed in Section~\ref{sec:TTI:BA}. 

\subsubsection{Analytic approach: $N^\fft32$ leading behavior in the M-theory limit}
\label{sec:TTI:summary:analytic}

As explained above, one should first solve the BAE~\eqref{BAE} in order to evaluate the TTI using the BA formulation \eqref{TTI:3}. To solve these equations explicitly, it is useful to rewrite them by taking the logarithm as (recall that $\sigma_i,\tilde\sigma_j\in\{\pm1\}$) 
\begin{equation}
	\begin{split}
\label{BAE:2}
		2\pi n_i&=\fft{1-\sigma_i}{2}\pi+ku_i+\ri\sum_{j=1}^N\bigg[\sum_{a=3,4}\text{Li}_1(e^{\ri(\tu_j-u_i+\pi\Delta_a)})-\sum_{a=1,2}\text{Li}_1(e^{\ri(\tu_j-u_i-\pi\Delta_a)})\bigg]~~(n_i\in\mathbb Z) \, ,\\
		2\pi\tn_j&=\fft{1-\tilde\sigma_j}{2}\pi+k\tu_j+\ri\sum_{i=1}^N\bigg[\sum_{a=3,4}\text{Li}_1(e^{\ri(\tu_j-u_i+\pi\Delta_a)})-\sum_{a=1,2}\text{Li}_1(e^{\ri(\tu_j-u_i-\pi\Delta_a)})\bigg]~~(\tn_j\in\mathbb Z) \, ,
	\end{split}
\end{equation}
which can also be obtained by extremizing the so-called Bethe potential
\begin{equation}
	\begin{split}
		\mathcal V(N,k,u,\tu,\Delta)&=\sum_{i=1}^N\Bigg[\fft{k}{2}(\tu_i^2 - u_i^2)-\pi\left(2\tn_i-\fft{1-\tilde\sigma_i}{2}\right)\tu_i+\pi\left(2n_i-\fft{1-\sigma_i}{2}\right)u_i\Bigg]\\
		&\quad+\sum_{i,j=1}^N\left[\sum_{a=3,4}\text{Li}_2(e^{\ri(\tu_j-u_i+\pi\Delta_a)})-\sum_{a=1,2}\text{Li}_2(e^{\ri(\tu_j-u_i-\pi\Delta_a)})\right]
	\end{split}\label{V}
\end{equation}
with respect to $u_i$ and $\tu_j$, respectively. The BAE~\eqref{BAE:2} have first been solved in the M-theory limit by applying a variational principle to the Bethe potential~\eqref{V} in \cite{Benini:2015eyy}, as we now briefly review. 

The first step is to use the leading order ansatz 
\begin{equation}
	u_i=\ri N^\fft12t_i+v_i \, , \qquad\tu_j=\ri N^\fft12t_j+\tv_j \, , \label{BAE:ansatz}
\end{equation}
in the M-theory limit. This discrete ansatz is then made continuous by introducing eigenvalue distributions $u,\tu:[t_{\ll},t_{\gg}]\subseteq\mathbb R\to\mathbb C$ and their real parts $v,\tv:[t_{\ll},t_{\gg}]\to\mathbb R$ satisfying
\begin{equation}
	u(t(i))=\ri N^\fft12t(i)+v(t(i))=u_i \, , \qquad \tu(t(j))=\ri N^\fft12t(i)+\tv(t(i))=\tu_j \, . \label{ansatz:conti}
\end{equation}
Here we have also introduced a continuous function $t:[1,N]\to[t_{\ll},t_{\gg}]$ that maps discrete indices $1,2,\ldots,N$ to a real interval $[t_{\ll},t_{\gg}]$. We then introduce the eigenvalue density $\rho:[t_{\ll},t_{\gg}]\to\mathbb R$ as
\begin{equation}
	\rho(t)=\fft{1}{N-1}\sum_{i=1}^N\delta(t-t_i)\quad\Leftrightarrow\quad di=(N-1)\rho(t)dt \, ,\label{rho}
\end{equation}
which, by construction, satisfies the normalization condition
\begin{equation}
	\int_{t_\ll}^{t_\gg}dt\,\rho(t)=1 \, .\label{rho:normal}
\end{equation}
Upon the continuation of the leading order ansatz~\eqref{BAE:ansatz} in terms of the eigenvalue distributions~\eqref{ansatz:conti} and the eigenvalue density~\eqref{rho}, the Bethe potential~\eqref{V} can be rewritten in terms of continuous functions $u(t)$, $\tu(t)$, and $\rho(t)$ in the M-theory limit for the set of integers $(n_i,\tn_i)=(1-i,i-N)$ as
\begin{align}
\label{V:conti}
		\mathcal V&=\ri N^\fft32\int_{t_\ll}^{t_\gg}dt\Bigg[k t\rho(t)\delta v(t)+\rho(t)^2\bigg(\sum_{a=3,4}g_+(\delta v(t)+\pi\Delta_a)-\sum_{a=1,2}g_-(\delta v(t)-\pi\Delta_a)\bigg)\Bigg] \nonumber \\
		&\quad+N\int_{t_\ll}^{t_\gg} dt\,\rho(t)\Bigg[\sum_{a=3,4}\text{Li}_2(e^{\ri(\delta v(t)+\pi\Delta_a)})-\sum_{a=1,2}\text{Li}_2(e^{\ri(\delta v(t)-\pi\Delta_a)})\Bigg]  \\ 
		&\quad-\ri N^\fft32\pi\mu\left[\int_{t_\ll}^{t\gg}dt\,\rho(t)-1\right]+\big(\rho(t),u(t),\tu(t)\text{\,--\,independent terms}\big)+ o(N^\fft32) \, , \nonumber
\end{align}
where we have defined $\delta v(t)$ and $g_\pm(u)$ as
\begin{equation}
	\delta v(t) = \tv(t)-v(t) \, ,\qquad g_\pm(u) = \fft16u^3\mp\fft{\pi}{2}u^2+\fft{\pi^2}{3}u \, . 
\end{equation}
To derive the large-$N$ limit of the Bethe potential (\ref{V:conti}) from (\ref{V}), we have also assumed identical sign ambiguities $\sigma_i=\tilde{\sigma}_i$ and
\begin{equation}
	0<\text{Re}[\tu_j-u_i+\pi\Delta_{3,4}]<2\pi\,,\quad-2\pi<\text{Re}[\tu_j-u_i-\pi\Delta_{1,2}]<0\quad~(i>j)\,.
\end{equation}
Observe that we have introduced a Lagrange multiplier $\mu$ in~\eqref{V:conti} to enforce the correct normalization of the eigenvalue density.

Solving the BAE~\eqref{BAE:2} with the discrete ansatz~\eqref{BAE:ansatz} in the M-theory limit becomes equivalent to finding $u(t)$, $\tu(t)$, $\rho(t)$, and the Lagrange multiplier $\mu$ that extremize the Bethe potential~\eqref{V:conti} over an appropriate real interval $[t_\ll,t_\gg]$. We skip the derivation and summarize the resulting leading order solution in the M-theory limit. For flavor chemical potentials satisfying
\begin{equation}
	\label{eq:delta-choice}
	\sum_{a=1}^4\Delta_a=2 \, , \qquad \Delta_a>0 \, , \qquad \Delta_1\leq\Delta_2 \, , \qquad \Delta_3\leq\Delta_4 \, ,
\end{equation}
the leading order solution in the M-theory limit reads
\begin{subequations}
	\begin{align}
		\delta v(t)&=\begin{cases}
			-\pi\Delta_3+e^{-N^\fft12Y_3(t)}\qquad \text{with} \;\; Y_3(t)=\fft{-kt\Delta_4-\mu}{\Delta_4-\Delta_3} & (t_\ll<t<t_<)\\[1mm]
			\fft{\pi\left(\mu(\Delta_1\Delta_2-\Delta_3\Delta_4)+kt\sum_{a<b<c}^4\Delta_a\Delta_b\Delta_c\right)}{2\mu-kt(\Delta_1\Delta_2-\Delta_3\Delta_4)} & (t_<<t<t_>)\\[1mm]
			\pi\Delta_1-e^{-N^\fft12Y_1(t)}\qquad\quad \text{with} \;\; Y_1(t)=\fft{kt\Delta_2-\mu}{\Delta_2-\Delta_1} & (t_><t<t_\gg)
		\end{cases}\label{BAE:sol:dv},\\[1mm]
		\rho(t)&=\begin{cases}
			\fft{\mu+kt\Delta_3}{\pi^2(\Delta_1+\Delta_3)(\Delta_2+\Delta_3)(\Delta_4-\Delta_3)} & (t_\ll<t<t_<)\\
			\fft{2\mu-kt(\Delta_1\Delta_2-\Delta_3\Delta_4)}{\pi^2(\Delta_1+\Delta_3)(\Delta_2+\Delta_3)(\Delta_1+\Delta_4)(\Delta_2+\Delta_4)} & (t_<<t<t_>)\\
			\fft{\mu-kt\Delta_1}{\pi^2(\Delta_1+\Delta_3)(\Delta_1+\Delta_4)(\Delta_2-\Delta_1)} & (t_><t<t_\gg)
		\end{cases},\label{BAE:sol:rho}
	\end{align}\label{BAE:sol}%
\end{subequations}
where the transition and end points are given by
\begin{equation}
	t_\ll=-\fft{\mu}{k\Delta_3} \, , \qquad t_<=-\fft{\mu}{k\Delta_4} \, , \qquad t_>=\fft{\mu}{k\Delta_2}\, , \qquad t_\gg=\fft{\mu}{k\Delta_1} \, , \label{range}
\end{equation}
and the Lagrange multiplier $\mu$ is determined by the normalization (\ref{rho:normal}) as\footnote{Compared to the convention of \cite{Benini:2015eyy}, we have $\mu^\text{(there)}=\pi\mu^\text{(here)}$ and $\Delta_a^\text{(there)}=\pi\Delta_a^\text{(here)}$.}
\begin{equation}
	\mu=\pi\sqrt{2k\Delta_1\Delta_2\Delta_3\Delta_4} \, .\label{mu}
\end{equation}
Note that the leading order solution~\eqref{BAE:sol} does not fix the sum of the real parts of the eigenvalue distributions $u(t)$ and $\tu(t)$, namely the quantity
$v(t)+\tv(t)$. 

Finally, substituting the leading order BAE solution~\eqref{BAE:sol},~\eqref{range} and~\eqref{mu} into the continuous version of the BA formula~\eqref{TTI:3},
\begin{equation}
\begin{split}
	\log Z_{S^1\times S^2}&=-N^\fft32\int_{t_\ll}^{t\gg}dt\,\rho(t)^2\left[\fft{2\pi^2}{3}+\sum_{a=3,4}(\mn_a-1)g_+'(\delta v(t)+\pi\Delta_a)\right.\\
	&\kern9em~~\left.+\sum_{a=1,2}(\mn_a-1)g_-'(\delta v(t)-\pi\Delta_a)\right]\\
	&\quad-N^\fft32\left[\mn_1\int_{t>}^{t_\gg}dt\,\rho(t)Y_1(t)+\mn_3\int_{t_\ll}^{t_<}dt\,\rho(t)Y_3(t)\right] + o(N^\fft32) \, , \label{TTI:conti}
\end{split}
\end{equation}
we obtain the following analytic result for the TTI in the M-theory limit:
\begin{equation}
	\log Z_{S^1\times S^2}(N,k,\Delta,\mn)=-\fft{\pi\sqrt{2\Delta_1\Delta_2\Delta_3\Delta_4}}{3}\sum_{a=1}^4\fft{\mn_a}{\Delta_a}k^\fft12N^\fft32+o(N^\fft32) \, .\label{TTI:M:known:leading}
\end{equation}
Note that the imaginary part of the logarithm of the TTI is defined modulo $2\pi \mathrm{i}\mathbb Z$ and thereby included in the $o(N^\fft32)$ terms in the M-theory limit. Also, as already mentionned below~\eqref{TTI:3}, we have implicitly assumed that the contributions from other BAE solutions to~\eqref{TTI:3} are subdominant compared to the leading large $N$ contribution in~\eqref{TTI:M:known:leading}.

A remark that will be useful later is that the Bethe potential~\eqref{V:conti} can be evaluated on~\eqref{BAE:sol} to yield\footnote{Strictly speaking this is true only if we ignore the holonomy independent terms in the last line of (\ref{V:conti}), see e.g.~\cite{Benini:2015eyy}.}
\begin{equation}
\label{eq:V-func}
\mathcal{V}(\Delta) = \frac{2\pi\mathrm{i}}{3}\,N^{\frac32}\,\mu(\Delta) + o(N^{\frac32}) \, ,
\end{equation}
where $\mu$ is given in~\eqref{mu}. At large $N$, the Bethe potential is related to the free energy of the ABJM theory on the round $S^3$~\cite{Hosseini:2016tor,Azzurli:2017kxo}, and the flavor chemical potentials $\Delta$ can be interpreted as a set of trial $R$-charges. We can then use F-maximization~\cite{Jafferis:2010un,Jafferis:2011zi} to obtain the exact $R$-charges singled out by the superconformal algebra. Extremizing~\eqref{eq:V-func} with the constraint~\eqref{eq:delta-choice} gives the values of $\Delta$ at the superconformal point (we distinguish the $*$-superscript from the complex conjugate represented by an overline throughout),
\begin{equation}
\label{eq:delta-conf}
\Delta^*_a = \frac12 \, ,  \qquad a = 1\ldots 4 \, .
\end{equation}
Combined with the constraint on the flavor magnetic fluxes in~\eqref{constraints}, this shows that the logarithm of the TTI in~\eqref{TTI:M:known:leading} is independent of $\mathfrak{n}$ at the superconformal point. We will see later in subsection \ref{sec:TTI:pert:all-order} that this property is in fact an exact statement to all orders in $1/N$ based on the property of the exact numerical BAE solution for $\Delta_a=\fft12$. 

Related to the special $\Delta^*$-configuration~\eqref{eq:delta-conf}, there is also a special configuration of flavor magnetic fluxes $\frak{n}^*$ that ensures a non-zero flux for the superconformal U(1) $R$-symmetry and vanishing flux for other U(1) flavor symmetries. In our conventions, this configuration is given by
\begin{equation}
\label{eq:n-conf}
\mathfrak{n}^*_a = \frac12 \, ,  \qquad a = 1\ldots 4 \, ,
\end{equation}
and corresponds to the so-called ``universal twist'' introduced in~\cite{Azzurli:2017kxo}. For some aspects of our evaluation of the TTI to all orders in a $1/N$ expansion, we will make use of these superconformal $(\Delta^*,\mathfrak{n}^*)$-configurations.

\subsubsection{Numerical approach: universal $\log N$ term}\label{sec:TTI:summary:numerical}

The leading order analytic result for the ABJM TTI~\eqref{TTI:M:known:leading} has been supplemented by a numerical analysis that evaluates~\eqref{TTI:3} using numerical solutions to the BAE~\eqref{BAE:2} in~\cite{Liu:2017vll}. The result reads
\begin{equation}
\begin{split}
	\log Z_{S^1\times S^2}(N,k,\Delta,\mn)&=-\fft{\pi\sqrt{2\Delta_1\Delta_2\Delta_3\Delta_4}}{3}\sum_{a=1}^4\fft{\mn_a}{\Delta_a}k^\fft12N^\fft32+f_{1/2}(k,\Delta,\mn)N^\fft12\\
	&\quad-\fft12\log N+f_0(k,\Delta,\mn)+\mathcal O(N^{-\fft12}).\label{TTI:M:known}
\end{split}
\end{equation}
The coefficient of the logarithmic contribution at large $N$ turned out to be universal, meaning that it does not depend on the flavor chemical potentials or on the magnetic fluxes. For the superconformal configuration~\eqref{eq:delta-conf}, the first subleading coefficient $f_{1/2}(k,\Delta,\mn)$ was also obtained explicitly,
\begin{equation}
	f_{1/2}(k,\Delta^*,\mn) = \fft{\pi\sqrt{2k}}{3}\left(\fft{k}{16}+\fft{2}{k}\right).\label{TTI:M:known:N12}
\end{equation}
Both~\eqref{TTI:M:known:N12} and the universal $\log N$ subleading correction were later confirmed analytically using a bulk supergravity analysis~\cite{Bobev:2020egg,Bobev:2021oku} and a one-loop calculation~\cite{Liu:2017vbl}, respectively.

Following the same numerical strategy but working in the IIA limit with large $N$ and fixed 't~Hooft coupling $\lambda=N/k$, the TTI has also been studied in~\cite{PandoZayas:2019hdb}. The result can be summarized as
\begin{align}
		\log Z_{S^1\times S^2}(N,k,\Delta,\mn)&=\left(-\fft{\pi\sqrt{2\Delta_1\Delta_2\Delta_3\Delta_4}}{3\lambda^\fft12}\sum_{a=1}^4\fft{\mn_a}{\Delta_a}+\sum_{s=2}^\infty\fft{g_{-s/2}(\Delta,\mn)}{\lambda^{s/2}}\right)N^2+\fft23\log N \nonumber \\[1mm]
		&\quad+\left(h(\Delta,\mn)\lambda^\fft12-\fft76\log\lambda+\mathcal O(\lambda^0)\right)+\mathcal O(N^{-2}) \, .\label{TTI:tHooft:known}
\end{align}
The coefficient of the logarithmic contribution is again universal. For the superconformal configuration~\eqref{eq:delta-conf},~\cite{PandoZayas:2019hdb} also obtained analytic expressions for some of the subleading coefficients as
\begin{equation}
	g_{-1}(\Delta^*,\mn) = 0 \, , \quad g_{-3/2}(\Delta^*,\mn) = \fft{\pi}{24\sqrt2} \, , \quad h(\Delta^*,\mn) = \fft{2\pi\sqrt2}{3} \, .
\end{equation}
This concludes our review of the known analytic and numerical results obtained from the BA formulation of the ABJM TTI on $S^1 \times S^2$.

\subsection{M-theory expansion of the topologically twisted index}
\label{sec:TTI:pert}

We now improve on the known results just reviewed by providing an analytic expression for the all-order $1/N$ expansion of the TTI in the M-theory limit, up to an $N$-independent constant term and non-perturbative corrections. Before delving into the details, we present our final result which reads
\begin{equation}
\begin{split}
	\log Z_{S^1\times S^2}(N,k,\Delta,\mn)&=-\fft{\pi\sqrt{2k\Delta_1\Delta_2\Delta_3\Delta_4}}{3}\sum_{a=1}^4\fft{\mn_a}{\Delta_a}\left(\hat N_\Delta^\fft32-\fft{\mathfrak c_a}{k}\hat N_\Delta^\fft12\right)\\
	&\quad-\fft12\log\hat N_\Delta+\hat f_0(k,\Delta,\mn)+\hat f_\text{np}(N,k,\Delta,\mn) \, ,
\end{split}\label{TTI:all-order:S^2}
\end{equation}
where $\hat N_\Delta$ and $\mathfrak c_a$ are given by
\begin{subequations}
\begin{align}
	\hat N_\Delta&=N-\fft{k}{24}+\fft{1}{12k}\sum_{a=1}^4\fft{1}{\Delta_a} \, , \label{N:shifted}\\[1mm]
	\mathfrak c_a&=\fft{\prod_{b\neq a}(\Delta_a+\Delta_b)}{8\Delta_1\Delta_2\Delta_3\Delta_4}\sum_{b\neq a}\Delta_b \, .\label{N12:coefficients}
\end{align}\label{TTI:all-order:S^2:detail}%
\end{subequations}
We refer the reader to Section~\ref{sec:TTI:pert:all-order} for the derivation of this formula using a numerical analysis. Before presenting the details, we first collect some general remarks on our result:
\begin{itemize}
	\item At present we do not have a closed form expression for the $N$-independent constant term $\hat f_0(k,\Delta,\mn)$, which we leave for future research. However, for the special flavor chemical potentials in~\eqref{eq:delta-conf}, we have obtained the first few terms in a large $k$ expansion as
	\begin{equation}
		\begin{split}
			\hat f_0(k,\Delta^*,\mn) &=-\fft{3\zeta(3)}{8\pi^2}\,k^2+\fft76\log k+\mathfrak f_0+\sum_{n=1}^5\left(\fft{2\pi}{k}\right)^{2n}\fft{\mathfrak f_{2n}}{3^{n+2}}+\mathcal O(k^{-12}) \, ,
		\end{split}\label{f0:superconformal}
	\end{equation}
	where
	\begin{equation}
		\begin{split}
			\mathfrak f_0&=-2.09684829977578309 \, , \\[1mm]
			\{\mathfrak f_{2n}|n=1,\ldots,5\}&=\left\{-\fft65,\fft{19}{70},-\fft{41}{175},\fft{279}{700},-\fft{964636}{875875}\right\} \, .
		\end{split}\label{f0:superconformal:coefficients}
	\end{equation}
	The expansion~\eqref{f0:superconformal} has a structure similar to the large $k$ expansion of the $N$-independent constant term in the $S^3$ partition function $\mathcal{A}(k) + \fft14\log\fft{k}{32}$, see~\eqref{eq:Z} and \eqref{eq:c-a-b-ABJM-round}. For generic $\Delta$, we were only able to numerically determine the first two terms in the large $k$ expansion of $\hat f_0(k,\Delta,\mn)$ as
	\begin{equation}
		\hat f_0(k,\Delta,\mn)=-\fft{\zeta(3)}{8\pi^2}\,k^2\sum_{a=1}^4\hat f_{0,2,a}(\Delta)\,\mn_a+\fft76\log k+\mathfrak{f}_0(\Delta,\mn)+\mathcal O(k^{-2}) \, , \label{f0:general}
	\end{equation}
	where
	\begin{align}
	\label{f0:general:lead}
			\hat f_{0,2,1}(\Delta)&=\Delta_1+\fft{\Delta_1\Delta_3}{\Delta_1+\Delta_4}+\fft{\Delta_1\Delta_4}{\Delta_1+\Delta_3}+\fft{\Delta_1\Delta_4(\Delta_2+\Delta_3)}{(\Delta_1+\Delta_4)^2}+\fft{\Delta_1\Delta_3(\Delta_2+\Delta_4)}{(\Delta_1+\Delta_3)^2} \nonumber \\
			&\quad-\fft{2\Delta_3\Delta_4}{(\Delta_1+\Delta_3)(\Delta_1+\Delta_4)}-\fft{\Delta_2^2(\Delta_1-\Delta_2)}{(\Delta_2+\Delta_3)(\Delta_2+\Delta_4)} \nonumber \\
			&\quad+\fft{\Delta_2\Delta_3(\Delta_1+\Delta_4)}{(\Delta_1+\Delta_3)(\Delta_2+\Delta_3)}+\fft{\Delta_2\Delta_4(\Delta_1+\Delta_3)}{(\Delta_1+\Delta_4)(\Delta_2+\Delta_4)} \, , \nonumber \\[1mm]
			\hat f_{0,2,2}(\Delta)&=\hat f_{0,2,1}(\Delta)|_{\Delta_1\leftrightarrow\Delta_2} \, , \\
			\hat f_{0,2,3}(\Delta)&=\hat f_{0,2,1}(\Delta)|_{\Delta_1\leftrightarrow\Delta_3,\Delta_2\leftrightarrow\Delta_4} \, , \nonumber \\
			\hat f_{0,2,4}(\Delta)&=\hat f_{0,2,1}(\Delta)|_{\Delta_1\leftrightarrow\Delta_4,\Delta_2\leftrightarrow\Delta_3} \, . \nonumber
	\end{align}
	Note that the $\log k$ term in the expansion~\eqref{f0:general} is universal. Another remark is that the full permutation symmetry between the $\Delta$ is explicitly broken by the function $\hat{f}_0(k,\Delta,\mn)$ down to~\eqref{TTI:sym}. This is similar to the symmetry breaking due to the function $\mathcal{A}(k,b,\Delta_a)$ in the $S^3$ partition discussed in Section~\ref{sec:sphere}. We refer the reader to Section~\ref{sec:TTI:pert:f0} for the derivation of~\eqref{f0:superconformal} and~\eqref{f0:general} using our numerical analysis.

	\item We have studied the non-perturbative corrections $\hat f_\text{np}(N,k,\Delta,\mn)$ in (\ref{TTI:all-order:S^2}), focusing on the superconformal $\Delta^*$-configuration and $k=1,2,3,4$. Extrapolated to all $k$, the result reads
	\begin{equation}
		\hat f_\text{np}(N,k,\Delta^*,\mn) = \sqrt{N}\,e^{-2\pi\sqrt{2N/k} \, + \, \mathcal O(N^0)} \, , \label{np:superconformal}
	\end{equation}
	in the M-theory limit. The numerical analysis and related comments on this result are presented in Section~\ref{sec:TTI:non-pert}. We leave a complete understanding of the non-perturbative corrections to the ABJM TTI in the large $N$ limit for future research.
	
	\item For real flavor chemical potentials, the numerical BAE solutions $\{x^\star_i,\tx^\star_i\}$ used to derive the all-order analytic expression (\ref{TTI:all-order:S^2}) satisfy
	\begin{equation}
		\overline{x^\star_i}=\tx^\star_i\,e^{-\fft{(1+(-1)^N)\pi i}{k}}\label{xtx:cc}
	\end{equation}
	under complex conjugation. The origin of the phase factor is explained in Section~\ref{sec:TTI:pert:all-order}.~Using (\ref{xtx:cc}) together with $\Delta\in\mathbb R$ and further assuming real flavor magnetic fluxes $\mn\in\mathbb R$, one can prove that the contribution from the BAE solutions $\{x^\star_i,\tx^\star_i\}$ to the TTI through the BA formula (\ref{TTI:3}) is real. Evaluating this contribution numerically we also confirmed that it is positive. Hence the logarithm of the all-order TTI (\ref{TTI:all-order:S^2}) is real for $\Delta,\mn\in\mathbb R$, up to contributions from other BAE solutions that we have ignored.
	
	\item For the superconformal $\Delta^*$ in~\eqref{eq:delta-conf}, the BAE solution $\{x^\star_i,\tx^\star_i\}$ also satisfies
	\begin{equation}
		x^\star_i\,x^\star_{N+1-i} = e^{\fft{(1+(-1)^N)\pi i}{k}}\, ,\qquad\tx^\star_i\,\tx^\star_{N+1-i} = e^{\fft{(1+(-1)^N)\pi i}{k}}\,.\label{xtx:sc}
	\end{equation}
	Based on this property, one can prove that the contribution from $\{x^\star_i,\tx^\star_i\}$ to the TTI is independent of the flavor magnetic fluxes $\mn$ that satisfy the constraint (\ref{constraints}). This promotes the earlier observation made at the leading order in Section~\ref{sec:TTI:summary:analytic} to an exact statement valid to all orders in the $1/N$ expansion. Hence the logarithm of the TTI (\ref{TTI:all-order:S^2}) does not depend on $\mn$ when evaluated at $\Delta = \Delta^*$, modulo the contributions from other BAE solutions.
	
	\item The generalization of (\ref{TTI:all-order:S^2}) to $S^1\times\Sigma_\mg$ is straightforward using~\eqref{TTI:Sigma:S^2} and reads
	\begin{equation}
		\begin{split}
			\log Z_{S^1\times\Sigma_\mg}(N,k,\Delta,\mn)&=-\fft{\pi\sqrt{2k\Delta_1\Delta_2\Delta_3\Delta_4}}{3}\sum_{a=1}^4\fft{\mn_a}{\Delta_a}\left(\hat N_\Delta^\fft32-\fft{\mathfrak c_a}{k}\hat N_\Delta^\fft12\right)\\
			&\quad-\fft{1-\mg}{2}\log\hat N_\Delta+\hat f_0(k,\Delta,\mn)+\hat f_\text{np}(N,k,\Delta,\mn) \, ,
		\end{split}\label{TTI:all-order:Sigma}
	\end{equation}
	where we have used the linearity of the TTI with respect to the flavor magnetic fluxes discussed above~\eqref{TTI:sym}. Recall that we assume $\mathfrak{g} \neq 1$ in our analysis.
\end{itemize}
%

\subsubsection{Deriving the all-order M-theory expansion}
\label{sec:TTI:pert:all-order}

Let us now explain how we obtain the all-order M-theory expansion of the ABJM TTI~\eqref{TTI:all-order:S^2} from a numerical analysis. The data that supports our finding is collected in Appendix~\ref{App:all-order}.

To begin with, we construct a numerical solution of the BAE~\eqref{BAE:2} for given $N$, $k$, $\Delta$-configuration, and the set of integers $(n_i,\tn_i)=(1-i,i-N)$ using \texttt{FindRoot} in \textit{Mathematica} at \texttt{WorkingPrecision} $=200$. We use the leading order solution~\eqref{BAE:sol} as initial conditions for $u_i,\tu_i$~\cite{Liu:2017vll} and fix the sign ambiguities as
\begin{equation}
	\sigma_i=\tilde\sigma_i=(-1)^N\,,\label{BA:sign:assump}
\end{equation}
which will be confirmed a posteriori. The leading order solution~\eqref{BAE:sol} does not yield initial conditions for the quantity $\text{Re}[u_i+\tu_i]=v_i+\tv_i$, so in addition we set\footnote{Note that (\ref{v+tv:ic}) matches the initial conditions used in~\cite{Liu:2017vll} only for even $N$. For odd $N$ their initial conditions has to be modified to (\ref{v+tv:ic}) when fixing the sign ambiguities as in (\ref{BA:sign:assump}).}
\begin{equation}
	v_i+\tv_i\Big|_\text{initial condition}=\fft{(1+(-1)^N)\pi}{k}\,.\label{v+tv:ic}
\end{equation}
Substituting the numerical solutions to the BAE produced by~\texttt{FindRoot} with the above choice of sign ambiguities and initial conditions back into~\eqref{BA:sign} and choosing the principal branch for square roots, we have checked that~\eqref{BA:sign:assump} is indeed consistent.

Substituting the numerical solution into the BA formulation \eqref{TTI:3} with a given $\mn$-configuration, we obtain the numerical value of the TTI for a given $N$, $k$, and $(\Delta,\mn)$-configuration. After repeating this process for $N=101\sim301$ in steps of 10, we fit the resulting data with respect to $N$ using \texttt{LinearModelFit} in \textit{Mathematica}. As a result, we obtain the following numerical M-theory expansion of the TTI,
\begin{align}
\label{TTI:M:expansion}
		\log Z_{S^1\times S^2}(N,k,\Delta,\mn)&=f^\text{(lmf)}_{3/2}(k,\Delta,\mn)N^\fft32+f^\text{(lmf)}_{1/2}(k,\Delta,\mn)N^\fft12+f^\text{(lmf)}_\text{log}(k,\Delta,\mn)\log N \nonumber \\
		&\quad+f^\text{(lmf)}_0(k,\Delta,\mn)+\sum_{s=1}^L f^\text{(lmf)}_{-s/2}(k,\Delta,\mn)N^{-\fft{s}{2}} \, ,
\end{align}
with numerical coefficients $f^\text{(lmf)}_X(k,\Delta,\mn)$ for a given $k$ and $(\Delta,\mn)$-configuration. The superscript ``(lmf)'' in the expansion coefficients indicates that they are numerical coefficients obtained with \texttt{LinearModelFit}. The upper bound $L$ for fitting is chosen as $L=16$ to minimize standard errors in estimating the numerical coefficients. 

Next, repeating the \texttt{LinearModelFit}~\eqref{TTI:M:expansion} with five different $\mn$-configurations satisfying the constraint~\eqref{constraints}, namely
\begin{equation}
\label{nas}
		\mathfrak n=(\fft12,\fft12,\fft12,\fft12) \, , \;\; (\fft14,\fft14,\fft34,\fft34) \, , \;\;(\fft14,\fft24,\fft34,\fft24) \, , \;\;(\fft14,\fft34,\fft14,\fft34) \, , \;\;(\fft18,\fft58,\fft38,\fft78) \, ,
\end{equation}
we obtain the numerical M-theory expansion of the TTI that is linear in the flavor magnetic fluxes $\mn$. The result reads
\begin{align}
\label{TTI:M:expansion:linear}
		\log Z_{S^1\times S^2}(N,k,\Delta,\mn)&=\sum_{a=1}^4\left(f^\text{(lmf)}_{3/2,a}(k,\Delta)N^\fft32+f^\text{(lmf)}_{1/2,a}(k,\Delta)N^\fft12\right)\mn_a+f^\text{(lmf)}_{\log}(k,\Delta,\mn)\log N \nonumber \\
		&\quad+f^\text{(lmf)}_0(k,\Delta,\mn)+\sum_{s=1}^L f^\text{(lmf)}_{-s/2}(k,\Delta,\mn)N^{-\fft{s}{2}} \,  .
\end{align}
Note that any four configurations out of the five in~\eqref{nas} are enough to determine the numerical coefficients in~\eqref{TTI:M:expansion:linear}, so one can use the fifth configuration as a consistency check on the linear structure of~\eqref{TTI:M:expansion:linear} with respect to the flavor magnetic fluxes. Recall that this linearity is in fact expected from the BA formulation \eqref{TTI:3}, assuming a dominant contribution from a particular BAE solution as discussed there. In~\eqref{TTI:M:expansion:linear}, we did not introduce a linear dependence of the numerical coefficients on the flavor magnetic fluxes beyond the $\mathcal{O}(N^\fft12)$ order as it will not be necessary in what follows.

We find that the $N^\fft32$ leading order coefficient and the coefficient of the universal $\log N$ term obtained in previous works (see~\eqref{TTI:M:known:leading} and~\eqref{TTI:M:known}) are accurately reproduced in our numerical M-theory expansion:
\begin{equation}
	f^\text{(lmf)}_{3/2,a}(k,\Delta)\simeq-\fft{\pi\sqrt{2k\Delta_1\Delta_2\Delta_3\Delta_4}}{3}\fft{1}{\Delta_a} \, , \qquad f^\text{(lmf)}_\text{log}(k,\Delta,\mn)\simeq-\fft12 \, , \label{f:known}
\end{equation}
for various $k$ and $(\Delta,\mn)$-configurations. Going beyond the known results~\eqref{f:known}, we observe that the negative integer powers of $N$ in our numerical expansion for various $k$ and $(\Delta,\mn)$-configurations can be resummed together with the universal $\log N$ contribution as
\begin{equation}
\begin{split}
	&f^\text{(lmf)}_\text{log}(k,\Delta,\mn)\log N+\sum_{s=1}^\infty f^\text{(lmf)}_{-s}(k,\Delta,\mn)N^{-s}\\
	&\simeq-\fft12\log\left(N-\fft{k}{24}+\fft{1}{12k}\left(\fft{1}{\Delta_1}+\fft{1}{\Delta_2}+\fft{1}{\Delta_3}+\fft{1}{\Delta_4}\right)\right)=-\fft12\log\hat N_\Delta \, , \label{resum}
\end{split}
\end{equation}
in terms of the shifted $N$ parameter introduced in~\eqref{N:shifted}. This resummation strongly motivates implementing \texttt{LinearModelFit} for numerical values of the TTI with respect to $\hat N_\Delta$ rather than $N$. Miraculously, we found that a \texttt{LinearModelFit} for the numerical values of the TTI with respect to $\hat N_\Delta$ terminates at order $\mathcal O(N^0)$ (see again Appendix \ref{App:all-order} for the data). In other words, we found that the following \texttt{LinearModelFit}
\begin{equation}
\begin{split}
	\log Z_{S^1\times S^2}(N,k,\Delta,\mn)+\fft12\log\hat N_\Delta&=\sum_{a=1}^4\left(\hat f^\text{(lmf)}_{3/2,a}(k,\Delta)\hat N_\Delta^\fft32+\hat f^\text{(lmf)}_{1/2,a}(k,\Delta)\hat N_\Delta^\fft12\right)\mn_a\\
	&\quad+\hat f^\text{(lmf)}_0(k,\Delta,\mn) \, ,
\end{split}\label{TTI:M:expansion:shifted}
\end{equation}
with only three fitting functions ($\hat N_\Delta^\fft32$, $\hat N_\Delta^\fft12$, and a constant) yields numerical coefficients with much lower standard errors compared to the previous fit~\eqref{TTI:M:expansion:linear}. In~\eqref{TTI:M:expansion:shifted}, we have pulled out the universal logarithmic contribution $-\fft12\log\hat N_\Delta$ in order to estimate the remaining numerical coefficients more accurately.

\begin{figure}[t!]
	\centering
	\includegraphics[width=0.46\textwidth]{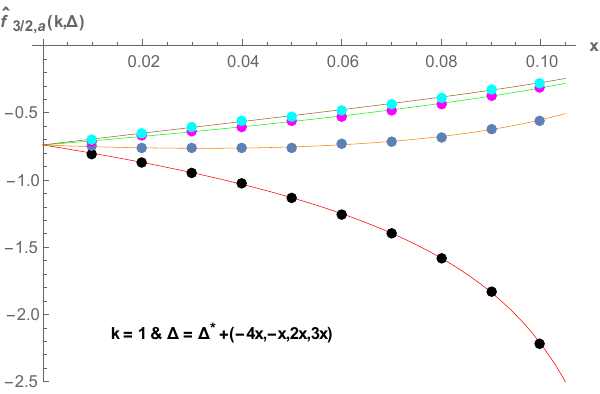}~~
	\includegraphics[width=0.46\textwidth]{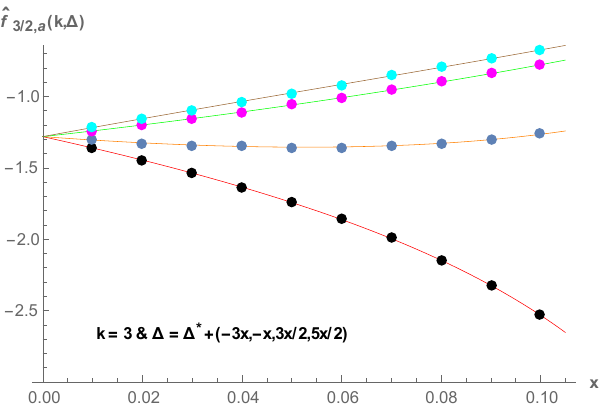}\\
	\includegraphics[width=0.46\textwidth]{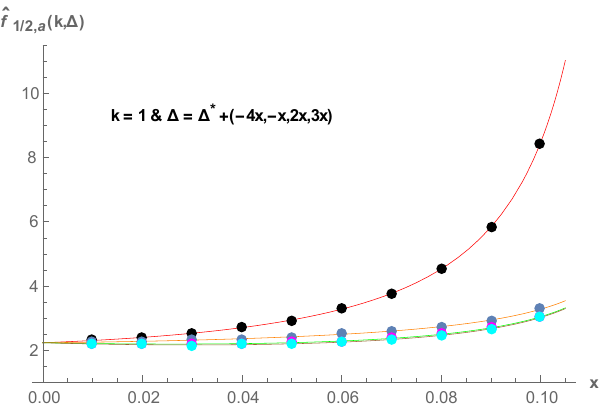}~~
	\includegraphics[width=0.46\textwidth]{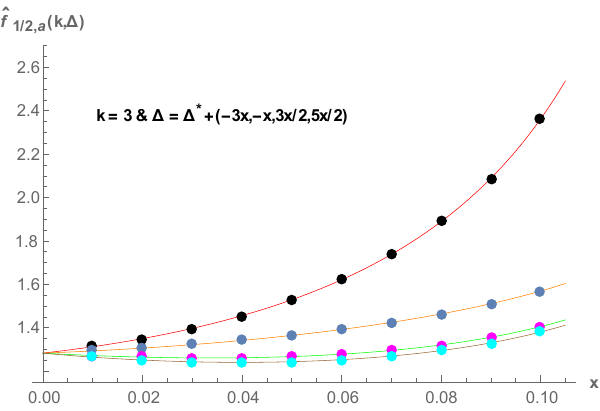}\\
	\caption{Black/blue/magenta/cyan points represent the numerical coefficients $\hat f^\text{(lmf)}_{X,a}(k,\Delta)$ in the fit (\ref{TTI:M:expansion:shifted}), and red/orange/green/brown lines represent the analytic expressions $\hat f_{X,a}(k,\Delta)$ on the RHS of~\eqref{TTI:M:expansion:shifted:coeffi} for $a\in\{1,2,3,4\}$, respectively. The symbol $\Delta^*$ stands for the superconformal configuration $\Delta_a=\fft12$ and $\texttt{x}$ parametrizes the deviation from it. In the left and right panels, we consider two different deviations of flavor chemical potentials from the superconformal configuration with $\texttt{x}\in\{\fft{1}{100},\fft{2}{100},\cdots,\fft{1}{10}\}$.}\label{Pert}
\end{figure}

As a final step, we study the numerical coefficients of the improved numerical M-theory expansion~\eqref{TTI:M:expansion:shifted} for various $k$ and $\Delta$-configurations. We find that they are accurately reproduced by the following analytic expressions,
\begin{equation}
\begin{split}
	\hat f^\text{(lmf)}_{3/2,a}(k,\Delta)&\simeq-\fft{\pi\sqrt{2k\Delta_1\Delta_2\Delta_3\Delta_4}}{3}\fft{1}{\Delta_a} \, , \\
	\hat f^\text{(lmf)}_{1/2,a}(k,\Delta)&\simeq\fft{\pi\sqrt{2k\Delta_1\Delta_2\Delta_3\Delta_4}}{3}\fft{1}{k\Delta_a}\fft{\prod_{b\neq a}(\Delta_a+\Delta_b)}{8\Delta_1\Delta_2\Delta_3\Delta_4}\sum_{b\neq a}\Delta_b \, .
\end{split}\label{TTI:M:expansion:shifted:coeffi}
\end{equation}
Figure~\ref{Pert} shows the comparison between the numerical coefficients obtained from the fit \eqref{TTI:M:expansion:shifted} and the analytic expressions on the RHS of~\eqref{TTI:M:expansion:shifted:coeffi}.
Substituting the expressions~\eqref{TTI:M:expansion:shifted:coeffi} back into the expansion~\eqref{TTI:M:expansion:shifted}, we arrive at the all-order M-theory expansion~\eqref{TTI:all-order:S^2}, where we have replaced the $N$-independent contribution $\hat f^\text{(lmf)}_0(k,\Delta,\mn)$ with the corresponding analytic symbol $\hat f_0(k,\Delta,\mn)$ and restored the non-perturbative corrections $\hat f_\text{np}(N,k,\Delta,\mn)$. The latter were so far ignored since they are exponentially suppressed compared to the power-law terms in the M-theory limit. We will explore $\hat f_0(k,\Delta,\mn)$ and $\hat f_\text{np}(N,k,\Delta,\mn)$ more carefully in the next subsections.

\subsubsection{The large $k$ expansion of $\hat f_0(k,\Delta,\mn)$}
\label{sec:TTI:pert:f0}

We now explain how we obtained the large $k$ expansion of the $N$-independent term $\hat f_0(k,\Delta,\mn)$ given in~\eqref{f0:superconformal} and~\eqref{f0:general} using our numerical analysis. We refer the reader to Appendix~\ref{App:f0} for the supporting numerical data.

We first construct a numerical solution of the BAE~\eqref{BAE:2} for a given $N$, $\lambda=N/k$, and $\Delta$-configuration using \texttt{FindRoot} in \textit{Mathematica} at \texttt{WorkingPrecision} $=200$, using the leading order solution~\eqref{BAE:sol} as initial conditions. Note that, in contrast to the analysis in the previous subsection, we now keep $\lambda$ fixed instead of $k$. Substituting the numerical solution into the BA formulation \eqref{TTI:3} with given $\mn$-configurations, we obtain the numerical value of the TTI for a given $N$, $\lambda$, and $(\Delta,\mn)$-configuration. We then subtract the known analytic terms in the all-order M-theory expansion~\eqref{TTI:all-order:S^2} from the numerical value of the TTI. This procedure therefore yields the sum of the $N$-independent contribution together with the non-perturbative corrections as
\begin{equation}
\begin{split}
	&\log Z_{S^1\times S^2}(N,k,\Delta,\mn)+\fft{\pi\sqrt{2k\Delta_1\Delta_2\Delta_3\Delta_4}}{3}\sum_{a=1}^4\fft{\mn_a}{\Delta_a}\left(\hat N_\Delta^\fft32-\fft{\mathfrak c_a}{k}\hat N_\Delta^\fft12\right)+\fft12\log\hat N_\Delta\\
	&=\hat f_0(k,\Delta,\mn)+\hat f_\text{np}(N,k,\Delta,\mn)\simeq \hat f_0(k,\Delta,\mn) \, , \label{f0np:evaluation}
\end{split}
\end{equation}
In the RHS of~\eqref{f0np:evaluation}, the non-perturbative corrections are expected to be exponentially suppressed compared to the $N$-independent term and we therefore ignore them. We will confirm that these corrections are indeed exponentially suppressed in Section~\ref{sec:TTI:non-pert} below.

After numerically evaluating~\eqref{f0np:evaluation} for various values of $N$, we fit the resulting data with respect to $k=N/\lambda$ using \texttt{LinearModelFit} in \textit{Mathematica}. As a result, we obtain the numerical large $k$ expansion of the $N$-independent term as
\begin{equation}
\begin{split}
	\hat f_0(k,\Delta,\mn)&=\hat f^\text{(lmf)}_{0,2}(\Delta,\mn)\,k^2+\hat f^\text{(lmf)}_{0,\log}(\Delta,\mn)\log k+\hat f^\text{(lmf)}_{0,0}(\Delta,\mn) \\
&\quad +\sum_{s=1}^L\hat f^\text{(lmf)}_{0,-2s}(\Delta,\mn)\,k^{-2s} \, ,
\end{split}\label{f0:expansion}
\end{equation}
with numerical coefficients $\hat f^\text{(lmf)}_{0,X}(\Delta,\mn)$ for a given $(\Delta,\mn)$-configuration. The upper bound $L$ for fitting is chosen to minimize standard errors in estimating these numerical coefficients.

\medskip

\noindent\textbf{Superconformal $(\Delta)$-configuration}

\medskip

As discussed below~\eqref{TTI:all-order:S^2}, the TTI is independent of $\mn$ for the superconformal configuration $\Delta = \Delta^*$ given in~\eqref{eq:delta-conf}. In this case, we can thus simplify our notation for the numerical coefficients in the expansion~\eqref{f0:expansion} by suppressing their arguments:
\begin{equation}
	\hat f^\text{(lmf)}_{0,X}(\Delta^*,\mn) = \hat f^\text{(lmf)}_{0,X}.
\end{equation}
We were able to deduce the following expressions for the numerical coefficients:
\begin{equation}
\begin{alignedat}{3}
		\hat f^\text{(lmf)}_{0,2}&\simeq-\fft{3\zeta(3)}{8\pi^2},\qquad&\hat f^\text{(lmf)}_{0,\log} &\simeq\fft76,\qquad& \hat f^\text{(lmf)}_{0,0}&\simeq-2.09684829977578309,\\
		\hat f^\text{(lmf)}_{0,-2}&\simeq-\fft{8\pi^2}{45},\qquad&\hat f^\text{(lmf)}_{0,-4}&\simeq\fft{152\pi^4}{2835},\qquad&\hat f^\text{(lmf)}_{0,-6}&\simeq-\fft{2624\pi^6}{42525},\\
		\hat f^\text{(lmf)}_{0,-8}&\simeq\fft{1984\pi^8}{14175},\qquad&\hat f^\text{(lmf)}_{0,-10}&\simeq-\fft{987787264\pi^{10}}{1915538625},
\end{alignedat}\label{f0:superconformal:coeffi:num}%
\end{equation}
where we have taken $N=101\sim551$ in steps of 10 and chosen $L=42$ in the fit~\eqref{f0:expansion}. Aside from~$\hat f^\text{(lmf)}_{0,0}$, these results are analytic and together they yield~\eqref{f0:superconformal}. We refer the reader to Appendix~\ref{App:f0:superconformal} for the numerical precision of the above estimates.

\medskip

\noindent\textbf{Generic $(\Delta,\mn)$-configuration}

\medskip

For generic $(\Delta,\frak{n})$-configurations, we focus on the $\mathcal{O}(k^2)$ leading order term and on the subleading $\log k$ term in the expansion~\eqref{f0:expansion}, namely $\hat f^\text{(lmf)}_{0,2}(\Delta,\mn)$ and $\hat f^\text{(lmf)}_{\log}(\Delta,\mn)$. From the structure of the $k^2$ term in the superconformal case~\eqref{f0:superconformal:coeffi:num}, and again invoking the linearity of the TTI with respect to the flavor magnetic fluxes $\mn$, we expect that $\hat f^\text{(lmf)}_{0,2}(\Delta,\mn)$ for generic $(\Delta,\mn)$-configurations takes the following form, 
\begin{equation}
	\hat f^\text{(lmf)}_{0,2}(\Delta,\mn)=-\fft{\zeta(3)}{8\pi^2}\sum_{a=1}^4\hat f^\text{(lmf)}_{0,2,a}(\Delta)\,\mn_a\simeq-\fft{\zeta(3)}{8\pi^2}\sum_{a=1}^4\hat f_{0,2,a}(\Delta)\,\mn_a \, ,\label{f0:expansion:lead}
\end{equation}
where $\hat f_{0,2,a}(\Delta)$ are some rational functions of $\Delta$. We first determine the numerical coefficients $\hat f^\text{(lmf)}_{0,2,a}(\Delta)$ by implementing the \texttt{LinearModelFit}~\eqref{f0:expansion} with the five $\mn$-configurations listed in~\eqref{nas} and for a given $\Delta$-configuration. Repeating this process for various $\Delta$-configurations, we deduce the rational functions $\hat f_{0,2,a}(\Delta)$ and the result is given in~\eqref{f0:general:lead}. In doing so, we were also able to confirm that
\begin{equation}
	\hat f^\text{(lmf)}_{0,\log}(\Delta,\mn)\simeq\fft76 \, , \label{f0:general:log}
\end{equation}
which is universal. The expansion (\ref{f0:expansion}) with the coefficients (\ref{f0:expansion:lead}), (\ref{f0:general:lead}), and (\ref{f0:general:log}) yields the analytic expression (\ref{f0:general}). To obtain the above results, we have used a number of $N$ values and corresponding upper bounds $L$ in the \texttt{LinearModelFit}. See Appendix \ref{App:f0:general} for more details and the numerical precision of the estimates (\ref{f0:general:lead}) and (\ref{f0:general:log}).

\subsection{Non-perturbative corrections to the topologically twisted index}
\label{sec:TTI:non-pert}

In this subsection we explore the non-perturbative corrections in~\eqref{TTI:all-order:S^2} numerically, focusing on the superconformal configuration $\Delta^*_a=\fft12$ for which the TTI does not depend on $\mn$. The starting point is~\eqref{f0np:evaluation}, which we repeat here for the superconformal configuration:
\begin{equation}
		\log Z_{S^1\times S^2}(N,k,\Delta^*,\mn) +\fft{\pi\sqrt{2k}}{3}\left(\hat N^\fft32-\fft3k\hat N^\fft12\right)+\fft12\log\hat N =\hat f_0(k)+\hat f_\text{np}(N,k) \, .
\label{f0np:evaluation:superconformal}
\end{equation}
Above, we have introduced the appropriate shifted $N$ for the superconformal case, 
\begin{equation}
\label{eq:N-shifted-min}
\hat N=N-\fft{k}{24}+\fft{2}{3k} \, ,
\end{equation} 
according to (\ref{N:shifted}). In contrast to the discussion above, here we do not discard the non-perturbative corrections $\hat f_\text{np}(N,k)$.

\begin{figure}[t!]
	\centering
	\includegraphics[width=0.72\textwidth]{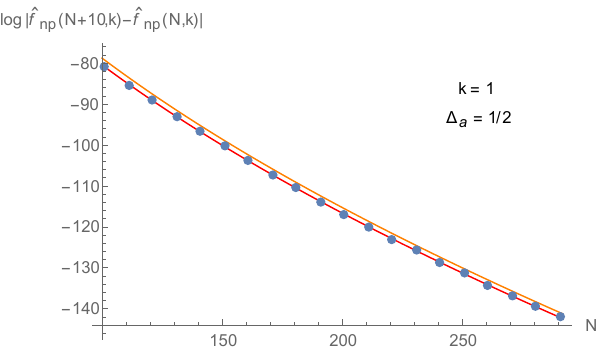}
	\caption{Numerical values of the difference between adjacent non-perturbative corrections (\ref{fnp}) (blue dots) versus the corresponding fitting curves (orange/red lines include the first three/five terms in the RHS of (\ref{fnp:expansion}), respectively).}\label{NonPert}
\end{figure}

Using the fact that $\hat f_0(k)$ does not depend on $N$, we can easily extract $\hat f_\text{np}(N,k)$ from~\eqref{f0np:evaluation:superconformal}. To be specific, we evaluate~\eqref{f0np:evaluation:superconformal} for $N=101\sim301$ in steps of 10 with fixed $k$ and then subtract the results with adjacent $N$ values. As a result, we generate the numerical values\footnote{The subleading correction in the RHS of (\ref{fnp}) is in general $\mathcal{O}(\log N)$ for the
	expansion (\ref{fnp:expansion}). We plan to study these corrections in more detail in future work.}
\begin{equation}
	\log\left|\hat f_\text{np}(N+10,k)-\hat f_\text{np}(N,k)\right|\simeq\log\left|\hat f_\text{np}(N,k)\right|+\mathcal O(\log N) \, , \label{fnp}
\end{equation}
for $N=101\sim291$ in steps of 10. The approximation in~\eqref{fnp} will be justified a posteriori below. The next step is to use \texttt{LinearModelFit} with respect to $N$, 
\begin{equation}
	\log\left|\hat f_\text{np}(N,k)\right|=\hat f^\text{(lmf)}_{\text{np},1/2}(k)N^\fft12+\hat f^\text{(lmf)}_{\text{np},\log}(k)\log N+\sum_{s=0}^L\hat f^\text{(lmf)}_{\text{np},-s}(k)N^{-s/2},\label{fnp:expansion}
\end{equation}
where the upper bound $L$ is chosen as $L=16$ to minimize standard errors in estimating the numerical coefficients. In Figure~\ref{NonPert}, we give the comparison between numerical values of the non-perturbative corrections~\eqref{fnp} and a fitting curve obtained from the RHS of~\eqref{fnp:expansion}. 
Repeating this process for $k=1,2,3,4$, we found
\begin{equation}
	\hat f^\text{(lmf)}_{\text{np},1/2}(k)\simeq-2\pi\sqrt{\fft2k} \, .\label{fnp:expansion:coeffi}
\end{equation}
Substituting the estimates (\ref{fnp:expansion:coeffi}) back into the expansion (\ref{fnp:expansion}), we obtain the non-perturbative behavior 
\begin{equation}
	\hat f_\text{np}(N,k)=e^{-2\pi\sqrt{2N/k} \, + \, \mathcal O(\log N)}\label{np:superconformal:2}
\end{equation}
given in (\ref{np:superconformal}). The following table provides error ratios for the estimates (\ref{fnp:expansion:coeffi}),
\begin{center}
	\footnotesize
	\begin{tabular}{ |c||c| } 
		\hline
		& $R_{\text{np},1/2}$  \\
		\hline\hline
		$k=1$ & $7.372{\times}10^{-10}$ \\
		\hline
		$k=2$ & $-9.384{\times}10^{-10}$ \\
		\hline
		$k=3$ & $-5.819{\times}10^{-11}$ \\ 
		\hline
		$k=4$ & $-9.650{\times}10^{-12}$ \\ 
		\hline
	\end{tabular}
\end{center}
where
\begin{equation}
	R_{\text{np},1/2}=\fft{\hat f^\text{(lmf)}_{\text{np},1/2}(k)-(-2\pi\sqrt{2/k})}{-2\pi\sqrt{2/k}}\, .
\end{equation}
Note that the non-perturbative behavior (\ref{np:superconformal:2}) indeed allows for the approximation in (\ref{fnp}) and the $\mathcal O(\log N)$ order difference does not affect the leading order estimate (\ref{fnp:expansion:coeffi}).

It is worth mentioning that the leading non-perturbative behavior of (\ref{np:superconformal:2}) in the M-theory limit, namely $\sim e^{-2\pi\sqrt{2N/k}}$, is precisely the same as the one for the $S^3$ partition function. The latter has been obtained by studying worldsheet (WS) instanton and D2 instanton contributions~\cite{Hatsuda:2012dt}. As shown in that paper, the leading non-perturbative behavior $\sim e^{-2\pi\sqrt{2N/k}}$ comes from a dominant WS instanton for $k>2$; for $k=1,2$, combined contributions from both instantons yield the leading non-perturbative behavior $\sim e^{-2\pi\sqrt{2N/k}}$. It would be very interesting if non-perturbative corrections of the form (\ref{np:superconformal:2}) could also be understood analytically for the TTI. We hope to come back to this point in the future.

We have also investigated non-perturbative corrections to the TTI for some generic $(\Delta,\mn)$-configuration following the same strategy as above. The leading non-perturbative behavior is always of order $\mathcal O(e^{-\#\sqrt N})$, even though we could not deduce an analytic expression for the coefficient $\#$ as a function of $\Delta$ and $\mn$. The coefficient of this leading non-perturbative correction, as well as the $\log N$ subleading behavior, are also not captured with reasonable precision by our numerics. We leave a more detailed investigation of this for the future.

\subsection{Type IIA expansion of the topologically twisted index}

Since the all-order M-theory expansion~\eqref{TTI:all-order:S^2} is exact up to non-perturbative corrections, one can obtain the type IIA expansion of the TTI by replacing $k$ with $k=N/\lambda$ and keeping $\lambda$ fixed. From the structure of the shifted $N$ parameter~\eqref{N:shifted}, it is then straightforward to see that the type IIA expansion of the TTI can be written in terms of the shifted 't~Hooft parameter $\hat\lambda = \lambda-\fft{1}{24}$ as in the case of the $S^3$ partition function in Section~\ref{sec:IIA-expand}. To be specific, for the superconformal configuration~\eqref{eq:delta-conf}, we have\footnote{We have used the same notation $F_{\tt g}(\lambda)$ for the genus-$\tt{g}$ free energies obtained from the $S^3$ partition function (\ref{eq:IIA-expand}) and from the $S^1\times S^2$ TTI (\ref{TTI:all-order:S^2:2A}), but they can be distinguished clearly from the context.}
\begin{equation}
	\log Z_{S^1\times S^2}(N,k,\Delta^*,\mn) = \sum_{\tt{g}\geq0}(2\pi \mathrm{i}\lambda)^{2\tt g-2}F_{\tt g}(\lambda)\,N^{2-2\tt g} \, , \label{TTI:all-order:S^2:2A}
\end{equation}
where the first few terms in the IIA expansion read
\begin{equation}
\label{eq:Fg-TTI}
	\begin{split}
		F_0(\lambda)&=\fft{4\pi^3\sqrt2}{3}\,\hat\lambda^\fft32+\fft{3\zeta(3)}{2} \, , \\
		F_1(\lambda)&=\fft{2\pi\sqrt2}{3}\,\hat\lambda^\fft12-\fft12\log\hat\lambda-\fft23\log\lambda+\mathfrak{f}_0 \, , \\
		F_2(\lambda)&=\fft{\hat\lambda^{-1}}{12\pi^2}-\fft{5\hat\lambda^{-\fft12}}{36\sqrt2\pi}+\fft{2}{45} \, , \\
		F_3(\lambda)&=\fft{\hat\lambda^{-2}}{144\pi^4}-\fft{\hat\lambda^{-\fft32}}{162\sqrt2\pi^3}+\fft{19}{5670} \, .
	\end{split}
\end{equation}
In $F_1(\lambda)$, the constant $\fft23\log N$ term has been discarded for the same reason explained below (\ref{eq:F1-ABJM-Marino}). We emphasize that these results are exact up to non-perturbative corrections in~$\lambda$, and that they do not depend on the flavor magnetic fluxes as follows from the comment below~\eqref{TTI:all-order:S^2}. To the best of our knowledge, no alternative derivation of~\eqref{eq:Fg-TTI} exists in the literature, in stark contrast with the case of the $S^3$ free energies discussed in Section~\ref{sec:IIA-expand}. As such, the precision of our numerical analysis for the TTI opens a new window into the properties of type IIA string theory. 

Proceeding in a similar fashion, for generic $(\Delta,\mn)$-configurations we find
\begin{equation}
\begin{split}
	F_0(\lambda,\Delta,\mn)&=\fft{4\pi^3\sqrt{2\Delta_1\Delta_2\Delta_3\Delta_4}}{3}\sum_{a=1}^4\fft{\mn_a}{\Delta_a}\hat\lambda^\fft32+\fft{\zeta(3)}{2}\sum_{a=1}^4\hat f_{0,2,a}(\Delta)\,\mn_a \, , \\
	F_1(\lambda,\Delta,\mn)&=\fft{\pi\sqrt{2\Delta_1\Delta_2\Delta_3\Delta_4}}{3}\sum_{a=1}^4\fft{\mn_a}{\Delta_a}\left(\mathfrak{c}_a-\fft18\sum_{b=1}^4\fft{1}{\Delta_b}\right)\hat\lambda^\fft12\\
	&\quad-\fft12\log\hat\lambda-\fft23\log\lambda+\mathfrak{f}_0(\Delta,\mn) \, ,
\end{split}
\end{equation}
where $\hat f_{0,2,a}(\Delta)$ is given explicitly in (\ref{f0:general:lead}). The $\mathfrak{f}_0(\Delta,\mn)$ term in the genus-1 free energy is introduced in (\ref{f0:general}) and can be obtained numerically for various $(\Delta,\mn)$-configurations but its closed form expression is left for future research. \\

Our results in the type IIA frame suggest that the genus-${\tt g}$ free energies for asymptotically locally AdS$_4\times \mathbb{CP}^3$ Euclidean string theory backgrounds of the type discussed in \cite{Bobev:2020pjk} can be obtained to all orders in $\alpha'$. It would be very interesting to understand how this calculation can be performed from the point of view of the worldsheet theory.

\section{Comments on holography}
\label{sec:holography}

Our results for the all-order $1/N$ expansion of the ABJM TTI derived in the previous section have important implications for holography which we now discuss. We begin by reviewing the so-called Euclidean Romans background~\cite{Romans:1991nq,Bobev:2020pjk}. This supersymmetric solution of the equations of motion of Euclidean 4d $\mathcal{N}=2$ minimal gauged supergravity reads
\begin{equation}
\label{eq:Romans}
\begin{split}
ds_4^2 =&\; U(r) d\tau^2 + \frac{dr^2}{U(r)} + r^2 ds^2_{\Sigma_\mathfrak{g}} \, , \quad U(r) = \Bigl(\frac{r}{L} + \frac{\kappa L}{2r}\Bigr)^2 - \frac{q^2}{4r^2} \, , \\
F =&\; \frac{q}{r^2}\,d\tau\wedge dr \pm \kappa L V_{\Sigma_\mathfrak{g}} \, .
\end{split}
\end{equation}
Here $ds^2_{\Sigma_\mathfrak{g}}$ denotes the metric on a constant curvature Riemann surface of genus $\mathfrak{g}$, normalized such that the curvature $\kappa$ is given by $\kappa = \{1,0,-1\}$ for genus $\mathfrak{g} = 0$, $\mathfrak{g} = 1$ and $\mathfrak{g} > 1$, respectively. Supersymmetry requires that the magnetic flux $p$ across $\Sigma_\mathfrak{g}$ has constant magnitude $|p| = |\kappa|L$, while the electric charge $q$ is a free parameter. The volume form on the Riemann surface is denoted by $V_{\Sigma_\mathfrak{g}}$ and satisfies
\begin{equation}
\int_{\Sigma_\mathfrak{g}} V_{\Sigma_\mathfrak{g}} = \begin{cases} 2|\mathfrak{g} - 1| \;\; &\text{for}\;\; \mathfrak{g} \neq 1 \\ 1 \quad &\text{for}\;\; \mathfrak{g} = 1\end{cases} \, .
\end{equation}
The function $U(r)$ has two zeroes located at
\begin{equation}
r_\pm = L\sqrt{-\frac{\kappa}{2} \pm \frac{|q|}{2L}} \, .
\end{equation}
Imposing that $r_+ \in \mathbb{R}_{>0}$ ensures that the spacetime smoothly caps off and is free of naked singularities. Thus we require that $|q| > \kappa L$, which implies that the limit $|q| \rightarrow 0$ cannot be taken in a smooth manner when the Riemann surface is a sphere or a torus. As $r \rightarrow r_+$, the metric becomes locally $\mathbb{R}^2 \times \Sigma_\mathfrak{g}$, and the polar angle $\tau$ for the $\mathbb{R}^2$ factor has periodicity $\beta = 2\pi L r_+|q|^{-1}$. This shows that for $\mathfrak{g}>1$ the Euclidean Romans solution admits an analytic continuation to the Lorentzian extremal Reissner-Nordstr\"om black hole when $|q| \rightarrow 0$. In this limit, the Lorentzian spacetime develops an infinite throat in the AdS$_2 \times \Sigma_{\mathfrak{g}>1}$ near-horizon region of the black hole.

The two-derivative Euclidean regularized on-shell action of the Romans solution was computed in~\cite{Azzurli:2017kxo,BenettiGenolini:2019jdz} and reads
\begin{equation}
I_{2\partial} = (1 - \mathfrak{g})\,\frac{\pi L^2}{2 G_N} \, .
\end{equation}
The right-hand side of this expression is nothing but the negative of the Bekenstein-Hawking entropy of the Lorentzian black hole solution. Going further, it was shown in~\cite{Bobev:2021oku} that this relation persists in a four-derivative theory of gauged supergravity, where the higher-derivative couplings corresponding to the square of the Weyl multiplet and the Gauss-Bonnet density are parametrized by two real constants $c_1$ and $c_2$. In terms of these, the correction to the Bekenstein-Hawking entropy $\Delta S$ can be computed using Wald's formula. The result is:
\begin{equation}
S_{\text{BH}} + \Delta S = - I_{4\partial} = (\mathfrak{g}-1)\left(\frac{\pi L^2}{2 G_N} + 32\pi^2(c_1 + c_2)\right) \, .
\end{equation}
Using the corrected holographic dictionary~\eqref{eq:L/G-all-order} and~\eqref{eq:c1c2-all-order}, we thus find that the corrected Bekenstein-Hawking entropy of the Lorentzian black hole can be expressed in terms of the boundary SCFT data $(N,k)$ as
\begin{equation}
S_{\text{BH}} + \Delta S = (\mathfrak{g} - 1)\left(\frac{\pi\sqrt{2k}}{3}\,N^{\frac32} - \frac{\pi(k^2 + 32)}{24\sqrt{2k}}\,N^{\frac12}\right) + \mathcal{O}(\log N) \, .
\end{equation}
This precisely agrees with the logarithm of the ABJM TTI~\eqref{TTI:all-order:Sigma} upon setting $\Delta_a = \frac12$ and expanding up to order $\mathcal{O}(\log N)$. The logarithmic correction to the entropy was computed in~\cite{Liu:2017vbl} and agrees with the $\log N$ term in $\log Z_{S^1 \times \Sigma_{\mathfrak{g}}}$. Thus, our all-order result for the TTI can be interpreted as encoding all perturbative corrections to the Bekenstein-Hawking entropy of the black hole obtained in the $|q| \rightarrow 0$ limit of the Euclidean Romans solution~\eqref{eq:Romans}. We note here that while the TTI is really an index, the agreement to sub-subleading order with a gravitational computation of the entropy suggests that no dangerous cancellations due to the insertion of $(-1)^F$ occur. This is similar to the situation for asymptotically flat black holes~\cite{Dabholkar:2010rm}.

The Euclidean Romans solution admits the following uplift to 11d~\cite{Gauntlett:2007ma}:
\begin{equation}
\label{eq:Romans-11}
\begin{split}
ds^2_{11} =&\; \fft14\,ds_4^2 + L^2\,ds^2_{\mathbb{CP}^3} + L^2\Bigl(d\psi + \sigma_{\mathbb{CP}^3} +\frac{1}{4L}\,A\Bigr)^2 \, , \\
G_4 =&\; \frac{3L^3}8\,\text{vol}_4 - \frac{L^2}{4}\,\star_4F\wedge J_{\mathbb{CP}^3} \, ,
\end{split}
\end{equation}
where the 11d metric is given in terms of the line element in~\eqref{eq:Romans} and $\sigma_{\mathbb{CP}^3}$ such that $d\sigma_{\mathbb{CP}^3} = 2\,J_{\mathbb{CP}^3}$ with $J_{\mathbb{CP}^3}$ the K\"ahler form on $\mathbb{CP}^3$. In (\ref{eq:Romans-11}) we kept the original convention of the Hodge star operation in the Lorentzian signature. The four form flux $G_4$ is given in terms of the 4d field strength $F = dA$ in~\eqref{eq:Romans}. The result for the ABJM TTI presented in~\eqref{TTI:all-order:Sigma}, upon setting $\Delta_a = \tfrac12$, can then be interpreted as an all-order prediction for the path-integral of M-theory on the background~\eqref{eq:Romans-11} up to exponentially suppressed corrections of the form~$\mathcal{O}(e^{-\sqrt{N}})$. \\

Going beyond the minimal supergravity case corresponding to the universal twist, we can introduce bulk scalar fields dual to mass deformations in the boundary ABJM theory. The corresponding Euclidean solutions of 4d $\mathcal{N}=2$ gauged supergravity have been discussed in~\cite{Bobev:2020pjk} where it was also shown that, for certain values of the solution parameters, there exist corresponding regular Lorentzian black holes coupled to scalar fields \cite{Gauntlett:2001qs,Cacciatori:2009iz}. The analysis of the parameter space is much more involved than in the minimal case and was mostly conducted numerically in~\cite{Bobev:2020pjk}. Nevertheless, once a proper range of Euclidean solution parameters has been identified, it is natural to expect that our all-order expression for the ABJM TTI for generic $(\Delta,\mn$)-configurations encodes all perturbative corrections to the Bekenstein-Hawking entropy of the Lorentzian black hole. Note however that, while the TTI is naturally written in an ensemble of fixed flavor chemical potentials $\Delta$, the entropy is expressed in the microcanonical ensemble of fixed charges. Thus, to fully extract the perturbatively exact thermodynamic entropy from the TTI, one needs to perform a Laplace transform to switch ensembles. This two-step procedure was implemented using the leading $\mathcal{O}(N^{\frac32})$ order for the TTI in~\cite{Benini:2015eyy,Benini:2016rke}, where the result was successfully matched to the Bekenstein-Hawking entropy of the corresponding Lorentzian black hole. It will be illuminating to concretely compute the subleading corrections to this entropy using our result~\eqref{TTI:all-order:Sigma}. \\

From a gauged supergravity point of view, confirming the predictions for the all-order perturbative corrections to the Bekenstein-Hawking entropy of asymptotically AdS$_4$ Lorentzian black holes (with or without scalar fields) appears to be a monumental task. The addition of higher-derivative couplings as in~\cite{Bobev:2020egg,Bobev:2021oku} can give access to the first subleading corrections, but this avenue gets increasingly more complex and unwieldy as we increase the order in derivatives. An alternative approach has recently been discussed in~\cite{Hristov:2021qsw,Hristov:2022lcw,Hristov:2018lod,Hristov:2019xku}, based on the method of supergravity localization. This technique was pioneered in~\cite{Dabholkar:2010uh,Dabholkar:2011ec} and used to exactly compute the entropy of asymptotically flat black holes in certain string compactifications (see e.g.~\cite{Dabholkar:2014ema,Murthy:2015zzy}), and later shown to generalize to gauged supergravity theories and asymptotically AdS black hole solutions therein~\cite{Hristov:2018lod,Hristov:2019xku}. But while the results of~\cite{Hristov:2021qsw,Hristov:2022lcw} have been shown to be compatible with our numerical studies of the TTI leading to~\eqref{TTI:all-order:Sigma}, it is worth emphasizing that this bulk localization approach remains conjectural at this stage. This is mostly due to the fact that a complete analysis of the localizing manifold in gauged supergravity, as well as the all-order form of the prepotential specifying the bulk action, are still lacking. A similar supergravity computation of the exact partition function of empty AdS$_4$ put forward in~\cite{Dabholkar:2014wpa}, but the match to the Airy-type partition function  of the ABJM theory on $S^3$ currently has the same conjectural status. In particular, the shift in $N$ due to the quantity $B$ in~\eqref{eq:Z} seems to lack a proper explanation. It will be invaluable to revisit the gauged supergravity localization framework to establish a precise match between the Airy-type partition functions or the TTI of the ABJM theory on the one hand, and the partition functions or thermodynamic entropy of various gauged supergravity backgrounds on the other, to all orders in perturbation theory. We hope to come back to this in the future, using the results presented in the present paper as a useful guide to extract such perturbatively exact results.

\section{Discussion}
\label{sec:discussions}

The Airy conjecture for the partition function of mass-deformed ABJM theory on the squashed three-sphere presented in Section~\ref{sec:sphere} contains a wealth of holographic information regarding the correlation functions in M-theory on backgrounds of the form $M_4 \times S^7/\mathbb{Z}_k$, or equivalently that of Type IIA string theory on the dual $M_4 \times \mathbb{CP}_3$. Here $M_4$ is either Euclidean AdS$_4$ or the 4d supergravity solutions dual to a the mass-deformed ABJM theory on a round or squashed $S^3$ studied in~\cite{AlonsoAlberca:2000cs,Martelli:2011fu,Freedman:2013oja}. We hope that these results can be applied in the context of the program pursued in \cite{Agmon:2017xes,Chester:2018aca,Binder:2018yvd,Binder:2019mpb,Agmon:2019imm} where supersymmetric localization results at large $N$ are used to determine or constrain the higher-derivative couplings in the effective action of string or M-theory.

Another holographic application of these SCFT results is to determine the corrections to the regularized on-shell action of the four-dimensional gauged supergravity theory obtained from a consistent truncation of the 11d theory \cite{Bobev:2020egg,Bobev:2020zov,Bobev:2021oku}. Such corrections can then be used to infer the form of the corrected prepotential of the supergravity theory and the finite $N$ corrections to the holographic dictionary relating bulk and boundary quantities, see~\eqref{eq:4der-STU} and~\eqref{eq:L/G-all-order}. The fact that we can access the higher-derivative corrections to the prepotential of the consistent truncation is particularly interesting, as it points to a structure akin to the one observed in string theory compactifications on Calabi-Yau manifolds. In the latter case it was understood long ago that the chiral superspace integrals controlling the effective action of 4d $\mathcal{N}=2$ ungauged supergravity could be studied using topological strings~\cite{Bershadsky:1993cx,Antoniadis:1993ze}. The prepotential of the four-dimensional supergravity theory in this case takes the form of an expansion
\begin{equation}
F(X,W) = \sum_{{\tt g} \geq 0} F_{\tt g}(X) W^{2{\tt g}} \, ,
\end{equation}
where $W$ is the chiral superfield of the $\mathcal{N}=2$ Weyl multiplet, and the $F_{\tt g}$ are the genus-$\tt g$ partition functions of topological string theory. Our analysis in Section~\ref{sec:expand} and Appendix~\ref{app:squashed-prepot} suggests that the situation in gauged supergravity involves a double expansion of the form above using both the Weyl multiplet as well as the superfield giving rise to the Gauss-Bonnet density in the 4d theory. The scalar field-dependent coefficients in this expansion can then be read off from our Airy conjecture. It would be most interesting to study this point in more detail. A dramatic consequence of such an analysis could be the formulation of an OSV-type formula~\cite{Ooguri:2004zv} for asymptotically AdS black holes. As mentioned in Section~\ref{sec:sphere}, the matrix models obtained from localization of the ABJM partition function can also be directly linked to a topological string theory on non-compact Calabi-Yau manifolds in certain favorable situations. This points to the possibility that topological string theory can be used to construct the effective action of gauged supergravities arising as consistent truncations, similar to the ungauged supergravities coming from KK reductions of string theory on compact CY manifolds. Independent of this potential relation to topological strings, it will be most interesting to continue the program of exploiting supersymmetric localization and holography in order to rigorously determine the corrections to the gauged supergravity prepotential in as many examples as possible, including also generalizations to different string and M-theory embeddings and gauged supergravity theories in other dimensions. 

Turning to our all-order result for the TTI in Section~\ref{sec:TTI}, the BA formulation~\eqref{TTI:3} shows that $Z_{S^1\times S^2}$ in the ABJM theory should schematically take the form
\begin{equation}
Z_{S^1 \times S^2} = e^{\mathcal{B}_1} + e^{\mathcal{B}_2} + \ldots \, ,
\end{equation}
where $\mathcal{B}_i$ refers to the contribution from the $i$-th solution to the BAE. Let $\mathcal{B}_1$ be the all-order result~\eqref{TTI:all-order:S^2}. If say, $\mathcal{B}_2$ would be a more dominant solution to the BAE in the large $N$ limit, it would mean that $\log Z_{S^1 \times S^2} \sim \mathcal{B}_2$ in this limit. The holographic agreement with bulk supergravity computations we reviewed at order $\mathcal{O}(N^{3/2})$, $\mathcal{O}(N^{1/2})$ and $\mathcal{O}(\log N)$ would then imply that $\mathcal{B}_2$ should be equal to $\mathcal{B}_1$ given in~\eqref{TTI:all-order:S^2}, at least up to this order, provided the Euclidean Romans background is the unique solution holographically dual to the TTI. There could be discrepancies at order $\mathcal{O}(N^{-p/2})$ for some $p > 0$, but those would be suppressed in the large $N$ limit. Therefore, holography suggests that the logarithm of the ABJM TTI in the large $N$ limit takes the schematic form
\begin{equation}
\label{eq:TTI-schem}
\log Z_{S^1 \times S^2} = \mathcal{B}_1 + \log\Bigl[1 + e^{\mathcal{B}_2 - \mathcal{B}_1} + \ldots\,\Bigr] \, ,
\end{equation}
where
\begin{equation}
\mathcal{B}_2 - \mathcal{B}_1 = \alpha_1\,N^{\tfrac32} + \alpha_2\,N^{\tfrac12} + \alpha_3 \log N + \alpha_4 + \ldots \, , \qquad \alpha_1,\alpha_2,\alpha_3 \leq 0 \, ,
\end{equation}
where the dots indicate terms that vanish in the large $N$ limit. The one caveat to this discussion is when $\alpha_4$ does not vanish, as it would then produce a constant contribution to~\eqref{eq:TTI-schem} and shift the constant $\hat{f}_0(k,\Delta,\mn)$ in $\mathcal{B}_1$. Since this pure constant is hard to extract from supergravity, holography cannot yet fix it unambiguously.

The structure just discussed also implies that our all-order result~\eqref{TTI:all-order:S^2} does not completely capture the integer degeneracies of the dual black hole microstates. Rather, to obtain a positive integer from the ABJM TTI, one must be sure to sum the contributions from all solutions to the BAE as
\begin{equation}
d(N) = e^{a_1 N^{3/2} + a_2 N^{1/2} + \ldots + \mathcal{O}(e^{-\sqrt{N}})} + e^{b_1 N^{3/2} + b_2 N^{1/2} + \ldots + \mathcal{O}(e^{-\sqrt{N}})} + \ldots \, ,
\end{equation}
where the constants $a_1,a_2$... can be read off from our result~\eqref{TTI:all-order:S^2} and $b_i < a_i$. In this sense, the results obtained in this work can be seen as a first step towards deriving the integer counting of the microstates for the black holes introduced in Section~\ref{sec:holography}. This is similar to the situation for asymptotically flat black holes in string theory, where the localization of a supergravity path-integral on the dominant saddle point AdS$_2 \times S^2$ gives rise to the ``almost integer'' degeneracies as per the terminology used in~\cite{Dabholkar:2010uh}. These almost integers were later promoted to integers once the contribution from orbifolded AdS$_2 \times S^2$ geometries were included, see~\cite{Dabholkar:2014ema,Iliesiu:2022kny}. Thus, our investigations point to the necessity and usefulness of a detailed study of all possible solutions to the BAE~\eqref{BAE} determining the TTI. 

Another intriguing aspect highlighted by our analysis is the central r\^ole played by the shifted $N$ introduced in~\eqref{N:shifted}, see~\eqref{eq:N-shifted-min} in the minimal case. In the context of the ABJM theory on the round $S^3$, a similar shift was shown in~\cite{Bergman:2009zh} (see also \cite{Aharony:2009fc}) to originate from an eight-derivative term in the 11d effective action of M-theory.\footnote{It is worth noting that the shift found in this paper does not account for the full $N-B(k)$ argument of the Airy function, where $B(k) = k/24 + 1/(3k)$ as reviewed in~\eqref{eq:c-a-b-ABJM-round}.} It would be interesting to understand the higher-dimensional significance of the $\hat{N}_\Delta$ and $\hat{N}$ quantities introduced in the present paper. We also note that, unlike in the case of the $S^3$ partition function, the shift of $N$ relevant for the ABJM TTI at the superconformal point can actually vanish for real positive $k$, i.e. we have $\hat{N} = N$ for $k=4$ as can be seen from~\eqref{eq:N-shifted-min}. We do not yet understand the significance (if any) of this observation and any possible relation to the recent analysis of the holography for the $k=4$ ABJM theory in \cite{Bobev:2021wir}.

A possible avenue to gain a better analytic understanding of the ABJM TTI is to leverage its relation to the $S^3$ partition function put forward in~\cite{Closset:2017zgf}. There, the authors showed on general grounds that the partition functions of 3d $\mathcal{N}=2$ SCFTs can be viewed as the expectation value of a certain ``fibering'' operator $\mathcal{F}$ in the topologically twisted theory on $S^1 \times S^2$. Since for the ABJM theory the former is of Airy-type and under good analytic control, one can hope to understand  $\langle \mathcal{F} \rangle$ in sufficient detail to be able to extract the ABJM TTI analytically. Further analysis along these lines could also produce perturbatively exact results for the partition function of ABJM theory on other 3-manifolds such as the Seifert manifold $\mathcal{M}_{p,\mathfrak{g}}$.

Finally, we note that the numerical techniques leveraged here in the context of the ABJM TTI can also be applied to other 3d $\mathcal{N}=2$ SCFTs. For a number of interesting theories, we have managed to obtain a perturbatively exact analytic answer for their TTI, which we will report on in a forthcoming publication~\cite{Bobev:2023lkx}. Moreover, in \cite{Bobev:2022wem} we will show how similar numerical techniques can also be leveraged to derive analytic answers for the superconformal index of 3d holographic SCFTs.

\section*{Acknowledgments}
We are grateful to Francesco Benini, Anthony Charles, Shai Chester, Fri\dh rik Freyr Gautason, Yasuyuki Hatsuda, Kiril Hristov, Vincent Min, Marcos Mari\~no, Dario Martelli, Leo Pando-Zayas, Krzysztof Pilch, Silviu Pufu, Jesse van Muiden, Yu Xin, and Alberto Zaffaroni for valuable discussions.  NB and JH are supported in part by an Odysseus grant G0F9516N from the FWO. NB, JH, and VR are also supported by the KU Leuven C1 grant ZKD1118 C16/16/005. VR would like to thank IPhT Saclay for its hospitality during the final stages of this project. 


\appendix

\section{Corrections to the STU prepotential}
\label{app:squashed-prepot}

In this Appendix, we use our Airy conjecture~\eqref{eq:c-a-b-ABJM-m-gen} in the M-theory limit to investigate the corrections to the STU model of 4d $\mathcal{N}=2$ gauged supergravity. Keeping the parameter $b$ generic, we expand the free energy at large $N$ and rearrange the expansion according to the degree of homogeneity in the real mass deformation parameters $\Delta_a$. The terms homogeneous of degree 2 and 0 take the following form, to all orders in the $1/N$ expansion:
\begin{equation}
\label{eq:F-homog-b}
\begin{split}
F(N,k,b,\Delta) =&\; \frac{\pi\sqrt{2k}}{3}\Bigl(b + \frac{1}{b}\Bigr)^2\Bigl(N - \frac{k}{24}\Bigr)^{3/2}\sqrt{\Delta_1\Delta_2\Delta_3\Delta_4} \\[1mm]
&\; - \frac{2\pi}{3\sqrt{2k}}\,\frac14\Bigl(b + \frac{1}{b}\Bigr)^2\Bigl(N - \frac{k}{24}\Bigr)^{1/2}F_{4\partial}^{\text{STU}}(\Delta) \\
&\; + \frac{\pi}{48\sqrt{2k}}\Bigl(b - \frac{1}{b}\Bigr)^2\Bigl(N - \frac{k}{24}\Bigr)^{1/2}\frac{(\sum_a\Delta_a)^2 - \sum_a \Delta_a^2}{\sqrt{\Delta_1\Delta_2\Delta_3\Delta_4}}\\[1mm]
&\; + \frac14\log\Bigl(N - \frac{k}{24}\Bigr) + \ldots \, ,
\end{split}
\end{equation}
where $F^{\text{STU}}_{4\partial}$ is defined in~\eqref{eq:4der-STU}, and the ellipsis denote terms of lower homogeneity degree. In this form, the free energy can be compared to the regularized on-shell action of the higher-derivative dual supergravity theory. As shown in~\cite{Bobev:2021oku}, the latter reads
\begin{equation}
\label{eq:I-HD}
I_{\text{HD}} = \frac14\Bigl(b + \frac{1}{b}\Bigr)^2\,\frac{\pi L^2}{2G_N} + 8\pi^2\Bigl(b + \frac{1}{b}\Bigr)^2\,c_2 - 8\pi^2\Bigl(b - \frac{1}{b}\Bigr)^2c_1 \, ,
\end{equation}
where $(c_1,c_2)$ are constant coefficients that control the two four-derivative invariants that consist of the supersymmetrization of the square of the Weyl tensor and of the Gauss-Bonnet density, respectively. Following the reasoning explained in Section~\ref{sec:expand}, we can use~\eqref{eq:F-homog-b} to read off the all-order expression of the dimensionless bulk quantities $L^2/G_N$, $c_1$ and $c_2$ in terms of the dual SCFT data $(N,k)$. Since~\eqref{eq:I-HD} was derived in minimal gauged supergravity, this can be done  by turning off the reall mass deformations and corresponds to setting $\Delta_a = \frac12$. The result for $L^2/G_N$ is given in~\eqref{eq:L/G-all-order}, while for the Wilsonian coefficients $c_1$ and $c_2$ we find
\begin{equation}
\label{eq:c1c2-all-order}
c_1 = -\frac{1}{32\pi\sqrt{2k}}\Bigl(N - \frac{k}{24}\Bigr)^{1/2} \, , \qquad c_2 = -\frac{1}{96\pi\sqrt{2k}}\Bigl(N - \frac{k}{24}\Bigr)^{1/2} \, .
\end{equation}
At leading order in the large $N$ expansion, these expressions are consistent with the results derived in~\cite{Bobev:2021oku}. We also note the interesting relation
\begin{equation}
c_1 = 3\,c_2 \, ,
\end{equation}
valid to all orders in the $1/N$ expansion. This was conjectured to hold at large $N$~\cite{Bobev:2021oku}, and our Airy conjecture confirms this expectation and extends it to an all-order statement. It would be most interesting to understand the origin of this relation between the Wilsonian coefficients of the 4d higher-derivative couplings from an 11d point of view. We also note in passing that the conclusions of~\cite{Bobev:2021oku} regarding the sign of certain corrections to the entropy of asymptotically AdS black holes remain valid in light of our Airy conjecture.

\section{Details of the IIA expansion}
\label{app:expand}

Given a partition function of Airy-type~\eqref{eq:Z}, we can use the asymptotic expansion of the Airy function~\eqref{eq:Ai-expand} to write the free energy $F(N,k,\mathfrak{a}) = -\log Z(N,k,\mathfrak{a})$ in a large $N$, fixed $k$ expansion. The first few terms are given by\footnote{Here and below, we suppress the dependence on the generic parameters $\mathfrak{a}$ in the $(\gamma,\alpha,\beta)$ and $\mathcal{A}$ quantities to lighten the notation.}
\begin{align}
\label{eq:F-Mth-explicit}
F(N,k,\mathfrak{a}) =&\; \frac{2\sqrt{k}}{3\sqrt{\gamma}}\,N^{3/2} - \frac{\alpha + \beta\,k^2}{\sqrt{k\,\gamma}}\,N^{1/2} + \frac14\log\Bigl(\frac{\gamma}{k}N\Bigr) + \frac12\log(4\pi) - \mathcal{A}(k) \nonumber \\
&\;+ \frac{(\alpha + \beta\,k^2)^2}{4\,k^{3/2}\sqrt{\gamma}}\,N^{-1/2} - \frac{\alpha + \beta\,k^2}{4\,k}\,N^{-1} + \frac{(\alpha + \beta\,k^2)^3 + 36\,u_1\,\gamma\,k^2}{24\,k^{5/2}\sqrt{\gamma}}\,N^{-3/2} \nonumber \\
&\;- \frac{(\alpha + \beta\,k^2)^2}{8\,k^2}\,N^{-2} + \frac{(\alpha + \beta\,k^2)\Bigl((\alpha + \beta\,k^2)^3 + 144\,u_1\,\gamma\,k^2\Bigr)}{64\,k^{7/2}\sqrt{\gamma}}\,N^{-5/2} \nonumber \\
&\;- \frac{2(\alpha + \beta\,k^2)^3 + 27\,(2\,u_2 - u_1^2)\,\gamma\,k^2}{24\,k^3}\,N^{-3} + \mathcal{O}\bigl(N^{-7/2}\bigr) \, .
\end{align}
From this expansion, we can read off the coefficient of the $\mathcal{O}\bigl(N^{3/2-n}k^{1/2+n}\bigr)$ terms, with $n \in \mathbb{N}$:
\begin{equation}
\frac{2}{3\sqrt{\gamma}}\,\frac{(-\beta)^n}{n!}\,\prod_{r=0}^{n-1}\Bigl(\frac32 - r\Bigr) \, .
\end{equation}
Similarly, the coefficient of the $\mathcal{O}\bigl(N^{3/2-n}k^{n-3/2}\bigr)$ terms is given by
\begin{equation}
\frac{2}{3\sqrt{\gamma}}\,\frac{\alpha\,n}{\beta}\,\frac{(-\beta)^n}{n!}\,\prod_{r=0}^{n-1}\Bigl(\frac32 - r\Bigr) \, .
\end{equation}
To obtain the IIA expansion of the free energy in the form~\eqref{eq:IIA-expand}, we now scale $k$ with $N$ while keeping $\lambda = N/k$ fixed. In this regime, the monomials above produce $\mathcal{O}\bigl(N^2\lambda^{-1/2 - n}\bigr)$ and $\mathcal{O}\bigl(N^0\lambda^{3/2 - n}\bigr)$ terms, respectively. They will thus contribute to the genus-0 and genus-1 free energies. Explicitly, the former receives a contribution
\begin{equation}
\sum_{n\geq0}\Bigl[\frac{2}{3\sqrt{\gamma}}\,\frac{(-\beta)^n}{n!}\,\prod_{r=0}^{n-1}\Bigl(\frac32 - r\Bigr)\Bigr]\,\lambda^{-1/2 - n} = \frac{2}{3\sqrt{\gamma}}\,\frac{(\lambda - \beta)^{3/2}}{\lambda^2} \, ,
\end{equation}
where we have given the resummed function of the 't Hooft coupling $\lambda$ from the series representation. The genus-1 free energy recieves a contribution of 
\begin{equation}
\sum_{n\geq0}\Bigl[\frac{2}{3\sqrt{\gamma}}\,\frac{\alpha\,n}{\beta}\,\frac{(-\beta)^n}{n!}\,\prod_{r=0}^{n-1}\Bigl(\frac32 - r\Bigr)\Bigr]\,\lambda^{3/2 - n} = -\frac{\alpha}{\sqrt{\gamma}}\,\sqrt{\lambda - \beta} \, .
\end{equation}
Additional contributions to the genus-0 and genus-1 free energies come from the $N^0 k^2$ and from the $N^{-n}k^{n}$ and $N^0k^0$ terms, respectively. We stress that some of these terms are implicit in the function $\mathcal{A}(k)$ in~\eqref{eq:F-Mth-explicit} which depends on the theory under consideration.\\

At higher genus, we find two types of contributions to the genus-$\tt{g}$ free energies from the expansion~\eqref{eq:F-Mth-explicit}. The first is from $\mathcal{O}\bigl(N^{3/2-n}k^{n+1/2 - 2\tt{g}}\bigr)$ terms, and the other is from $\mathcal{O}\bigl(N^{-n}k^{n+2 - 2\tt{g}}\bigr)$ terms. Both are of order $\mathcal{O}\bigl(N^{2 - 2\tt{g}}\bigr)$ when scaling $k$ and keeping $\lambda$ fixed. Evidently, as the genus increases, the coefficients of these terms increase in complexity. However, by collecting and rearranging the terms in the expansion~\eqref{eq:F-Mth-explicit}, we find that they can be written in a rather compact form as
\begin{equation}
\begin{split}
\mathcal{O}\bigl(N^{3/2-n}k^{n+1/2 - 2\tt{g}}\bigr) \; \longrightarrow \;&\; \mathcal{F}_{{\tt{g}},n}\,\frac{2}{3\sqrt{\gamma}}\,\frac{(-\beta)^n}{n!}\,\prod_{r=0}^{n-1}\Bigl(\frac{3}{2} - r\Bigr)\,\lambda^{-n - 1/2 + 2\tt{g}} \\
\mathcal{O}\bigl(N^{-n}k^{n+2 - 2\tt{g}}\bigr) \; \longrightarrow \;&\; \,\mathcal{G}_{{\tt{g}},n}\,\lambda^{-n-2+2\tt{g}} \, , 
\end{split}
\end{equation}
The quantities~$\mathcal{F}$ and~$\mathcal{G}$ are the functions of $(\gamma,\alpha,\beta)$ given in the main text, see~\eqref{eq:cal-F} and~\eqref{eq:cal-G}, and they are each entirely specified by a set of rational functions of degree zero in the $u_n$ coefficients\footnote{We can use the fact that $u_0 = 1$ to enforce homogeneity of degree zero.} entering the Airy asymptotics~\eqref{eq:Ai-expand}. These functions can be systematically extracted from the expansion~\eqref{eq:F-Mth-explicit} to arbitrarily high order. We give the first six below as an example. For the constants specifying~$\mathcal{F}$, we have
\begin{align}
	\mathcal{P}^{(0)}(u) =&\; 1 \, , \nonumber \\[1mm]
	\mathcal{P}^{(1)}(u) =&\; 6\,U_1 \, , \nonumber \\[1mm]
	\mathcal{P}^{(2)}(u) =&\; \frac{12}{35}\,\Bigl(U_1^3 - 3\,U_1\,U_2 + 3\,U_3\Bigr) \, , \\[1mm]
	\mathcal{P}^{(3)}(u) =&\; \frac{72}{25025}\,\Bigl(U_1^5 - 5\,U_1^3\,U_2 + 5\,U_1^2\,U_3 + 5\,U_1(U_2^2 - U_4) - 5\,U_2\,U_3 + 5\,U_5\Bigr) \nonumber \\[1mm]
	\mathcal{P}^{(4)}(u) =&\; \frac{432}{56581525}\Bigl(U_1^7 - 7\,U_1^5\,U_2 + 7\,U_1^4\,U_3 + 7\,U_1^3(2U_2^2 - U_4) - 7\,U_1^2(3\,U_2\,U_3 - U_5) \nonumber \\
	&\qquad\qquad\;\; - 7\,U_1(U_2^3 - U_3^2 - 2\,U_2\,U_4 + U_6) + 7\,U_2^2\,U_3 - 7\,U_3\,U_4 - 7\,U_2\,U_5 + 7\,U_7\Bigr) \nonumber \\[1mm]
	\mathcal{P}^{(5)}(u) =&\; \frac{288}{32534376875}\Bigl(U_1^9 - 9\,U_1^7\,U_2 + 9\,U_1^6\,U_3 + 9\,U_1^5(3U_2^2 - U_4) - 9\,U_1^4(5\,U_2\,U_3 - U_5) \nonumber \\
	&\qquad\qquad\qquad - U_1^3(30\,U_2^3 - 18\,U_3^2 - 36\,U_2\,U_4 + 9\,U_6) \nonumber \\
	&\qquad\qquad\qquad + 9\,U_1^2(6\,U_2^2\,U_3 - 3\,U_3\,U_4 - 3\,U_2\,U_5 + U_7) \nonumber \\
	&\qquad\qquad\qquad + 9\,U_1\bigl(U_2^4 - 3\,U_2^2\,U_4 + U_4^2 + 2\,U_3\,U_5 - U_2(3\,U_3^2 - 2\,U_6) - U_8\bigr) \nonumber \\
	&\qquad\qquad\qquad - 9\,U_2^3\,U_3 + 3\,U_3^3 + 9\,U_2^2\,U_5 - 9\,U_4\,U_5 - 9\,U_3\,U_6 + 18\,U_2\,U_3\,U_4 - 9\,U_2\,U_7 + 9\,U_9\Bigr) \nonumber \, ,
\end{align}
where we have defined~$U_i = u_i/u_0$. Using the definition~\eqref{eq:u}, these quantities evaluate to rational numbers:
\begingroup
\begin{center}
	\renewcommand*{\arraystretch}{1.3}
	\begin{tabular}{| c || c | c | c | c | c | c |}
		\hline
		$m$ & 0 & 1 & 2 & 3 & 4 & 5 \\
		\hline
		$\mathcal{P}^{(m)}$ & 1 & $\frac{5}{12}$ & $\frac{221}{6048}$ & $\frac{16565}{10378368}$ & $\frac{51281261}{1126343522304}$ & $\frac{12188095}{12659897327616}$ \\
		\hline
	\end{tabular}
\end{center}
\endgroup
For the constants specifying~$\mathcal{G}$, we have
\begin{align}
	\mathcal{Q}^{(0)}(u) =&\; 0 \, , \nonumber \\[1mm]
	\mathcal{Q}^{(1)}(u) =&\; -\frac14 \, , \nonumber \\[1mm]
	\mathcal{Q}^{(2)}(u) =&\; \frac{9}{16}\,\Bigl(U_1^2 - 2\,U_2\Bigr) \, , \nonumber \\[1mm]
	\mathcal{Q}^{(3)}(u) =&\; \frac{27}{2560}\,\Bigl(U_1^4 - 4\,U_1^2\,U_2 + 4\,U_1\,U_3 + 2\,U_2^2 - 4\,U_4\Bigr) \, , \\[1mm]
	\mathcal{Q}^{(4)}(u) =&\; \frac{27}{573440}\Bigl(U_1^6 - 6\,U_1^4\,U_2 + 6\,U_1^3\,U_3 + U_1^2(9\,U_2^2 - 6\,U_4) - 6\,U_1(U_2\,U_3 - U_5) \nonumber \\
	&\qquad\qquad - 2\,U_2^3 + 3\,U_3^2 + 6\,U_2\,U_4 - 6\,U_6\Bigr) \nonumber \\
	\mathcal{Q}^{(5)}(u) =&\; \frac{81}{1009254400}\Bigl(U_1^8 - 8\,U_1^6\,U_2 + 8\,U_1^5\,U_3 + 4\,U_1^4(5\,U_2^2 - 2\,U_4) - 8\,U_1^3(4\,U_2\,U_3 - U_5) \nonumber \\
	&\qquad\qquad\qquad - 4\,U_1^2(4\,U_2^3 - 3\,U_3^2 - 6\,U_2\,U_4 + 2\,U_6) \nonumber \\
	&\qquad\qquad\qquad + 8\,U_1(3\,U_2^2\,U_3 - 2\,U_3\,U_4 - 2\,U_2\,U_5 + U_7) \nonumber \\
	&\qquad\qquad\qquad + 2\,U_2^4 - 8\,U_2^2\,U_4 - 8\,U_2(U_3^2 - U_6) + 4\,U_4^2 + 8\,U_3\,U_5 - 8\,U_8\Bigr) \nonumber \, .
\end{align}
Using~\eqref{eq:u}, we obtain
\begingroup
\begin{center}
	\renewcommand*{\arraystretch}{1.3}
	\begin{tabular}{| c || c | c | c | c | c | c |}
		\hline
		$m$ & 0 & 1 & 2 & 3 & 4 & 5 \\
		\hline
		$\mathcal{Q}^{(m)}$ & 0 & $-\frac{1}{4}$ & $-\frac{5}{128}$ & $-\frac{113}{49152}$ & $-\frac{3935}{49545216}$ & $-\frac{3229117}{1674231939072}$ \\
		\hline
	\end{tabular}
\end{center}
\endgroup

\section{All-order M-theory expansion of the topologically twisted index}
\label{App:all-order}

In this Appendix, we provide numerical data that supports the all-order M-theory expansion (\ref{TTI:all-order:S^2}) with (\ref{TTI:all-order:S^2:detail}). To begin with, we estimate the numerical coefficients
\begin{equation}
	\hat f^\text{(lmf)}_{3/2}(k,\Delta,\mn),\qquad\hat f^\text{(lmf)}_{1/2}(k,\Delta,\mn),\qquad \hat f^\text{(lmf)}_0(k,\Delta,\mn),\label{f:lmf}
\end{equation}
through the \texttt{LinearModelFit} given in (\ref{TTI:M:expansion:shifted}), which we repeat here as (with $N=101\sim301$ step=10)
\begin{equation}
	\begin{split}
		\log Z_{S^1\times S^2}(N,k,\Delta,\mn)+\fft12\log\hat N_\Delta&=\hat f^\text{(lmf)}_{3/2}(k,\Delta,\mn)\hat N_\Delta^\fft32+\hat f^\text{(lmf)}_{1/2}(k,\Delta,\mn)\hat N_\Delta^\fft12\\
		&\quad+\hat f^\text{(lmf)}_0(k,\Delta,\mn).
	\end{split}\label{TTI:M:expansion:shifted:3}
\end{equation}
We compare the first two numerical coefficients in (\ref{f:lmf}) with the corresponding analytic expressions
\begin{equation}
\begin{split}
	\hat f_{3/2}(k,\Delta,\mn)&=-\fft{\pi\sqrt{2k\Delta_1\Delta_2\Delta_3\Delta_4}}{3}\sum_{a=1}^3\fft{\mn_a}{\Delta_a},\\ \hat f_{1/2}(k,\Delta,\mn)&=\fft{\pi\sqrt{2k\Delta_1\Delta_2\Delta_3\Delta_4}}{3}\sum_{a=1}^3\fft{\mathfrak c_a\mn_a}{k\Delta_a},
\end{split}
\end{equation}
from (\ref{TTI:all-order:S^2}). For a precise comparison, we provide the error ratio
\begin{equation}
	R_{X}(k,\Delta,\mn)\equiv\fft{\hat f_{X}^\text{(lmf)}(k,\Delta,\mn)-\hat f_{X}(k,\Delta,\mn)}{\hat f_{X}(k,\Delta,\mn)}\quad~(X=3/2,1/2)
\end{equation}
for various $k$ values and $(\Delta,\mn)$-configurations below. Since we do not have an analytic expression corresponding to the third numerical coefficient in (\ref{f:lmf}), we provide the standard error for the numerical estimate $\hat f_0^\text{(lmf)}(k,\Delta,\mn)$ instead, namely $\sigma_0$. Small values of error ratios $R_{3/2},R_{1/2}$ and a standard error $\sigma_0$ will be strong evidence for the all-order M-theory expansion of TTI (\ref{TTI:all-order:S^2}).

\subsection{$k=1$ with various $(\Delta,\mn)$-configurations}
Here we provide data with $k=1$ and various $(\Delta,\mn)$-configurations. For flavor magnetic fluxes, we take the five different configurations listed in (\ref{nas}). For flavor chemical potentials, we investigate the various configurations listed in (\ref{M:case1}-\ref{M:case4}). Since there are numerous $\Delta$-configurations listed below, we provide error ratios $R_{3/2},R_{1/2}$ and a standard error $\sigma_0$ only for one example for each \textbf{Case} below. But we have confirmed that all the other configurations listed in (\ref{M:case1}-\ref{M:case4}) also yield small error ratios and a standard error, which strongly support the all-order M-theory expansion of TTI (\ref{TTI:all-order:S^2}) with $k=1$. 

In the following data, one can observe that the precision tends to be better for $\Delta$-configurations closer to the superconformal configuration $\Delta_a=\fft12$.

\noindent\textbf{Case 1. $\Delta=(\Delta_1,\Delta_1,1-\Delta_1,1-\Delta_1)$}
%
\begin{equation}
	\Delta_1\in\bigg\{\fft13,\fft14,\fft15,\fft25,\fft37,\fft38,\fft{5}{12}\bigg\}\label{M:case1}
\end{equation}
For $\Delta_1=\fft13$,
\begin{center}
	\footnotesize
	\begin{tabular}{ |c||c|c|c|c| } 
		\hline
		& $R_{3/2}$ & $R_{1/2}$ & $\hat f_0^\text{(lmf)}$ & $\sigma_0$ \\
		\hline\hline
		\text{1st in }(\ref{nas}) & $4.914{\times}10^{-40}$ & $9.543{\times}10^{-38}$ & $-3.5077320285248926271$ & $1.582{\times}10^{-36}$  \\
		\hline
		\text{2nd in }(\ref{nas}) & $6.707{\times}10^{-40}$ & $1.149{\times}10^{-37}$ & $-3.4212557473122823646$ & $1.799{\times}10^{-36}$  \\
		\hline
		\text{3rd in }(\ref{nas}) & $5.729{\times}10^{-40}$ & $1.049{\times}10^{-37}$ & $-3.4644938879185874958$ & $1.690{\times}10^{-36}$  \\ 
		\hline
		\text{4th in }(\ref{nas}) & $4.914{\times}10^{-40}$ & $9.543{\times}10^{-38}$ & $-3.5077320285248926271$ & $1.582{\times}10^{-36}$  \\ 
		\hline
		\text{5th in }(\ref{nas}) & $5.729{\times}10^{-40}$ & $1.049{\times}10^{-37}$ & $-3.4644938879185874958$ & $1.690{\times}10^{-36}$  \\ 
		\hline
	\end{tabular}
\end{center}
%

\noindent\textbf{Case 2. $\Delta=(\Delta_1,\fft12,\fft12,1-\Delta_1)$}
%
\begin{equation}
	\Delta_1\in\bigg\{\fft13,\fft37,\fft38,\fft25,\fft{5}{12},\fft{5}{14}\bigg\},\label{M:case2}
\end{equation}
For $\Delta_1=\fft25$,
\begin{center}
	\footnotesize
	\begin{tabular}{ |c||c|c|c|c| } 
		\hline
		& $R_{3/2}$ & $R_{1/2}$ & $\hat f_0^\text{(lmf)}$ & $\sigma_0$ \\
		\hline\hline
		\text{1st in }(\ref{nas}) & $7.660{\times}10^{-35}$ & $1.640{\times}10^{-32}$ & $-3.1202038806984986501$ & $2.427{\times}10^{-31}$  \\
		\hline
		\text{2nd in }(\ref{nas}) & $6.749{\times}10^{-34}$ & $1.395{\times}10^{-31}$ & $-3.0980843847549988914$ & $2.019{\times}10^{-30}$  \\
		\hline
		\text{3rd in }(\ref{nas}) & $4.543{\times}10^{-34}$ & $9.535{\times}10^{-32}$ & $-3.1037931874737728519$ & $1.389{\times}10^{-30}$  \\
		\hline
		\text{4th in }(\ref{nas}) & $6.749{\times}10^{-34}$ & $1.395{\times}10^{-31}$ & $-3.0980843847549988914$ & $2.019{\times}10^{-30}$  \\
		\hline
		\text{5th in }(\ref{nas}) & $9.989{\times}10^{-34}$ & $2.027{\times}10^{-31}$ & $-3.0870246367832490120$ & $2.908{\times}10^{-30}$  \\
		\hline
	\end{tabular}
\end{center}
%
\noindent\textbf{Case 3. $\Delta=(\Delta_1,\Delta_2,1-\Delta_2,1-\Delta_1)$}
%
\begin{equation}
	(\Delta_1,\Delta_2)\in\bigg\{(\fft14,\fft13),(\fft15,\fft13),(\fft15,\fft25),(\fft14,\fft25),(\fft25,\fft37),(\fft13,\fft{5}{12}),(\fft25,\fft{5}{12})\bigg\},\label{M:case3}
\end{equation}
For $(\Delta_1,\Delta_2)=(\fft25,\fft37)$,
\begin{center}
	\footnotesize
	\begin{tabular}{ |c||c|c|c|c| } 
		\hline
		& $R_{3/2}$ & $R_{1/2}$ & $\hat f_0^\text{(lmf)}$ & $\sigma_0$ \\
		\hline\hline
		\text{1st in }(\ref{nas}) & $-3.427{\times}10^{-31}$ & $-7.266{\times}10^{-28}$ & $-3.1576123219158870630$ & $1.058{\times}10^{-27}$  \\
		\hline
		\text{2nd in }(\ref{nas}) & $5.608{\times}10^{-31}$ & $1.118{\times}10^{-28}$ & $-3.1211538580093534064$ & $1.581{\times}10^{-27}$  \\
		\hline
		\text{3rd in }(\ref{nas}) & $1.087{\times}10^{-31}$ & $2.235{\times}10^{-29}$ & $-3.1366757085547572184$ & $3.196{\times}10^{-28}$  \\
		\hline
		\text{4th in }(\ref{nas}) & $-3.269{\times}10^{-31}$ & $-6.862{\times}10^{-29}$ & $-3.1503045578384887568$ & $9.935{\times}10^{-28}$  \\
		\hline
		\text{5th in }(\ref{nas}) & $1.122{\times}10^{-31}$ & $2.286{\times}10^{-29}$ & $-3.1320753258852219285$ & $3.257{\times}10^{-28}$  \\
		\hline
	\end{tabular}
\end{center}
%
\noindent\textbf{Case 4. $\Delta=(\Delta_1,\Delta_2,\Delta_3,\Delta_4)$}
%
\begin{equation}
	\begin{split}
		\Delta&\in\bigg\{(\fft13,\fft13,\fft13,1),(\fft18,\fft68,\fft48,\fft58),(\fft28,\fft78,\fft38,\fft48),(\fft{3}{10\pi},\fft{4}{10\pi},\fft{5}{10\pi},2-\fft{12}{10\pi}),\\
		&\qquad(\fft39,\fft59,\fft29,\fft89),(\fft{3}{10},\fft{5}{10},\fft{4}{10},\fft{8}{10}),(\fft{5}{11},\fft{7}{11},\fft{2}{11},\fft{8}{11}),(\fft1\pi,\fft2\pi,\fft{3}{2\pi},2-\fft{9}{2\pi})\bigg\}\\
		&\quad~\cup\bigg\{(\fft12-4x,\fft12-x,\fft12+2x,\fft12+3x)\,\Big|\,x=\fft{1}{100},\fft{2}{100},\cdots,\fft{1}{10}\bigg\}.\label{M:case4}
	\end{split}
\end{equation}
For $\Delta=(\fft{5}{11},\fft{7}{11},\fft{2}{11},\fft{8}{11})$, 
\begin{center}
	\footnotesize
	\begin{tabular}{ |c||c|c|c|c| } 
		\hline
		& $R_{3/2}$ & $R_{1/2}$ & $\hat f_0^\text{(lmf)}$ & $\sigma_0$ \\
		\hline\hline
		\text{1st in }(\ref{nas}) & $3.824{\times}10^{-21}$ & $6.469{\times}10^{-19}$ & $-4.3954684057902448724$ & $1.070{\times}10^{-17}$  \\
		\hline
		\text{2nd in }(\ref{nas}) & $2.782{\times}10^{-21}$ & $5.015{\times}10^{-19}$ & $-4.6340289673765309638$ & $8.923{\times}10^{-18}$  \\
		\hline
		\text{3rd in }(\ref{nas}) & $3.312{\times}10^{-21}$ & $6.009{\times}10^{-19}$ & $-4.6319262057634415054$ & $1.071{\times}10^{-17}$  \\
		\hline
		\text{4th in }(\ref{nas}) & $7.386{\times}10^{-21}$ & $1.069{\times}10^{-18}$ & $-4.1461502319880991918$ & $1.606{\times}10^{-17}$  \\
		\hline
		\text{5th in }(\ref{nas}) & $6.378{\times}10^{-21}$ & $9.699{\times}10^{-19}$ & $-4.2654305127812422375$ & $1.517{\times}10^{-17}$  \\
		\hline
	\end{tabular}
\end{center}
We have also investigated the case with $k=3$ and 
\begin{equation}
	\Delta\in\bigg\{(\fft12-3x,\fft12-x,\fft12+\fft32x,\fft12+\fft52x)\,\Big|\,x=\fft{1}{100},\fft{2}{100},\cdots,\fft{1}{10}\bigg\},
\end{equation}
and found a precise agreement between numerical coefficients and analytic expressions as the above listed configurations.

\subsection{$k=1,2,3,4$ with a few different $(\Delta,\mn)$-configurations}
Here we provide data with different $k$ values for some $(\Delta,\mn)$-configurations. We found that the error ratios $R_{3/2},R_{1/2}$ and the standard error $\sigma_0$ are larger for higher values of $k$ but they are still small enough to support the all-order M-theory expansion of TTI (\ref{TTI:all-order:S^2}). 
\noindent\textbf{$\Delta=(\fft12,\fft12,\fft12,\fft12)$ \& $\mn=(\fft12,\fft12,\fft12,\fft12)$}
%
\begin{center}
	\footnotesize
	\begin{tabular}{ |c||c|c|c|c| } 
		\hline
		& $R_{3/2}$ & $R_{1/2}$ & $\hat f_0^\text{(lmf)}$ & $\sigma_0$ \\
		\hline\hline
		k=1 & $2.436{\times}10^{-39}$ & $5.319{\times}10^{-37}$ & $-3.0459513105331823845$ & $7.834{\times}10^{-36}$  \\
		\hline
		k=2 & $9.935{\times}10^{-28}$ & $4.336{\times}10^{-25}$ & $-1.7865975337335498966$ & $4.310{\times}10^{-24}$  \\
		\hline
		k=3 & $1.250{\times}10^{-23}$ & $8.185{\times}10^{-21}$ & $-1.3863730440038858190$ & $6.330{\times}10^{-20}$  \\ 
		\hline
		k=4 & $1.340{\times}10^{-20}$ & $1.171{\times}10^{-17}$ & $-1.3065895525823577338$ & $7.488{\times}10^{-17}$  \\ 
		\hline
	\end{tabular}
\end{center}
%
\noindent\textbf{$\Delta=(\fft37,\fft12,\fft12,\fft47)$ \& $\mn=(\fft18,\fft58,\fft38,\fft78)$}
%
\begin{center}
	\footnotesize
	\begin{tabular}{ |c||c|c|c|c| } 
		\hline
		& $R_{3/2}$ & $R_{1/2}$ & $\hat f_0^\text{(lmf)}$ & $\sigma_0$ \\
		\hline\hline
		k=1 & $2.018{\times}10^{-35}$ & $4.204{\times}10^{-33}$ & $-3.0605573608897936910$ & $6.062{\times}10^{-32}$  \\
		\hline
		k=2 & $4.034{\times}10^{-25}$ & $1.680{\times}10^{-22}$ & $-1.8044287925557806895$ & $1.608{\times}10^{-21}$  \\
		\hline
		k=3 & $9.523{\times}10^{-22}$ & $5.953{\times}10^{-19}$ & $-1.4095876138872784032$ & $4.399{\times}10^{-18}$  \\ 
		\hline
		k=4 & $4.482{\times}10^{-19}$ & $3.740{\times}10^{-16}$ & $-1.3372227288462853143$ & $2.270{\times}10^{-15}$  \\ 
		\hline
	\end{tabular}
\end{center}
%
\noindent\textbf{$\Delta=(\fft13,\fft{5}{12},\fft{7}{12},\fft23)$ \& $\mn=(\fft{3}{10},\fft{5}{10},\fft{4}{10},\fft{8}{10})$}
%
\begin{center}
	\footnotesize
	\begin{tabular}{ |c||c|c|c|c| } 
		\hline
		& $R_{3/2}$ & $R_{1/2}$ & $\hat f_0^\text{(lmf)}$ & $\sigma_0$ \\
		\hline\hline
		k=1 & $-1.937{\times}10^{-27}$ & $-3.740{\times}10^{-25}$ & $-3.2887852060053039583$ & $5.360{\times}10^{-24}$  \\
		\hline
		k=2 & $4.873{\times}10^{-20}$ & $1.884{\times}10^{-17}$ & $-1.9220513886027755028$ & $1.724{\times}10^{-16}$  \\
		\hline
		k=3 & $2.895{\times}10^{-18}$ & $1.683{\times}10^{-15}$ & $-1.4826350102616867360$ & $1.140{\times}10^{-14}$  \\ 
		\hline
		k=4 & $-1.133{\times}10^{-16}$ & $-8.796{\times}10^{-14}$ & $-1.3787212191091059259$ & $4.830{\times}10^{-13}$  \\ 
		\hline
	\end{tabular}
\end{center}
%
\noindent\textbf{$\Delta=(\fft1\pi,\fft2\pi,\fft{3}{2\pi},2-\fft{9}{2\pi})$ \& $\mn=(\fft18,\fft58,\fft38,\fft78)$}
%
\begin{center}
	\footnotesize
	\begin{tabular}{ |c||c|c|c|c| } 
		\hline
		& $R_{3/2}$ & $R_{1/2}$ & $\hat f_0^\text{(lmf)}$ & $\sigma_0$ \\
		\hline\hline
		k=1 & $5.781{\times}10^{-29}$ & $1.077{\times}10^{-26}$ & $-3.2260672004299636583$ & $1.516{\times}10^{-25}$  \\
		\hline
		k=2 & $2.806{\times}10^{-20}$ & $1.046{\times}10^{-17}$ & $-1.9014822992681926381$ & $9.434{\times}10^{-17}$  \\
		\hline
		k=3 & $2.264{\times}10^{-17}$ & $1.269{\times}10^{-14}$ & $-1.4927147973472065263$ & $8.562{\times}10^{-14}$  \\ 
		\hline
		k=4 & $2.106{\times}10^{-15}$ & $1.577{\times}10^{-12}$ & $-1.4203457465152409628$ & $8.564{\times}10^{-12}$  \\ 
		\hline
	\end{tabular}
\end{center}
%

\section{The large-$k$ expansion of $\hat f_0(k,\Delta,\mn)$}
\label{App:f0}
In this Appendix, we provide numerical data that supports the large-$k$ expansion of the $N$-independent $\mathcal O(1)$ order contribution $\hat f_0(k,\Delta,\mn)$ given in (\ref{f0:superconformal}) and (\ref{f0:general}).

\subsection{Superconformal $\Delta$-configuration}\label{App:f0:superconformal}
For the superconformal $\Delta$-configuration $\Delta_a=\fft12$, we compare numerical expansion coefficients $\hat f^\text{(lmf)}_{0,X}$ in the \texttt{LinearModelFit} (\ref{f0:expansion}) with the corresponding analytic expressions (\ref{f0:superconformal:coeffi:num}). For a better precision, however, first we improve numerical expansion coefficients by subtracting known analytic parts before implementing \texttt{LinearModelFit} as follows:
\begin{equation}
\begin{split}
	\hat f_0(k)&=\hat f^\text{(lmf)}_{0,2}k^2+\hat f^\text{(lmf)}_{0,\log}\log k+\#+\sum_{s=1}^L\#k^{-2s},\\
	\hat f_0(k)-\fft76\log k&=\#k^2+\hat f^\text{(lmf)}_{0,0}+\hat f^\text{(lmf)}_{0,-2}k^{-2}+\sum_{s=2}^{L+1}\#k^{-2s},\\
	\hat f_0(k)-\fft76\log k+\fft{8\pi^2}{45k^2}&=\#k^2+\#+\hat f^\text{(lmf)}_{0,-4}k^{-4}+\sum_{s=3}^{L+2}\#k^{-2s},\\
	\hat f_0(k)-\fft76\log k-\sum_{s=1}^{s_0-1}\hat f_{0,-2s}k^{-2s}&=\#k^2+\#+\hat f^\text{(lmf)}_{0,-2s_0}k^{-2s_0}+\sum_{s=s_0+1}^{L+s_0}\#k^{-2s}.
\end{split}\label{f0:superconformal:coeffi:improved}
\end{equation}
Observe that we did not subtract the analytic expression for the $k^2$-leading order, $-\fft{3\zeta(3)}{8\pi^2}k^2$, since it rather decreases precision for estimating higher order coefficients. Here we take $N=101\sim551$ (step=10) with $L=42$ for the \texttt{LinearModelFit}.

After the improvement of numerical expansion coefficients $\hat f^\text{(lmf)}_{0,X}$ through (\ref{f0:superconformal:coeffi:improved}), we provide the error ratios
\begin{equation}
	R_{0,X}\equiv\fft{\hat f^\text{(lmf)}_{0,X}-\hat f_{0,X}}{\hat f_{0,X}}\quad~(X=2,\log,-2,\cdots,-10),
\end{equation}
where the analytic coefficients $\hat f_{0,X}$ are given as (\ref{f0:superconformal:coeffi:num}). For $\hat f^\text{(lmf)}_{0,0}$ without a known analytic expression, we provide a standard error $\sigma_{0,0}$ in estimating $\hat f^\text{(lmf)}_{0,0}$ instead.
\begin{center}
	\footnotesize
	\begin{tabular}{ |c|c|c|c|c| } 
		\hline
		$R_{0,2}$ & $R_{0,\log}$ & \multicolumn{2}{c}{$\hat f^\text{(lmf)}_{0,0}$} \vline & $\sigma_{0,0}$ \\
		\hline
		$6.065{\times}10^{-18}$ & $2.056{\times}10^{-15}$ & \multicolumn{2}{c}{$-2.09684829977578309$} \vline & $1.930{\times}10^{-18}$  \\
		\hline\hline
		$R_{0,-2}$ & $R_{0,-4}$ & $R_{0,-6}$ & $R_{0,-8}$ & $R_{0,-10}$\\
		\hline
		$-1.048{\times}10^{-13}$ & $-3.848{\times}10^{-12}$ & $-5.946{\times}10^{-11}$ & $-5.998{\times}10^{-10}$ & $-4.389{\times}10^{-9}$ \\
		\hline
	\end{tabular}
\end{center}
%

\subsection{Generic $\Delta$-configurations}\label{App:f0:general}
For generic $\Delta$-configurations, we compare the $k^2$-leading and $\log k$-subleading numerical coefficients, namely $\hat f^\text{(lmf)}_{0,2,a}(\Delta,\mn)$ and $\hat f^\text{(lmf)}_{0,\log}(\Delta,\mn)$ determined by (\ref{f0:expansion}) and (\ref{f0:expansion:lead}), with the corresponding analytic expressions $\hat f_{0,2,a}(\Delta,\mn)$ and $\hat f_{0,\log}(\Delta,\mn)=\fft76$ given in (\ref{f0:general:lead}) and (\ref{f0:general:log}) respectively. For a precise comparison, we provide error ratios
\begin{equation}
	R_{0,2,a}\equiv\fft{\hat f^\text{(lmf)}_{0,2,a}(\Delta,\mn)-\hat f_{0,2,a}(\Delta,\mn)}{\hat f_{0,2,a}(\Delta,\mn)}~~(a=1,2,3,4),\quad~R_{0,\log}\equiv\fft{\hat f^\text{(lmf)}_{0,\log}-7/6}{7/6},
\end{equation}
for various $\Delta$-configurations listed in (\ref{tHooft:case1}-\ref{tHooft:case8}). Since there are numerous $\Delta$-configurations listed below, we provide error ratios $R_{0,2,a},R_{0,\log}$ only for one example for each \textbf{Case}. But we have confirmed that all the other configurations listed in (\ref{tHooft:case1}-\ref{tHooft:case8}) also yield small error ratios, which strongly support analytic expressions for the first two leading coefficients in the large-$k$ expansion (\ref{f0:general}) with (\ref{f0:general:lead}).

In the following data, we have fixed a 't~Hooft parameter as $\lambda=30$ except for \textbf{Case 1} where we have taken $\lambda=35$. The upperbound $L$ for the \texttt{LinearModelFit} (\ref{f0:expansion}) is chosen appropriately for each \textbf{Case} to minimize standard errors.

\noindent\textbf{Case 1. $\Delta=(\Delta_1,\Delta_1,1-\Delta_1,1-\Delta_1)$ \& $N=101\sim301~(\text{step }10)$ \& $L=17$}
%
\begin{equation}
	\Delta_1\in\bigg\{\fft13,\fft14,\fft15,\fft25\bigg\}.\label{tHooft:case1}
\end{equation}
For $\Delta_1=\fft25$,
\begin{center}
	\footnotesize
	\begin{tabular}{ |c|c|c|c|c| } 
		\hline
		$R_{0,2,1}$ & $R_{0,2,2}$ & $R_{0,2,3}$ & $R_{0,2,4}$ & $R_{0,\log}$ \\
		\hline
		$1.944{\times}10^{-7}$ & $1.944{\times}10^{-7}$ & $6.419{\times}10^{-8}$ & $6.419{\times}10^{-8}$ & $6.365{\times}10^{-6}$ \\
		\hline
	\end{tabular}
\end{center}
%

\noindent\textbf{Case 2. $\Delta=(\Delta_1,1-\Delta_1,\Delta_1,1-\Delta_1)$ \& $N=101\sim351~(\text{step }10)$ \& $L=22$}
%
\begin{equation}
	\Delta_1\in\bigg\{\fft25,\fft37,\fft{5}{12}\bigg\}.\label{tHooft:case2}
\end{equation}
For $\Delta_1=\fft{5}{12}$,
\begin{center}
	\footnotesize
	\begin{tabular}{ |c|c|c|c|c| } 
		\hline
		$R_{0,2,1}$ & $R_{0,2,2}$ & $R_{0,2,3}$ & $R_{0,2,4}$ & $R_{0,\log}$ \\
		\hline
		$2.800{\times}10^{-10}$ & $7.156{\times}10^{-11}$ & $2.800{\times}10^{-10}$ & $7.156{\times}10^{-11}$ & $2.116{\times}10^{-8}$ \\
		\hline
	\end{tabular}
\end{center}
%

\noindent\textbf{Case 3. $\Delta=(\Delta_1,\Delta_1,\Delta_1,2-3\Delta_1)$ \& $N=101\sim351~(\text{step }10)$ \& $L=22$}
%
\begin{equation}
	\Delta_1\in\bigg\{\fft13,\fft25,\fft38,\fft{5}{12},\fft37\bigg\}.\label{tHooft:case3}
\end{equation}
For $\Delta_1=\fft38$,
\begin{center}
	\footnotesize
	\begin{tabular}{ |c|c|c|c|c| } 
		\hline
		$R_{0,2,1}$ & $R_{0,2,2}$ & $R_{0,2,3}$ & $R_{0,2,4}$ & $R_{0,\log}$ \\
		\hline
		$2.072{\times}10^{-9}$ & $7.156{\times}10^{-9}$ & $1.457{\times}10^{-9}$ & $7.088{\times}10^{-10}$ & $1.782{\times}10^{-7}$ \\
		\hline
	\end{tabular}
\end{center}
%

\noindent\textbf{Case 4. $\Delta=(\Delta_1,1/2,1/2,1-\Delta_1)$ \& $N=101\sim351~(\text{step }10)$ \& $L=22$}
%
\begin{equation}
	\Delta_1\in\bigg\{\fft25,\fft38,\fft{5}{12},\fft37,\fft{7}{16},\fft49\bigg\}.\label{tHooft:case4}
\end{equation}
For $\Delta_1=\fft37$,
\begin{center}
	\footnotesize
	\begin{tabular}{ |c|c|c|c|c| } 
		\hline
		$R_{0,2,1}$ & $R_{0,2,2}$ & $R_{0,2,3}$ & $R_{0,2,4}$ & $R_{0,\log}$ \\
		\hline
		$2.085{\times}10^{-10}$ & $7.388{\times}10^{-11}$ & $6.716{\times}10^{-11}$ & $5.397{\times}10^{-11}$ & $1.193{\times}10^{-8}$ \\
		\hline
	\end{tabular}
\end{center}
%

\noindent\textbf{Case 5. $\Delta=(\Delta_1,1-\Delta_1,1/2,1/2)$ \& $N=101\sim351~(\text{step }10)$ \& $L=22$}
%
\begin{equation}
	\Delta_1\in\bigg\{\fft25,\fft{5}{12},\fft37,\fft{7}{16},\fft49\bigg\}.\label{tHooft:case5}
\end{equation}
For $\Delta_1=\fft{7}{16}$,
\begin{center}
	\footnotesize
	\begin{tabular}{ |c|c|c|c|c| } 
		\hline
		$R_{0,2,1}$ & $R_{0,2,2}$ & $R_{0,2,3}$ & $R_{0,2,4}$ & $R_{0,\log}$ \\
		\hline
		$1.415{\times}10^{-10}$ & $3.894{\times}10^{-11}$ & $5.159{\times}10^{-11}$ & $5.159{\times}10^{-11}$ & $8.627{\times}10^{-9}$ \\
		\hline
	\end{tabular}
\end{center}
%

\noindent\textbf{Case 6. $\Delta=(\Delta_1,\Delta_2,1-\Delta_2,1-\Delta_1)$ \& $N=101\sim351~(\text{step }10)$ \& $L=22$}
%
\begin{equation}
	(\Delta_1,\Delta_2)\in\bigg\{(\fft25,\fft37),(\fft25,\fft{7}{16}),(\fft25,\fft49),(\fft37,\fft{7}{16}),(\fft37,\fft49),(\fft{7}{16},\fft49)\bigg\}.\label{tHooft:case6}
\end{equation}
For $(\Delta_1,\Delta_2)=(\fft25,\fft49)$,
\begin{center}
	\footnotesize
	\begin{tabular}{ |c|c|c|c|c| } 
		\hline
		$R_{0,2,1}$ & $R_{0,2,2}$ & $R_{0,2,3}$ & $R_{0,2,4}$ & $R_{0,\log}$ \\
		\hline
		$2.890{\times}10^{-10}$ & $1.334{\times}10^{-10}$ & $7.079{\times}10^{-11}$ & $6.933{\times}10^{-11}$ & $1.754{\times}10^{-8}$ \\
		\hline
	\end{tabular}
\end{center}
%

\noindent\textbf{Case 7. $\Delta=(\Delta_1,1-\Delta_1,\Delta_2,1-\Delta_2)$ \& $N=101\sim401~(\text{step }10)$ \& $L=27$}
%
\begin{equation}
	(\Delta_1,\Delta_2)\in\bigg\{(\fft25,\fft37),(\fft25,\fft{7}{16}),(\fft25,\fft49),(\fft37,\fft{7}{16}),(\fft37,\fft49),(\fft{7}{16},\fft49)\bigg\}.\label{tHooft:case7}
\end{equation}
For $(\Delta_1,\Delta_2)=(\fft37,\fft716)$,
\begin{center}
	\footnotesize
	\begin{tabular}{ |c|c|c|c|c| } 
		\hline
		$R_{0,2,1}$ & $R_{0,2,2}$ & $R_{0,2,3}$ & $R_{0,2,4}$ & $R_{0,\log}$ \\
		\hline
		$4.555{\times}10^{-12}$ & $1.111{\times}10^{-12}$ & $3.556{\times}10^{-12}$ & $1.121{\times}10^{-12}$ & $4.695{\times}10^{-10}$ \\
		\hline
	\end{tabular}
\end{center}
%

\noindent\textbf{Case 8. $\Delta=(\Delta_1,\Delta_2,\Delta_3,\Delta_4)$ \& $N=101\sim401~(\text{step }10)$ \& $L=27$}
%
\begin{equation}
	\Delta\in\bigg\{(\fft{4}{10},\fft{5}{10},\fft{4}{10},\fft{7}{10}),(\fft{3}{10},\fft{5}{10},\fft{4}{10},\fft{8}{10}),(\fft{15}{40},\fft{17}{40},\fft{21}{40},\fft{27}{40}),(\fft1\pi,\fft2\pi,\fft{3}{2\pi},2-\fft{9}{2\pi})\bigg\}.\label{tHooft:case8}
\end{equation}
For $\Delta=(\fft{15}{40},\fft{17}{40},\fft{21}{40},\fft{27}{40})$,
\begin{center}
	\footnotesize
	\begin{tabular}{ |c|c|c|c|c| } 
		\hline
		$R_{0,2,1}$ & $R_{0,2,2}$ & $R_{0,2,3}$ & $R_{0,2,4}$ & $R_{0,\log}$ \\
		\hline
		$5.123{\times}10^{-11}$ & $1.913{\times}10^{-11}$ & $1.005{\times}10^{-11}$ & $9.723{\times}10^{-12}$ & $3.511{\times}10^{-9}$ \\
		\hline
	\end{tabular}
\end{center}
%


\bibliography{ABJM-TTI}
\bibliographystyle{JHEP}

\end{document}